\documentclass[a4paper,11pt]{article}

\usepackage{hyperref}

\usepackage{graphicx,amsmath,amssymb,epsf}
\usepackage{bm}
\usepackage{color,cite}
\usepackage{epstopdf}

\usepackage{tikz}
\usetikzlibrary{shapes,arrows,positioning}

\makeatletter
\@addtoreset{equation}{section}
\makeatother

\newcommand{\be}{\begin{equation}}
\newcommand{\ee}{\end{equation}}
\newcommand{\ben}{\begin{equation}}
\newcommand{\een}{\end{equation}}
\newcommand{\bea}{\setlength\arraycolsep{2pt} \begin{eqnarray}}
\newcommand{\eea}{\end{eqnarray}}
\newcommand{\nnr}{\nonumber \\}

\newcommand{\eq}[1]{(\ref{#1})}

\newcommand{\fr}{\frac}
\newcommand{\tf}{\tfrac}

\newcommand{\wtd}{\widetilde}

\newcommand{\df}{\textrm{d}}
\newcommand{\expe}[1]{\textrm{e}^{#1}}
\newcommand{\pd}{\partial}
\newcommand{\sr}{\sqrt}

\newcommand{\gc}{\gamma}
\newcommand{\gC}{\Gamma}
\newcommand{\gd}{\delta}
\newcommand{\gD}{\Delta}
\newcommand{\gep}{\epsilon}

\newcommand{\gz}{\zeta}
\newcommand{\gq}{\theta}

\newcommand{\gl}{\lambda}
\newcommand{\gL}{\Lambda}

\newcommand{\gs}{\sigma}
\newcommand{\gS}{\Sigma}

\newcommand{\gvf}{\varphi}
\newcommand{\gw}{\omega}

\newcommand{\gf}{\phi}

\newcommand{\im}{\textrm{i}}
\newcommand{\Imag}{\textrm{Im} \,}
\newcommand{\Real}{\textrm{Re} \,}
\newcommand{\SL}{\textrm{SL}}
\newcommand{\SO}{\textrm{SO}}
\newcommand{\Sp}{\textrm{Sp}}

\newcommand{\SU}{\textrm{SU}}

\DeclareMathOperator{\Tr}{Tr}

\newcommand{\bbI}{\mathbb{I}}

\newcommand{\bbR}{\mathbb{R}}

\newcommand{\cA}{\mathcal{A}}
\newcommand{\cF}{\mathcal{F}}
\newcommand{\cL}{\mathcal{L}}
\newcommand{\cM}{\mathcal{M}}
\newcommand{\cN}{\mathcal{N}}
\newcommand{\cQ}{\mathcal{Q}}
\newcommand{\cV}{\mathcal{V}}

\newcommand{\so}{\mathfrak{so}}

\newcommand{\tzeta}{\wtd{\zeta}}

\newcommand{\p}{\partial}
\newcommand{\eps}{\epsilon}

\newcommand{\three}{\textrm{(3d)}}

\DeclareFontFamily{U}{mathx}{\hyphenchar\font45}
\DeclareFontShape{U}{mathx}{m}{n}{
      <5> <6> <7> <8> <9> <10>
      <10.95> <12> <14.4> <17.28> <20.74> <24.88>
      mathx10
      }{}
\DeclareSymbolFont{mathx}{U}{mathx}{m}{n}
\DeclareMathAccent{\widecheck}{0}{mathx}{"71}

\begin{document}

\thispagestyle{empty}

\begin{flushright}
\end{flushright}

\begin{center}
{\Large \bf
Black holes in $\mathcal N = 8$ supergravity\vspace{8pt}\\
from $\SO(4, 4)$ hidden symmetries}
 \vspace*{0.5cm}\\
 {David D. K. Chow and Geoffrey Comp\`ere
 }
\end{center}
 \vspace*{0.1cm}

 \begin{center}
 {\it
 Physique Th\'eorique et Math\'ematique,\\
Universit\'e Libre de Bruxelles and International Solvay Institutes\\
Campus Plaine C.P.\ 231, B-1050 Bruxelles, Belgium
 }
\vspace*{0.5cm}\\
{\tt david.chow@ulb.ac.be} $\quad$ {\tt gcompere@ulb.ac.be}
 \end{center}

\bigskip

\begin{abstract}
{\normalsize \noindent
We detail the construction of the most general asymptotically flat, stationary, rotating, non-extremal, dyonic black hole of the four-dimensional $\mathcal N = 2$ supergravity coupled to 3 vector multiplets that describes the $STU$ model.  It generates through U-dualities the most general asymptotically flat, stationary black hole of $\cN = 8$ supergravity. We develop the solution generating technique based on $\SO(4,4)/\SL(2, \bbR)^4$ coset model symmetries, with an emphasis on the 4-fold permutation symmetry of the gauge fields. We indicate how previously known non-extremal and extremal solutions of the $STU$ model are recovered as limiting cases. Several properties of the general black hole solution are discussed, including its thermodynamics, the quadratic mass formula, the Bogomolny--Prasad--Sommerfield limit, the slow and fast extremal rotating limits, its properties in regards to the Kerr/conformal field theory correspondence, its Killing tensors and the separability of geodesic motion and probe scalars.
}
\vspace{12pt}

\vspace{12pt}
 \end{abstract}

\newpage

\tableofcontents


\section{Introduction}


Black holes are some of the most important non-perturbative objects of quantum gravity.  To understand their fundamental properties, such as their microscopic description, it is essential to have explicit black hole solutions and understand all their classical properties, such as their thermodynamics.  In four-dimensional Einstein--Maxwell theory, the Kerr--Newman solution represents a general stationary, asymptotically flat black hole.  More general theories, such as those arising from string theory, admit more general families of black hole solutions.  One of the most studied string theory compactifications down to 4 dimensions is the reduction of M-theory on $T^7$, which is described in the low-energy regime by maximal $\cN = 8$ supergravity \cite{Cremmer:1978ds, Cremmer:1979up}.  The bosonic sector, which is relevant for classical solutions, includes Einstein--Maxwell theory as a truncation, and also includes several other well-studied theories of gravity coupled to vectors and scalars.  A number of black hole solutions of $\cN = 8$ supergravity and its truncations have been discovered over the last 35 years, but the most general family had proved elusive.  In this paper, we give a derivation of the most general stationary, asymptotically flat black hole of $\cN = 8$ supergravity in a specific U-duality frame, as announced in \cite{Chow:2013tia}.

$\cN = 8$ supergravity admits a consistent truncation to an $\cN = 2$ supergravity coupled to three vector multiplets, which is known as the $STU$ model \cite{Cremmer:1984hj,Duff:1995sm} ($S$, $T$ and $U$ are sometimes used to denote its three complex scalar fields).  The $STU$ supergravity is particularly useful because a suitable 5-charge solution of $STU$ supergravity suffices to generate the general black hole of $\cN = 8$ supergravity through U-dualities \cite{Sen:1994eb, Cvetic:1996zq}.  Solutions of $STU$ supergravity can also be used to generate solutions of pure $\cN > 2$ supergravities and heterotic supergravity \cite{Cvetic:1996zq}.  Such U-dualities only act on the matter fields, while leaving the four-dimensional metric invariant.

While $\cN = 8$ supergravity admits an $\textrm{E}_{7(7)}(\mathbb R)$ symmetry of its field equations, the $STU$ supergravity action has an $\SL(2,\mathbb R)^3$ symmetry, and also symmetry under permutations of the three $\SL(2, \bbR)$ factors, which is commonly referred to as the ``$S$-$T$-$U$'' triality symmetry \cite{Duff:1995sm}.  Upon dimensional reduction along time, the classical symmetry of the action enhances to $\SO(4,4)$, which contains an $\SL(2,\mathbb R)^4$ subgroup.  The extra $\SL(2, \bbR)$ is associated with the Ehlers $\SL(2, \bbR)$ that arises from reduction of Einstein gravity \cite{Ehlers:1957aa, Ehlers:1959aa}.

$\cN = 8$ supergravity has been of considerable interest recently thanks to the identification of elegant ultraviolet cancellations in perturbation theory, see e.g.\ \cite{Bern:2009kd}.  Using Kawai--Lewellen--Tye (KLT) relations \cite{Kawai:1985xq}, amplitudes in $\cN = 8$ supergravity are related to amplitudes in $\cN = 4$ super-Yang--Mills theory.  The latter theory is finite \cite{Mandelstam:1982cb, Howe:1984aa}, prompting speculation that $\cN = 8$ supergravity might be finite.  However, pure $\cN = 8$ supergravity cannot be decoupled from string theory \cite{Green:2007zzb}, contrary to $\cN = 4$ super-Yang--Mills theory \cite{Maldacena:1997re}.

The entropy of extremal  black holes in $\mathcal N = 8$ supergravity is related to qubit entanglement measures in quantum information systems, as reviewed in \cite{Borsten:2008wd, Borsten:2012fx}.  There have been in particular studies of the $STU$ supergravity, which corresponds to entanglement of three qubits \cite{Duff:2006uz, Kallosh:2006zs, Levay:2006kf} and four qubits \cite{Levay:2010ua, Borsten:2010db, Borsten:2011is}.  More generally, extremal black hole entropy in $\cN = 8$ supergravity corresponds to tripartite entanglement of seven qubits \cite{Duff:2006ue}.

Ungauged $\cN = 8$ supergravity can be generalized to gauged $\cN = 8$ supergravities.  Whereas the ungauged theory admits a Minkowski vacuum solution, the gauged theories admit anti-de Sitter (AdS) vacuum solutions, so are relevant for studying the AdS/CFT correspondence \cite{Maldacena:1997re}.  The gauged theories have attracted recent interest because the original $\cN = 8$, $\SO(8)$ gauged supergravity \cite{deWit:1981eq, deWit:1982ig}, previously thought to be unique, has been generalized to a one-parameter family of $\cN = 8$ gauged supergravities \cite{Dall'Agata:2012bb}.  Black holes in the ungauged $\cN = 8$ supergravity provide a starting point for finding black holes of the gauged $\cN = 8$ supergravity.  Systematic solution generating techniques, which work for the ungauged theory, fail for the gauged theories.  Therefore, finding solutions of the gauged theory requires guesswork based on solutions of the ungauged theory. For some recent results in this direction, see \cite{Chow:2013gba,Gnecchi:2013mja,Lu:2014fpa} and references therein.

It is conceptually straightforward to find complicated charged black hole solutions of interest, such as the most general black hole of $\cN =8$ supergravity, given the existence of well-known algorithms and suitable uncharged black hole solutions, but it can be a difficult algebraic task.  A common method of generating charged, stationary black holes is to dimensionally reduce the theory on the time coordinate to give Euclidean 3-dimensional gravity coupled to matter.  After Hodge dualizing three-dimensional vectors to scalars, the resulting bosonic matter Lagrangian typically consists of a coset model of scalar fields, which admits symmetries forming a real Lie algebra.  A solution can then be generated starting from an initial seed solution and acting on it with coset model symmetries.  In this paper we will detail the coset model based on $\SO(4,4)$ symmetries and use it to obtain general black holes.  The four-dimensional $STU$ supergravity has four gauge fields on an equal footing.  In this paper, we will present a formulation of the $\SO(4,4)$ coset model that keeps the permutation symmetry between the four gauge fields manifest.

The conceptual foundations of coset model symmetries have been known for years \cite{Marcus:1983hb, Breitenlohner:1987dg}. The main interest of such symmetries, when considering spacelike reductions down to 4 dimensions only, is their role as symmetries of string theory after quantization \cite{Sen:1995aa} (see e.g.\ \cite{Giveon:1994fu, Youm:1997hw, Vafa:1997pm,Obers:1998fb} for reviews).  Attempts have been made to similarly understand symmetries appearing in timelike reductions, which led to string theories in mixed time signatures \cite{Hull:1998br, Hull:1998vg}, but it is not clear if such theories can be quantized. In the case of reductions down to 3 dimensions, it has been conjectured that the classical symmetry group is quantized in string theory \cite{Hull:1994ys,Julia:85aa} but only partial indications have been obtained in this direction \cite{Damour:2002cu,Englert:2003zs}. In this paper we will only treat symmetries classically as a solution generating technique.  A classification of the symmetries appearing in torus reductions of various maximal supergravities (on both space and time) has been performed \cite{Cremmer:1997ct, Cremmer:1998px, Breitenlohner:1998cv, Cremmer:1999du}.  Explicit algorithms for particular cosets have been developed extensively over the years, starting from the pioneering work on Einstein gravity \cite{Ehlers:1957aa, Ehlers:1959aa}, understood in terms of an $\SL(2, \bbR)$ coset \cite{Geroch:1971aa}, and on Einstein--Maxwell theory \cite{Harrison:1968aa}, understood in terms of an $\SU(2,1)$ coset \cite{Kinnersley:1973aa, Kinnersley:1977pg, Neugebauer:1969wr, Mazur:1983vi}.  Other theories considered are Kaluza--Klein theory, understood in terms of an $\SL(3, \bbR)$ coset \cite{Maison:1979kx}; the particular Einstein--Maxwell--dilaton--axion theory used for generating solutions of $\mathcal N = 4$ supergravity written in terms of a $\Sp(4,\mathbb R)$ coset \cite{Galtsov:1994pd, Gal'tsov:1994bn, Gal'tsov:1995na, Clement:1996nh, Gal'tsov:1996cm, Galtsov:1995zu, Gal'tsov:1997kp, Chen:1999rv}; 5d minimal supergravity, which admits $\textrm{G}_{2 (2)}$ symmetries in \cite{Mizoguchi:1998wv, Cremmer:1999du, Mizoguchi:1999fu, Possel:2003rk, Bouchareb:2007ax, Clement:2007qy, Berkooz:2008rj, Tomizawa:2008qr, Compere:2009zh}; and $STU$ supergravity in 4 and 5 dimensions, which admits $\SO(4,4)$ symmetries \cite{Chong:2004na, Gal'tsov:2008nz, Gal'tsov:2008sh, Bossard:2009we, Gal'tsov:2009da}.  For the full $\cN = 8$ supergravity, reduction to 3 Euclidean dimensions gives the maximal $\cN = 16 $ supergravity theory \cite{Marcus:1983hb} with 128 scalars parameterizing the coset $\textrm{E}_{8 (8)}/\SO^*(16)$ \cite{Julia:1980gr, Nicolai:1986jk,Breitenlohner:1987dg}. 

The stationary asymptotically flat black hole which generates, under U-dualities, all single-centered, stationary black holes of $\cN = 8$ supergravity has been presented in \cite{Chow:2013tia}. The main purpose of this article is to present the details of the solution and its generation from $\SO(4,4)$ hidden symmetries. The solution generalizes previously known subcases \cite{demianskinewman, Galtsov:1994pd, Rasheed:1995zv, Cvetic:1995kv, Cvetic:1996kv, Matos:1996km, Larsen:1999pp, LozanoTellechea:1999my, Chong:2004na, Giusto:2007tt, Bena:2009ev, Compere:2010fm}; see also \cite{Cvetic:1995bj,Behrndt:1996hu,Bertolini:2000ei,Gimon:2007mh,Bellucci:2008sv, Dall'Agata:2010dy, Song:2011ii, Bossard:2012ge, Ortin:2012gg} for extremal branches.  It admits 8 independent electromagnetic charges (4 electric and 4 magnetic), in addition to mass and angular momentum (the generalization with Newman--Unti-Tamborino (NUT) charge is considered as well).  Since there are 4 gauge fields on an equal footing, it is simpler to present explicitly the more general solution with 8 independent charges rather than a 5-charge solution.  Moreover, keeping the NUT charge on the same footing as the mass allows for a simplifying $\SO(2)$ symmetry that can be broken as a final step to specialize to asymptotically flat black holes.

Many physical properties of the general solution are as expected from  its known subcases, such as the Kerr--Newman black hole.  There are generically two horizons.  The asymptotically flat solution obeys the first law of thermodynamics and the Smarr relation.  The formal first law of thermodynamics and Smarr relation at the inner horizon also hold.  The product of areas of the outer and inner horizons is quantized, i.e.\ independent of the mass.  This product is proportional to the sum of the angular momentum squared and the Cayley hyperdeterminant, which is a quartic invariant of the electromagnetic charges.  Rotating extremal limits exist, with both fast and slow rotation.  The black hole entropy takes the expected chiral Cardy form in these extremal cases and the near-horizon limits have the expected $\SL(2,\mathbb R)$ enhanced symmetry. Supersymmetric black holes with finite horizon area are recovered in a specific non-rotating extremal limit.

We show that in a different conformal frame, the metric belongs to a class of spacetimes admitting a Killing--St\"{a}ckel tensor, similar to all other known charged generalizations of the Kerr black hole \cite{Chow:2008fe}.  Consequently, the geodesics of the conformally related metric are completely integrable, and the null geodesics in Einstein frame are completely integrable.  The massless Klein--Gordon equation is separable around the general stationary asymptotically flat black hole of $\cN = 8$ supergravity obtained from our solution by U-dualities.

The entropy of extremal black holes in $\cN = 8$ supergravity is known to have a simple expression \cite{Kallosh:1996uy} in terms of the Cartan--Cremmer--Julia quartic $\textrm{E}_{7(7)}$ invariant, which is constructed solely from the electromagnetic charges.  Here, we derive the formula for entropy of the non-extremal black hole, and show that it cannot be expressed as a function of the usual $\textrm{E}_{7(7)}$ invariants, namely the quartic invariant, the mass and angular momentum. Instead, the entropy of the generic non-extremal stationary asymptotically flat black hole of $\cN=8$ supergravity depends upon an additional $\textrm{E}_{7(7)}$ invariant that remains to be understood.  We identify this quantity for black holes in the U-duality frame of the $STU$ model in terms of auxiliary parameters that are also used to parameterize the conserved charges of the black hole.  In specific subcases including the dyonic Kerr--Newman black hole and the dyonic, rotating Kaluza--Klein black hole, we are able to provide the explicit expression for the invariant and therefore the entropy in terms of conserved charges.

The rest of the paper is organized as follows.  We present the relevant supergravity theories in Section \ref{action}.  We outline the solution generating technique based on $\SO(4,4)$ symmetries in Section \ref{gentec}, and then apply it to the particular case of a Kerr--Taub--NUT seed solution in Section \ref{charBH}.  We summarize the general resulting solution in Section \ref{gensol}, and present its physical properties in Section \ref{phys}.  Then we discuss particular limits of the general solution, recovering known non-extremal solutions in Section \ref{nonextremal} and finding some extremal limits in Section \ref{part}.  In Section \ref{Killingtensors}, we consider a more general class of metrics, discuss Killing tensors and the separability of geodesic motion and the Klein--Gordon equation.  We conclude in Section \ref{conclusion}.


\section{\texorpdfstring{$STU$}{STU} supergravity}
\label{action}


Four-dimensional maximal $\mathcal N = 8$ supergravity can be obtained from $T^7$ reduction of 11-dimensional supergravity, via 10-dimensional type IIA supergravity.  The bosonic fields of $\cN = 8$ supergravity are the metric, 28 $\textrm{U}(1)$ gauge fields, and 70 scalar fields parameterizing $\textrm{E}_{7 (7)} / \SU(8)$.  To obtain a generating solution for the most general black hole of $\cN = 8$ supergravity, global symmetries of the field equations (classical U-dualities) imply that it suffices to truncate to a theory with only 4 gauge fields \cite{Cvetic:1996zq}.  The relevant supergravity theory, sometimes called the $STU$ model, is an $\cN = 2$ supergravity coupled to 3 vector multiplets.  Each vector multiplet contains a gauge field, a dilaton, and an axion.  The fourth gauge field belongs to the $\cN = 2$ supergravity multiplet.  Together, the bosonic fields are the metric, four $\textrm{U} (1)$ gauge fields $A^I$, three dilatons $\gvf_i$ and three axions $\chi_i$.  We label the gauge fields by $I = 1, 2, 3, 4$, and label the dilatons and axions by $i = 1, 2, 3$.  It is convenient to denote\footnote{The literature has various conventions; our previous papers \cite{Chow:2013tia, Chow:2013gba} stated $x_i = - \chi_i$, but used only $\chi_i$. The convention in \cite{Virmani:2012kw} is $x_i = -\chi_i$.} 
\begin{align}
x_i & = \chi_i , & y_i & = \expe{-\gvf_i} ,
\end{align}
which can be united as a complex scalar
\be
z_i = x_i + \im y_i .
\ee
The scalars parameterize $(\SL(2, \bbR)/\textrm{U}(1))^3$.  These complex scalars are sometimes denoted $S, T, U$, hence the name ``$STU$ supergravity''.

Since we are in 4 dimensions, the gauge fields $A^I$ may be dualized to dual gauge fields $\widetilde{A}_I$.  The field strengths are $F^I = \df A^I$ and the dual field strengths are $\wtd{F}_I = \df \wtd{A}_I$.  We use the terminology ``electric'' and ``magnetic'' according to the nature of the gauge fields $A^I$.  Note that other literature often uses the terms ``electric'' and ``magnetic'' differently, depending on the choice of duality frame.

We choose a duality frame so that there is a 4-fold symmetry of the gauge fields $A^I$.  One way to understand this is that the original gauged generalization of the theory, the original maximal $\cN = 8$, $\SO(8)$ gauged supergravity, arises from $S^7$ reduction of 11-dimensional supergravity \cite{deWit:1982ig}.  An abelian truncation then gives $\cN = 2$, $\textrm{U}(1)^4$ gauged supergravity \cite{Cvetic:1999xp}.  The four $\textrm{U}(1)$ gauge fields originate from the $\textrm{U}(1)^4$ Cartan subgroup of the full $\SO(8)$ gauge group, explaining why the four gauge fields $A^I$ are on an equal footing.  Taking the ungauged limit then gives the $STU$ supergravity.  Furthermore, setting all the gauge fields equal as $A^1 = A^2 = A^3 = A^4$, with vanishing scalars, recovers Einstein--Maxwell theory.

The Lagrangian in terms of $(A^1, \wtd{A}_2, \wtd{A}_3, A^4)$ is relatively short,
\begin{align}
\cL_4 & = R \star 1 - \fr{1}{2} \sum_{i = 1}^3 (\star \df \gvf_i \wedge \df \gvf_i + \expe{2 \gvf_i} \star \df \chi_i \wedge \df \chi_i) - \fr{1}{2} \expe{-\gvf_1} (\expe{\gvf_2 + \gvf_3} \star \cF^1 \wedge \cF^1 + \expe{\gvf_2 - \gvf_3} \star \wtd{\cF}_2 \wedge \wtd{\cF}_2 , \nnr
& \quad + \expe{- \gvf_2 + \gvf_3} \star \wtd{\cF}_3 \wedge \wtd{\cF}_3 + \expe{- \gvf_2 - \gvf_3} \star \cF^4 \wedge \cF^4) + \chi_1 (F^1 \wedge F^4 + \wtd{F}_2 \wedge \wtd{F}_3) ,
\label{Lagrangian}
\end{align}
where $\cF^I$ and $\wtd{\cF}_I$ are field strengths modified by ``transgression'' terms, given by
\begin{align}
\cF^1 & = F^1 + \chi_3 \wtd{F}_2 + \chi_2 \wtd{F}_3 - \chi_2 \chi_3 F^4 , & \cF^4 & = F^4 , \nnr
\wtd{\cF}_2 & = \wtd{F}_2 - \chi_2 F^4 , & \wtd{\cF}_3 & = \wtd{F}_3 - \chi_3 F^4 .
\end{align}
Note that the parity-odd terms can also be written as $\chi_1 (\cF^1 \wedge \cF^4 + \wtd{\cF}_2 \wedge \wtd{\cF}_3)$.  After relabelling and changing the signs of some axions, this matches the Lagrangian of \cite{Cvetic:1999xp, Chong:2004na}\footnote{Our field strengths are related to the hatted field strengths of \cite{Chong:2004na} by $\cF^1 = \widehat F_{{2}}$, $\wtd{\cF}_2 = \widehat F_{1}$, $\wtd{\cF}_3 = \widehat{\cF}^1$, $\cF^4 = \widehat{\cF}^2$ and the signs of $\chi_1$ and $\chi_3$ are flipped while the one of $\chi_2$ is kept fixed.}.  A further advantage of this formulation is that it comes directly from $T^2$ reduction of a 6-dimensional supergravity, given in Section \ref{6d}.

It is also useful to write the Lagrangian \eqref{Lagrangian} in the general form
\be
\cL_4 = \df ^4 x \, \sqrt{-g} [R - \tfrac{1}{2} f_{AB}(\Phi)
	\partial_\mu \Phi^A \, \partial^\mu \Phi^B - \tfrac{1}{4} k_{IJ}(\Phi) \mathbf{F}^I_{\mu\nu} \mathbf{F}^{J\mu\nu}
	+ \tfrac{1}{4} h_{IJ} (\Phi) \epsilon^{\mu\nu\rho\sigma} \mathbf{F}^I_{\mu\nu} \mathbf{F}^J_{\rho\sigma} ] ,
\label{generalaction}
\ee
where $\Phi^A = (\varphi_1,\varphi_2,\varphi_3,\chi_1,\chi_2,\chi_3)$ are the scalar fields, and $\mathbf{A}^I = (A^1, \widetilde A_2, \widetilde A_3, A^4)$ are the $\textrm{U}(1)$ gauge fields, with field strengths $\mathbf{F}^I = \df \mathbf{A}^I$. The kinetic coefficients are
\begin{align}
f_{AB} & = \text{diag}(1, 1, 1, \expe{2\varphi_1}, \expe{2\varphi_2}, \expe{2\varphi_3}), & h_{IJ} & = - \frac{\chi_1}{2}
\begin{pmatrix}
0 & 0 & 0 & 1\\
0 & 0 & 1 & 0\\
0 & 1 & 0 & 0\\
1 & 0 & 0 & 0
\end{pmatrix} ,
\end{align}
and $k_{IJ}$ is a longer expression that can be easily deduced from the Lagrangian \eqref{Lagrangian}.


\subsection{Symmetries: \texorpdfstring{$\SL(2, \bbR)$}{SL(2,R)} and triality}


We define the three matrices of scalars $\mathcal M_i$ as (see e.g. \cite{Duff:1995sm})
\be
\mathcal M_i = \frac{1}{y_i}
\begin{pmatrix}
1 & x_i \\
x_i & x_i^2 + y_i^2
\end{pmatrix}
=
\begin{pmatrix}
\expe{\gvf_i} &  \chi_i \expe{\gvf_i} \\
 \chi_i \expe{\gvf_i} & \expe{-\gvf_i} + \chi_i^2 \expe{\gvf_i}
\end{pmatrix}
.
\ee
The scalar matrix $\mathcal M_i$ transforms under the classical $\SL(2,\mathbb R)_1 \times \SL(2,\mathbb R)_2 \times \SL(2,\mathbb R)_3$ U-dualities in the trivial representation for two out of the three $\SL(2,\mathbb R)$ groups.  For the non-trivial corresponding $\SL(2,\mathbb R)_i$ group, it transforms as
\be
\mathcal M_i \rightarrow \omega_i^T \mathcal M_i \omega_i ,
\label{Mimatrix}
\ee
where $\omega_i \in \SL(2, \bbR)_i$, given by
\begin{align}
\omega_i & = 
\begin{pmatrix}
d & b \\
c & a
\end{pmatrix}
, & a d - b c & = 1 .
\end{align}
In the quantum theory, $a,b,c,d$ are integers.  The scalar kinetic terms of the Lagrangian may be written as
\be
\cL_{\textrm{scalar}} = - \fr{1}{2} \sum_{i = 1}^3 (\star \df \gvf_i \wedge \df \gvf_i + \expe{2 \gvf_i} \star \df \chi_i \wedge \df \chi_i) = \fr{1}{4} \sum_{i = 1}^3 \Tr (\star \df \cM_i^{-1} \wedge \df \cM_i) ,
\ee
which is manifestly invariant under $\SL(2, \bbR)^3$ and under permutation of the three pairs of scalars.  Note that if the scalars $(\varphi_i,\chi_i)$, $i=1,2,3$ vanish at infinity, then $\cM_i = \bbI + O(1/r)$.

More generally, one can show that the equations of motion of the Lagrangian \eqref{Lagrangian} can be written in a form manifestly invariant under $\SL(2, \bbR)^3$ and under permutation of the three copies of $\SL(2, \bbR)$. The symmetry is however not manifest in the action \eqref{Lagrangian}. However, there exist three actions that each make manifest a pair of $\SL(2, \bbR)$ symmetries and that only differ by dualizations of gauge fields \cite{Duff:1995sm}.  In this sense, the theory described by \eqref{Lagrangian} admits a triality symmetry. 


\subsection{Dualization}


There are several other formulations of $STU$ supergravity that appear in the literature, corresponding to different duality frames.  To obtain these, we need relations between gauge fields $F^I$ and dual gauge fields $\wtd{F}_I$, for each $I$.  We introduce the dual gauge potential as a Lagrange multiplier to enforce the Bianchi identity for the original gauge field strength, and then vary with respect to the original field strength. To dualize $F^1$ to $\wtd{F}_1$, we add to the Lagrangian \eq{Lagrangian} an extra term
\be
- \wtd{A}_1 \wedge \df F^1 = - \wtd{F}_1 \wedge F^1 + \df (\wtd{A}_1 \wedge F^1) .
\ee
Varying the modified Lagrangian with respect to $F^1$, we see that $F^1$ and $\wtd{F}^1$ are related by
\be
\wtd{F}_1 - \chi_1 F^4 = -\expe{-\gvf_1 + \gvf_2 + \gvf_3} \star \cF^1 .
\label{F1dual}
\ee
Similarly, $F^4$ and $\wtd{F}_4$ are related by
\be
\wtd{F}_4 - \chi_1 F^1 = \expe{-\gvf_1} (- \expe{- \gvf_2 - \gvf_3} \star \cF^4 + \chi_2 \chi_3 \expe{\gvf_2 + \gvf_3} \star \cF^1 + \chi_2 \expe{\gvf_2 - \gvf_3} \star \wtd{\cF}_2 + \chi_3 \expe{- \gvf_2 + \gvf_3} \star \wtd{\cF}_3) .
\label{F4dual}
\ee
To dualize $\wtd{F}_2$ to $F^2$, we instead add to the Lagrangian \eq{Lagrangian} an extra term
\be
A^2 \wedge \df \wtd{F}_2 = F^2 \wedge \wtd{F}_2 - \df (A^2 \wedge \wtd{F}_2) ,
\ee
and similarly for dualizing $\wtd{F}_3$ to $F^3$.  We see that $F^2$ and $F^3$ are related to $\wtd{F}_2$ and $\wtd{F}_3$ by
\begin{align}
F^2 + \chi_1 \wtd{F}_3 & = \expe{-\gvf_1 + \gvf_2}(\expe{- \gvf_3} \star \wtd{\cF}_2 + \chi_3 \expe{\gvf_3} \star \cF^1) , \nnr
F^3 + \chi_1 \wtd{F}_2 & = \expe{-\gvf_1 + \gvf_3}(\expe{- \gvf_2} \star \wtd{\cF}_3 + \chi_2 \expe{\gvf_2} \star \cF^1) .
\label{F23dual}
\end{align}

To obtain a dual Lagrangian, we take the original Lagrangian modified by adding the extra term, and then substitute in the algebraic relation between a gauge field strength and its dual.  Applying the procedure to replace $F^1$ in favour of $\wtd{F}_1$, we obtain the Lagrangian in terms of $(\wtd{A}_1, \wtd{A}_2, \wtd{A}_3, A^4)$,
\begin{align}
\cL_4 & = R \star 1 - \fr{1}{2} \sum_{i = 1}^3 (\star \df \gvf_i \wedge \df \gvf_i + \expe{2 \gvf_i} \star \df \chi_i \wedge \df \chi_i) - \fr{1}{2} \expe{ -\gvf_1 - \gvf_2 - \gvf_3} \star F^4 \wedge F^4 \nnr
& \quad - \fr{1}{2} \sum_{i = 1}^3 \expe{2 \gvf_i - \gvf_1 - \gvf_2 - \gvf_3} \star (\wtd{F}_i - \chi_i F^4) \wedge (\wtd{F}_i - \chi_i F^4) + \chi_1 \chi_2 \chi_3 F^4 \wedge F^4 \nnr
& \quad - (\chi_1 \chi_2 \wtd{F}_3 + \chi_2 \chi_3 \wtd{F}_1 + \chi_3 \chi_1 \wtd{F}_2) \wedge F^4 + \chi_1 \wtd{F}_2 \wedge \wtd{F}_3  +  \chi_2 \wtd{F}_3 \wedge \wtd{F}_1 + \chi_3 \wtd{F}_1 \wedge \wtd{F}_2 .
\label{tilde123Lagrangian}
\end{align}
An advantage of this Lagrangian is that there is a manifest symmetry between 3 gauge fields, and it fits into a more general prepotential formalism for $\cN = 2$ supergravity coupled to vector multiplets, as discussed later in Section \ref{Prepotential}.

The Lagrangian \eq{tilde123Lagrangian} gives duality relations involving $(\wtd{F}_1, \wtd{F}_2, \wtd{F}_3, F^4)$, namely
\begin{align}
\expe{\gvf_1 - \gvf_2 - \gvf_3} \star (\wtd{F}_1 - \chi_1 F^4) & = F^1 + \chi_3 \wtd{F}_2 + \chi_2 \wtd{F}_3 - \chi_2 \chi_3 F^4 , \nnr
\expe{\gvf_2 - \gvf_3 - \gvf_1} \star (\wtd{F}_2 - \chi_2 F^4) & = F^2 + \chi_1 \wtd{F}_3 + \chi_3 \wtd{F}_1 - \chi_3 \chi_1 F^4 , \nnr
\expe{\gvf_3 - \gvf_1 - \gvf_2} \star (\wtd{F}_3 - \chi_3 F^4) & = F^3 + \chi_2 \wtd{F}_1 + \chi_1 \wtd{F}_2 - \chi_1 \chi_2 F^4 ,
\label{dualF123}
\end{align}
and
\begin{align}
\wtd{F}_4 & = - \expe{-\gvf_1 - \gvf_2 - \gvf_3} \star F^4 + \sum_{i = 1}^3 \expe{2 \gvf_i - \gvf_1 - \gvf_2 - \gvf_3} \chi_i \star (\wtd{F}_i - \chi_i F^4) + 2 \chi_1 \chi_2 \chi_3 F^4 \nnr
& \quad - (\chi_2 \chi_3 \wtd{F}_1 + \chi_3 \chi_1 \wtd{F}_2 + \chi_1 \chi_2 \wtd{F}_3) .
\label{dualF4bis}
\end{align}
Alternatively, these can be obtained from the duality relations involving $(F^1, \wtd{F}_2, \wtd{F}_3, F^4)$ that arise from the first Lagrangian \eq{Lagrangian}.


\subsection{Prepotential formalism}
\label{Prepotential}


Any $\cN = 2$ supergravity coupled to vector multiplets can be derived from a prepotential in a certain duality frame.   We first define the gauge field and dual gauge field
\begin{align}
A^0 & \equiv - \wtd{A}_4 , & \wtd{A}_0 & \equiv A^4 .\label{defAA}
\end{align}
In this formalism, $STU$ supergravity has complex scalars $X^\gL$, $\gL = 0, 1, 2, 3$ and gauge fields $\widetilde{F}_\gL = \df \widetilde{A}_\gL$ 
for $\gL = 0, 1, 2, 3$.  The Lagrangian is
\be
\cL_4 = R \star 1 - 2 g_{i \overline{j}} \star \df X^i \wedge \df \overline{X}^{\overline{j}} + \tf{1}{2} \widetilde{F}_\gL \wedge \widetilde{G}^{\gL} ,
\ee
where $g_{i \overline{j}} = \pd_i \pd_{\overline{j}} K$ is a K\"{a}hler metric derived from a K\"{a}hler potential $K$, and $\widetilde{G}^{\gL}$ depends on $\widetilde{F}_\gL$ and its dual.  The prepotential is 
\be
F(X)=- \frac{X^1 X^2 X^3 }{ X^0}.
\ee
One may define complex scalars $z_i = X^i/X^0$, fix the gauge $X^0 = 1$, and relate $z_i = x_i + \im y_i = \chi_i + \im \expe{-\gvf_i}$. For more details, see e.g.\ \cite{Virmani:2012kw}.   



\subsection{Truncations}

\begin{figure}[!hbt]
\includegraphics{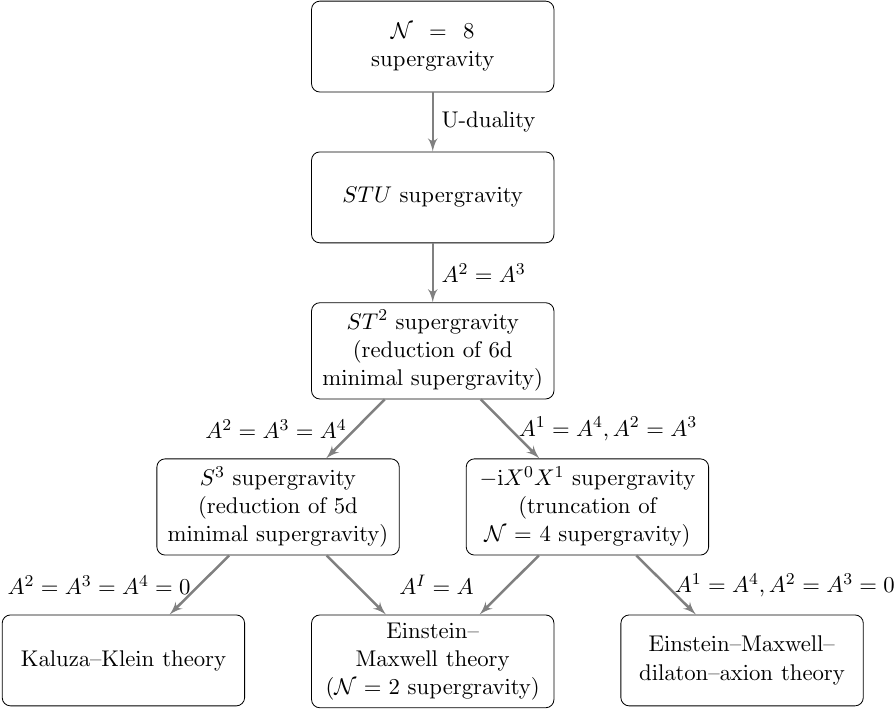}
\caption{Bosonic truncations of $\cN = 8$ supergravity}
\label{truncations}
\end{figure}

Some special cases of our general black hole solutions are already known in the literature.  Most of these are solutions of theories that are consistent bosonic truncations of the $STU$ model, and some of these are bosonic truncations of other supergravity theories.  We therefore review these truncations (see also \cite{Clement:2013fc}).  The relationships between these truncations are indicated in Figure \ref{truncations}.

\subsubsection{\texorpdfstring{$S T^2$}{ST2} supergravity}


There is a consistent truncation of $STU$ supergravity to an $\cN = 2$ supergravity coupled to two vector multiplets. We refer to it as $ST^2$ supergravity, since it involves setting the complex scalars $T=U$ in $STU$ supergravity.  There are 3 independent gauge fields, 2 dilatons and 2 axions.  It is obtained by setting $A^2 = A^3$, $\gvf_2 = \gvf_3$ and $\chi_2 = \chi_3$, which implies that $\wtd{A}_2= \wtd{A}_3$.  The theory can be obtained by reduction of 5-dimensional supergravity coupled to a vector multiplet, as discussed in Section \ref{5d}.  This theory admits $\SO(2,2) \sim \SL(2,\mathbb R) \times \SL(2,\mathbb R)$ symmetries which get enhanced to $\SO(4,3)$ upon dimensional reduction to 3 dimensions \cite{Lavrinenko:1998hf,Clement:2013fc}.


\subsubsection{\texorpdfstring{$S^3$}{S3} supergravity}


There is a consistent truncation of $STU$ supergravity to an $\cN = 2$ supergravity coupled to one vector multiplet.  This is sometimes known as $S^3$ supergravity (or $T^3$ supergravity), since the truncation of the $STU$ supergravity includes setting the three complex scalars equal, $S = T = U$.  There are 2 independent gauge fields, 1 dilaton and 1 axion.  It is obtained by setting equal the fields in each of the three vector multiplets of $STU$ supergravity, namely $A/\sr{3} \equiv A^1 = A^2 = A^3$, $\gvf/\sr{3} \equiv \gvf_1 = \gvf_2 = \gvf_3$ and $\chi/\sr{3} \equiv \chi_1 = \chi_2 = \chi_3$.  The $(\wtd{A}_1, \wtd{A}_2, \wtd{A}_3, A^4)$ Lagrangian \eq{tilde123Lagrangian} becomes
\begin{align}
\cL_4 & = R \star 1 - \frac{1}{2} \star \df \gvf \wedge \df \gvf - \frac{1}{2} \expe{2 \gvf/\sr{3}} \star \df \chi \wedge \df \chi - \frac{1}{2} \expe{-\gvf/\sr{3}} \star (\wtd{F} - \chi F^4) \wedge (\wtd{F} - \chi F^4) \nnr
& \quad - \fr{1}{2} \expe{- \sr{3} \gvf} \star F^4 \wedge F^4 + \fr{\chi}{\sr{3}} \bigg( \wtd{F} \wedge \wtd{F} - \chi \wtd{F} \wedge F^4 + \fr{\chi^2}{3} F^4 \wedge F^4 \bigg) .
\label{S3Lagrangian}
\end{align}
It can be obtained by reduction of 5-dimensional minimal supergravity, as discussed in Section \ref{5d}. The 3 dimensional action obtained by dimensional reduction has $G_{2(2)}$ symmetries. 


\subsubsection{Kaluza--Klein theory}


A further consistent bosonic truncation of $S^3$ supergravity is Kaluza--Klein theory, i.e.\ the reduction to 4 dimensions of 5-dimensional Einstein gravity.  This comes from \eq{S3Lagrangian} by taking $\wtd{A} = 0$ and $\chi = 0$.  Relabelling $A^4 \rightarrow A$, the Lagrangian is
\be
\cL_4 = R \star 1 - \tf{1}{2} \star \df \gvf \wedge \df \gvf - \tf{1}{2} \expe{- \sr{3} \gvf} \star F \wedge F .
\label{KKLagrangian}
\ee
The symmetry group obtained upon dimensional reduction to 3 dimensions is $\SL(3,\mathbb R)$. 


\subsubsection{\texorpdfstring{$-\im X^0 X^1$}{-i X0 X1} supergravity}


A different set of consistent truncations from $STU$ supergravity comes from setting the 4 gauge fields pairwise equal.  From the Lagrangian \eq{Lagrangian}, we set $A^1 = A^4$, $\wtd{A}_2 = \wtd{A}_3$, and $\gvf_2 = \gvf_3 = \chi_2 = \chi_3 = 0$, giving the Lagrangian
\be
\cL_4 = R \star 1 - \tf{1}{2} \star \df \gvf \wedge \df \gvf - \tf{1}{2} \expe{2 \gvf} \star \df \chi \wedge \df \chi - \expe{-\gvf} (\star F^1 \wedge F^1 + \star \wtd{F}_2 \wedge \wtd{F}_2) + \chi (F^1 \wedge F^1 + \wtd{F}_2 \wedge \wtd{F}_2) ,
\label{iX0X1Lagrangian}
\ee
where $\gvf \equiv \gvf_1$ and $\chi \equiv \chi_1$.  This is the bosonic truncation of an $\cN = 2$ supergravity coupled to one vector mutiplet. This theory is also known in the literature as the EM$_2$DA theory \cite{Gal'tsov:1996cm,Gal'tsov:1997kp}. An important use is to generate solutions of $\cN = 4$ supergravity, since it is a truncation of the $\SU(4)$ formulation of $\cN = 4$ supergravity \cite{Cremmer:1977tt}.  By dualizing $\wtd{F}_2$ to $F^2$, or equivalently making a symplectic transformation, the theory is equivalent to that obtained from a prepotential $F(X) = - \im X^0 X^1$ \cite{Ceresole:1995jg}.  A truncation of the $\SO(4)$ formulation of $\mathcal N = 4$ supergravity \cite{Das:1977uy, Cremmer:1977aa} corresponds to the dual formulation \cite{Kallosh:1992ii}. Upon dimensional reduction to 3 dimensions, the theory admits $\SU(2,2) \sim \SO(4,2)$ symmetries. 


\subsubsection{Einstein--Maxwell--dilaton--axion theory}


A further consistent bosonic truncation of the $-\im X^0 X^1$ supergravity has just one gauge field.  We take $\wtd{A}_2 = 0$ in \eq{iX0X1Lagrangian}, so the Lagrangian is
\be
\cL_4 = R \star 1 - \tf{1}{2} \star \df \gvf \wedge \df \gvf - \tf{1}{2} \expe{2 \gvf} \star \df \chi \wedge \df \chi - \expe{-\gvf} \star F \wedge F + \chi F \wedge F ,
\label{EMDALagrangian}
\ee
where $F = F^1$.  This is sometimes known as Einstein--Maxwell--dilaton--axion (EMDA) theory or dilaton--axion gravity.  Again, the theory is used when generating solutions of $\cN = 4$ supergravity. Upon dimensional reduction to 3 dimensions, the theory admits $Sp(4,\mathbb R) \sim SO(3,2)$ symmetries. 


\subsubsection{Einstein--Maxwell theory}


Einstein--Maxwell theory corresponds to setting the gauge fields equal, $A = A^1 = A^2 = A^3 = A^4$, and the scalars trivial, $\gvf_i = \chi_i = 0$.  The Lagrangian is
\be
\cL_4 = R \star 1 - 2 \star F \wedge F .
\label{EMLagrangian}
\ee
It is the bosonic sector of pure $\cN = 2$ supergravity. Upon dimensional reduction to 3 dimensions, the theory admits $SU(2,1)$ symmetries. 


\subsection{Oxidation to higher dimensions}


Some special cases of our general black hole solutions have been discussed in the literature with a higher-dimensional interpretation.  For example, a 4-dimensional black hole can be regarded as a 5-dimensional homogeneous black string.  Also, the embedding in 10-dimensional or 11-dimensional supergravity allows for a microscopic interpretation of black holes in terms of string theory or M-theory and its web of dual theories.  We therefore quickly review several oxidations of 4-dimensional $STU$ supergravity into higher-dimensional theories.  A review of the lift to 5 and 6 dimensions, including truncations and a generalization to an $\SO(5, 4)$ coset model, is \cite{Clement:2013fc}.


\subsubsection{Uplift to 5 dimensions}
\label{5d}


The Lagrangian \eq{tilde123Lagrangian} has a direct uplift to a  5-dimensional $\cN = 2$ supergravity coupled to 2 vector multiplets, also known as the $STU$ model or 5-dimensional  $\textrm{U}(1)^3$ supergravity \cite{Cremmer:1984hj,Duff:1995sm}.  This 5-dimensional theory has 3 gauge fields $\wtd{A}_i$, $i = 1, 2, 3$ on an equal footing. The Lagrangian is
\be
\cL_5 = R \star 1- \frac{1}{2} \sum_{i = 1}^3 h_i^{-2} (\star \df h_i \wedge \df h_i + \star \wtd{F}_i \wedge \wtd{F}_i) + \wtd{F}_1 \wedge \wtd{F}_2 \wedge \wtd{A}_3,\label{LU13}
\ee
subject to the constraint that $h_1 h_2 h_3=1$.  A common parameterization of the scalars is
\begin{align}
h_1 & = \expe{-\gvf'_1 / \sr{6} - \gvf'_2/\sr{2}} , & h_2 & = \expe{-\gvf'_1 / \sr{6} + \gvf'_2/\sr{2}} , & h_3 & = \expe{2 \gvf'_1 / \sr{6}} .
\end{align}
Another parameterization of the scalars, which is useful for lifting to 6 dimensions, is
\begin{align}
h_1 & = \expe{2 \phi_2/\sr{6}} , & h_2 & = \expe{\phi/\sr{2} - \phi_2/\sr{6}} , & h_3 & = \expe{-\phi/\sr{2} - \phi_2/\sr{6}} .
\end{align}
The scalar kinetic terms with these parameterizations are
\be
\fr{1}{2} \sum_{i = 1}^3 h_i^{-2} \star \df h_i \wedge \df h_i = \fr{1}{2} \sum_{i = 1}^2 \star \df \gvf'_i \wedge \df \gvf'_i = \fr{1}{2} (\star \df \phi \wedge \df \phi + \star \df \phi_2 \wedge \df \phi_2) .
\ee

We may dualize the third gauge field $\wtd{A}_3$ to a 2-form potential $B$.  The usual dualization procedure gives $\wtd{F}_3 = \df \wtd{A}_3 = - h_1^{-2} h_2^{-2} \star \mathcal H$, where $\df \mathcal H = - \wtd{F}_1 \wedge \wtd{F}_2$, and the Lagrangian is
\be
\cL_5 = R \star 1 - \fr{1}{2} \sum_{i = 1}^3 h_i^{-2} \star \df h_i \wedge \df h_i - \fr{1}{2} \sum_{i = 1}^2 h_i^{-2} \star \wtd{F}_i \wedge \wtd{F}_i - \fr{1}{2} h_1^{-2} h_2^{-2} \star \mathcal H \wedge \mathcal H .
\label{L5H}
\ee

The Kaluza--Klein reduction ansatz is
\begin{align}
\df s^2_5 & = f^{-1} \, \df s^2 + f^2 (\df z_5 - A^4)^2 , & \wtd{A}_{\textrm{(5d)}}{_i} & = \wtd{A}_i + \chi_i (\df z_5 - A^4) .
\end{align}
Three of the four gauge fields $\wtd{A}_i$ are manifestly on an equal footing; the fourth gauge field $A^4$ is the graviphoton.  	Redefining $f h_i = \expe{-\varphi_i}$, the Lagrangian \eq{tilde123Lagrangian} is recovered.

There are some notable consistent truncations.  Setting $\wtd A_2 = \wtd A_3$ and $h_2 = h_3$ gives an $\cN = 2$ supergravity coupled to 1 vector multiplet.  If we set $h_1 = \expe{2 \gvf/\sr{6}}$, then the Lagrangian is
\be\label{LU13trunc}
\cL_5 = R \star 1 - \tf{1}{2} \star \df \gvf \wedge \df \gvf - \tf{1}{2} \expe{-4 \gvf/\sr{6}} \star \wtd{F}_1 \wedge \wtd{F}_1 - \expe{2 \gvf/\sr{6}} \star \wtd{F}_2 \wedge \wtd{F}_2 + \wtd{F}_2 \wedge \wtd{F}_2 \wedge \wtd{A}_1 .
\ee
Reduction to 4 dimensions gives the $S T^2$ supergravity.  A further consistent truncation is to set all gauge fields equal, $\wtd{A}_i = \wtd{A}$, and trivial scalars $h_i = 1$.  This gives the minimal pure $\cN = 2$ supergravity, whose bosonic Lagrangian is
\be
\cL_5 = R \star 1 - \tf{3}{2} \star \wtd{F} \wedge \wtd{F} + \wtd{F} \wedge \wtd{F} \wedge \wtd{A} .
\ee
Reduction to 4 dimensions gives the $S^3$ supergravity \eq{S3Lagrangian}.


\subsubsection{Uplift to 6 dimensions}
\label{6d}


The 4-dimensional theory \eq{Lagrangian} has a higher-dimensional origin in minimal 6-dimensional $\cN = (2, 0)$ supergravity coupled to a tensor multiplet. The Lagrangian is
\be
\cL_6 = R \star 1 - \tf{1}{2} \star \df \phi \wedge \df \phi - \tfrac{1}{2} \expe{-\sqrt{2} \phi} \star H \wedge H ,
\label{L6}
\ee
where $H = \df B$ is a 3-form field strength.  

Directly reducing $\cL_6$ on $T^2$, and then dualizing the 4-dimensional 2-form potential $B$ to an axion $\chi_1$ leads to the $(A^1, \wtd{A}_2, \wtd{A}_3, A^4)$ Lagrangian \eq{Lagrangian}.  If instead the 2-form $B$ is dualized to a vector in 5 dimensions, and then reduced to 4 dimensions, then we obtain the $(\wtd{A}_1, \wtd{A}_2, \wtd{A}_3, A^4)$ Lagrangian \eq{tilde123Lagrangian}.  Either way, there is the same intermediate 5-dimensional $STU$ supergravity theory in some duality frame.  

Kaluza--Klein reduction of the 6-dimensional theory \eq{L6} directly gives the Lagrangian in terms of $(\wtd{A}_1, \wtd{A}_2, H)$.  We make the reduction ansatz (see e.g. \cite{Pope:Notes})
\begin{align}
\df s^2_{(6 d)} & = \expe{\phi_2/\sr{6}} \df s^2 + \expe{-3 \phi_2/\sr{6}} (\df z_6 + \wtd{A}_1)^2 , & B_{(6 d)} = B + \wtd{A}_2 \wedge (\df z_6 + \wtd{A}_1) ,
\end{align}
decomposing the field strengths as
\begin{align}
H_{(6 d)} & = \mathcal H + \wtd{F}_2 \wedge (\df z_6 + \wtd{A}_1) , & \mathcal H & = \df B - \wtd{A}_2 \wedge \wtd{F}_1 , & \wtd{F}_i & = \df \wtd{A}_i .
\end{align}
This gives 5-dimensional $STU$ supergravity in the form \eq{L5H}.  The 5-dimensional fields $\wtd{F}_1$ and $\phi_2$ come from reduction of the Einstein--Hilbert term; $H$ and $\wtd{F}_2$ come from reduction of the 6-dimensional $H$.

There is a consistent truncation of the 6-dimensional theory \eq{L6} to the minimal pure $\cN = (2, 0)$ supergravity by setting $\phi = 0$ and imposing the constraint that $H$ is self-dual, 
\be
H = \star H .
\ee
The theory is obtained from the Lagrangian
\be
\cL_6 = R \star 1 - \tf{1}{2} \star H \wedge H ,
\ee
with the self-duality condition imposed on the resulting field equations.  Upon dimensional reduction to 5 dimensions, the latter condition is equivalent to $\wtd{A}_2 = \wtd{A}_3$ and $h_2=h_3$.  The resulting 5-dimensional theory is therefore given by \eqref{LU13trunc}.


\subsubsection{Uplift to 10 dimensions}
\label{10d}

The 6-dimensional supergravity action \eqref{L6} naturally uplifts to a consistent truncation of Type IIB supergravity on $T^4$. The non-trivial 10-dimensional fields are the metric $g_{\mu\nu}$, the Ramond-Ramond two-form $C$ and the dilaton $\Phi$. The reduction ansatz is 
\begin{align}
\df s_{10}^2 & = \df s_6^2 + \expe{\phi / \sqrt{2}} (\df X_1^2 + \df X_2^2 + \df X_3^2 + \df X_4^2), & \Phi & = \frac{\phi}{\sqrt{2}}, & C & \equiv B. 
\end{align}


\subsubsection{Uplift to 11 dimensions}
\label{11d}


The 5-dimensional $STU$ supergravity can be embedded in 11-dimensional supergravity as follows. The action of 11-dimensional supergravity is
\be
\cL_{11} = R \star 1 - \tfrac{1}{2}\star \cF \wedge  \cF -\tfrac{1}{6} \cF \wedge \cF \wedge \mathcal A,\label{L11}
\ee
where $\mathcal A$ is the 3-form and $\cF = \df \cA$ its 4-form field strength. We Kaluza--Klein reduce on $T^6$ as (see e.g.\ \cite{Emparan:2006mm})
\begin{align}
\df s^2_{11} & = \df s^2_5 + h_1 (\df X_1^2 + \df X_2^2) + h_2 (\df X_3^2 + \df X_4^2) + h_3 (\df X_5^2 + \df X_6^2 ) , \nnr
\mathcal A & = \widetilde A_1 \wedge \df X_1 \wedge \df X_2 + \widetilde A_2 \wedge \df X_3 \wedge \df X_4 + \widetilde A_3 \wedge \df X_5 \wedge \df X_6 ,
\end{align}
with the constraint that $h_1 h_2 h_3 =1$ in order that $T^6$ has constant volume. The 11-dimensional action \eqref{L11} then reduces to the 5-dimensional action \eqref{LU13}.


\section{Generating technique}
\label{gentec}


Ungauged supergravity theories have global symmetries that can be used for solution generating techniques.  When considering solutions with Killing vectors, one may dimensionally reduce the theories, leading to enhanced symmetries.  If a 4-dimensional solution has a timelike Killing vector field, then we may perform a timelike dimensional reduction to a 3-dimensional theory.  It has been generally shown that, if the 4-dimensional theory is gravity coupled to scalars parameterizing a symmetric space $\overline{G}/\overline{K}$ (a feature of all supergravity theories with enough supersymmetry) and vectors transforming in a representation of $\overline{G}$, then the 3-dimensional theory is a theory of gravity coupled to scalars that parameterize a larger symmetric space $G/K$ \cite{Breitenlohner:1987dg}.  In particular, the 3-dimensional symmetric space is $\SO(4, 4)/\SL(2, \bbR)^4$ for the $STU$ model.  These coset model techniques are described in for example \cite{Breitenlohner:1987dg, Maison:2000fj, Bossard:2009at}, and in the particular case of $\SO (4,4)$ in \cite{Chong:2004na, Gal'tsov:2008nz, Bossard:2009we}.  The reduction down to three dimensions was already worked out explicitly for the $STU$ model in terms of the so-called $c^*$-map, as done for example in \cite{Gaiotto:2007ag}.

There are other solution generating techniques available, but the reduction to 3 dimensions is particularly efficient.  For example, an alternative method is to lift to higher dimensions, perform boosts to add charges, reduce back to 4 dimensions, apply permutations of gauge fields and electromagnetic duality, and repeat, but this requires multiple steps.  Reduction to 3 dimensions is advantageous because the solution generating technique is essentially a one-step process once an appropriate group element has been identified.


\subsection{Reduction to 3 dimensions}
\label{reduction3d}


After Hodge dualizing 3-dimensional vectors to scalars, the 3-dimensional theory corresponding to the $STU$ model is a theory of euclidean-signature gravity coupled to 16 scalars: a scalar $U$ corresponding to $g_{t t}$; a scalar $\gs$ dual to the Kaluza--Klein vector; 8 electromagnetic scalars $\gz^I$ and $\wtd{\gz}_I$; 3 dilatons $y_i = \expe{-\gvf_i}$; and 3 axions $x_i =  \chi_i$.  The 8 scalars $\{ U, \sigma, x_i, y_i \}$ arising from the 4-dimensional metric and scalars have the usual positive sign kinetic terms, whereas the 8 scalars $\{ \gz^I, \wtd{\gz}_I \}$ arising from the 4-dimensional vectors have negative sign kinetic terms. The scalars parameterize a symmetric space $G/K = \SO(4, 4) / \SL(2, \bbR)^4$.

Let us first present the 3-dimensional Lagrangian in terms of the 16 scalars, before explaining the relationship to 4-dimensional fields.  The set of 3-dimensional (pseudo)scalar fields is $\varphi^a = \{ U, \sigma, x_i, y_i, \zeta^I, \wtd{\zeta}_I \}$. They parameterize the target space of the coset model whose Lagrangian is
\be
\cL_3 = R \star_3 1 - \tf{1}{2} G_{ab} \, \partial_\mu \varphi^a \, \partial^\mu \varphi^b \star_3 1 .
\ee
The 3-dimensional moduli space metric $G_{a b}$ is of the form
\begin{align}
\df s^2_{G/K} & = \sum_i \fr{\df x_i^2 + \df y_i^2}{y_i^2} + 4 \, \df U^2 + \fr{\expe{-4 U}}{4} \bigg( \df \sigma + \sum_I (\wtd{\gz}_I \, \df \gz^I - \gz^I \, \df \wtd{\gz}_I) \bigg) ^2 \nnr
& \qquad - \expe{- 2 U} \sum_{I, J} \bigg( \fr{Y Y_I Y_J}{X_{I J}} \df \gz^I \, \df \gz^J + \fr{Y}{X_{I J} Y_I Y_J} \, \df \wtd{\gz}_I \, \df \wtd{\gz}_J + \fr{X X_{I J} Y_I}{Y_J} 2 \, \df \gz^I \, \df \wtd{\gz}_J \bigg) .
\label{scalarmetric}
\end{align}
$X_{I J}$ is symmetric, $X_{I J} = X_{(I J)}$, and obeys the ``self-duality'' conditions $X_{1 2} = X_{3 4}$, $X_{1 3} = X_{2 4}$, $X_{2 3} = X_{1 4}$, with
\begin{align}
X_{1 2} & = \fr{\sr{(x_1^2 + y_1^2) (x_2^2 + y_2^2)}}{x_1 x_2} , & X_{1 3} & = \fr{\sr{(x_1^2 + y_1^2) (x_3^2 + y_3^2)}}{x_1 x_3} , \nnr
X_{2 3} & = \fr{\sr{(x_2^2 + y_2^2) (x_3^2 + y_3^2)}}{x_2 x_3} , & X_{1 1} & = X_{2 2} = X_{3 3} = X_{4 4} = 1 .
\end{align}
The remaining functions are
\begin{align}
Y_i & = \fr{\sr{x_i^2 + y_i^2}}{[(x_1^2 + y_1^2) (x_2^2 + y_2^2) (x_3^2 + y_3^2)]^{1/4}} , & Y_4 & = - [(x_1^2 + y_1^2) (x_2^2 + y_2^2) (x_3^2 + y_3^2)]^{1/4} , \nnr
X & = \fr{x_1 x_2 x_3}{y_1 y_2 y_3} , & Y & = \fr{\sr{(x_1^2 + y_1^2) (x_2^2 + y_2^2) (x_3^2 + y_3^2)}}{y_1 y_2 y_3} .
\end{align}
They obey  the constraints 
\begin{align}
Y_1 Y_2 Y_3 Y_4 & = - 1 , & X^2 X_{1 2} X_{1 3} X_{1 4} & = Y^2 , & \fr{1}{X^2} & = \bigg( \fr{Y^2}{X^2 X_{1 2}^2} - 1 \bigg) \bigg( \fr{Y^2}{X^2 X_{1 3}^2} - 1 \bigg) \bigg( \fr{Y^2}{X^2 X_{1 4}^2} - 1 \bigg) .
\end{align}

From varying with respect to $\sigma$, we have the field equation
\be
\df \bigg[ \expe{-4 U} \star_3 \bigg( \df \sigma + \sum_I (\wtd{\gz}_I \, \df \gz^I - \gz^I \, \df \wtd{\gz}_I)  \bigg)  \bigg] = 0 .
\ee
We may therefore dualize the scalar $\sigma$ in favour of a 1-form potential $\omega_3$ through the relation
\be
\df \omega_3 = - \fr{\expe{-4 U}}{2} \star_3 \bigg( \df \sigma + \sum_I (\wtd{\gz}_I \, \df \gz^I - \gz^I \, \df \wtd{\gz}_I)  \bigg) .\label{d1}
\ee
Similarly, we may dualize the electromagnetic scalars $\gz^I$ and $\wtd{\gz}_I$ to 1-form potentials $A^I_\three$ and $\wtd{A}_{I \three}$ through
\begin{align}
\df A^I_\three & = - \gz^I \, \df \omega_3 + \expe{-2 U} \star_3 \sum_J \bigg( \fr{Y}{X_{I J} Y_I Y_J} \df \wtd{\gz}_J + \fr{X X_{I J} Y_J}{Y_I} \df \gz^J \bigg) , \nnr
\df \wtd{A}_{I \three} & = - \wtd{\gz}_I \, \df \omega_3 - \expe{-2U} \star_3 \sum_J \bigg( \fr{Y Y_I Y_J}{X_{I J}} \df \gz^J + \fr{X X_{I J} Y_I}{Y_J} \df \wtd{\gz}_J \bigg) .
\label{d2}
\end{align}

The 4-dimensional fields of $STU$ supergravity are reconstructed as follows. The metric is
\ben
\label{metricansatz}
\df s^2 = - \expe{2 U} (\df t + \gw_3)^2 + \expe{- 2 U} \, \df s^2_3 ,
\een
and the gauge fields and dual gauge fields are
\begin{align}
A^I & = \gz^I (\df t + \omega_3) + A^I_\three , & \wtd{A}_I & = \wtd{\gz}_I (\df t + \omega_3) + \wtd{A}_{I \three} .
\label{vectoransatz}
\end{align}
The dilatons $\gvf_i$ and axions $\chi_i$ are the same in both 3 and 4 dimensions.  

We have presented the 3-dimensional theory in a manner that emphasizes the 4-fold symmetry of the gauge fields.  Other treatments in the literature dualize various 4-dimensional gauge fields, so use different notations in 3 dimensions. Consistently with \eqref{defAA}, we define
\begin{align}
\gz^0 & \equiv - \wtd{\gz}_4 , & \wtd{\gz}_0 & \equiv \gz^4 ,
\label{relI}
\end{align}
and $\zeta^\Lambda = (\zeta^0,\zeta^1,\zeta^2,\zeta^3)$, $\wtd{\zeta}_\Lambda = (\wtd{\zeta}_0,\wtd{\zeta}_1,\wtd{\zeta}_2,\wtd{\zeta}_3)$. Then the scalar metric $G_{a b}$ takes the form (see e.g.\ \cite{Bossard:2009we,Virmani:2012kw})
\begin{align}
\df s^2_{G/K} & = \sum_i \fr{\df x_i^2 + \df y_i^2}{y_i^2} + 4 \, \df U^2 + \fr{\expe{- 4 U}}{4} (\df \sigma - \gz^\gL \, \df \wtd{\gz}_\gL + \wtd{\zeta}_\gL \, \df \gz^\gL)^2 + \expe{- 2 U} [(\Imag \cN)^{\gL \gS} \, \df \wtd\gz_\gL \, \df \wtd\gz_\gS \nonumber \\
& \quad + ((\Imag \cN)^{-1})_{\gL \gS} (\df {\gz}^\gL - (\Real \cN)^{\gL \gC} \, \df \wtd \gz_\gC) (\df {\gz}^\gS - (\Real \cN)^{\gS \gD} \, \df \wtd\gz_\gD)] .
\end{align}
The period matrix $\cN_{\Lambda \Sigma}$ is symmetric and given by (see e.g. \cite{Gimon:2009gk})\footnote{Our conventions relates to the ones of \cite{Bossard:2009we,Virmani:2012kw} as $(\zeta_{\textrm{ours}}^\gL, \widetilde \zeta^{\textrm{ours}}_\gL) = (- \widetilde \zeta^{\textrm{theirs}}_\gL, \zeta_{\textrm{theirs}}^\gL)$. Also, with respect to our conventions $y_i$ and $F$ are defined in \cite{Gimon:2009gk} with an opposite sign.}
\be
\cN =
\begin{pmatrix}
-2 x_1 x_2 x_3 - \im y_1 y_2 y_3 \big( 1 + \sum_{i=1}^3 \frac{x_i^2}{y_i^2} \big) & x_2 x_3 + \im \frac{x_1 y_2 y_3}{y_1} & x_1 x_3 + \im \frac{x_2 y_1 y_3}{y_2} & x_1 x_2 + \im \frac{x_3 y_1 y_2}{y_3} \\
x_2 x_3 + \im \frac{x_1 y_2 y_3}{y_1} & - \im \frac{y_2 y_3}{y_1} & - x_3 & -x_2 \\
x_1 x_3 + \im \frac{x_2 y_1 y_3}{y_2} & - x_3 & - \im \frac{y_1 y_3}{y_2} & - x_1 \\
x_1 x_2 + \im \frac{x_3 y_1 y_2}{y_3} & -x_2 & -x_1 & -\im \frac{y_1 y_2}{y_3} \end{pmatrix}
.
\ee
Note that if the scalars vanish at infinity, then $\cN = - \im \, \bbI + O(1/r)$.  As shown in \cite{Gaiotto:2007ag}, the pseudoscalar $\sigma$ dual to $\omega_3$ is given by
\be
\label{sigmadual}
\df \omega_3 = - \tf{1}{2} \expe{-4U} \star_3 (\df \sigma + \wtd\zeta_{\Lambda} \, \df {\zeta}^{\Lambda} - {\zeta}^{\Lambda} \, \df \wtd\zeta_{\Lambda}) .
\ee
The dualization relations for the 3-dimensional gauge fields and dual gauge fields are
\begin{align}
\label{zetadual}
\df A^\gL_{\three} & = - {\gz}^\gL \, \df \gw_3 - \expe{-2U} \star_3 [ (\Imag \cN)^{\gL \gS} \df \wtd\gz_\gS + (\Real \cN)^{\gL \gC} ((\Imag \cN)^{-1})_{\gC \gS} (\df {\gz}^\gS -(\Real \cN)^{\gS \gD} \, \df \wtd\gz_\gD) ] , \nnr
\df \widetilde{A}_{\Sigma \three} & = - \wtd\zeta_{\Sigma} \, \df \omega_3 + \expe{-2U} \star_3 ((\Imag \cN)^{-1})_{\Sigma \Lambda} (\df \zeta^\Lambda - (\Real \cN)^{\Lambda \Sigma} \, \df \wtd\zeta_{\Sigma}) .
\end{align}
These dualities are equivalent to the dualities \eqref{d1} and \eqref{d2}.  

To match the notation of \cite{Chong:2004na}, which essentially dualizes two of the gauge fields, apply the previous changes of $x_i$ and $y_i$, and let
\begin{align}
(\gz^1, \wtd{\gz}_1) & = (\sigma_2, -\psi_2) , & (\gz^2, \wtd{\gz}_2) & = (\psi_1, \sigma_1) , & (\gz^3, \wtd{\gz}_3) & = (\psi_3, \sigma_3) , & (\gz^4, \wtd{\gz}_4) & = (\sigma_4, -\psi_4) ,
\end{align}
and
\be
\gs = - 2 \chi_4 - \gz^1 \wtd{\gz}_1 + \gz^2 \wtd{\gz}_2 + \gz^3 \wtd{\gz}_3 - \gz^4 \wtd{\gz}_4 .
\ee


\subsection{Parameterizing \texorpdfstring{$\mathfrak{so} (4,4)$}{so(4,4)}}


We choose an explicit parameterization of the Lie algebra $\so (4, 4)$ as given in \cite{Bossard:2009we}.  However, to make the 4-fold permutation symmetry of the gauge fields manifest, we make some notational changes.  We have the 4 Cartan generators
\begin{align}
H_0 & = E_{3 3} + E_{4 4} - E_{7 7} - E_{8 8} ,& H_1 & = E_{3 3} - E_{4 4} - E_{7 7} + E_{8 8} , \nnr
H_2 & = E_{1 1} + E_{2 2} - E_{5 5} - E_{6 6} ,& H_3 & = E_{1 1} - E_{2 2} - E_{5 5} + E_{6 6} ,
\label{Cartangenerators}
\end{align}
12 positive-root generators
\begin{align}
E_0 & = E_{4 7} - E_{3 8} ,& E_1 & = E_{8 7} - E_{3 4} ,& E_2 & = E_{2 5} - E_{1 6} ,& E_3 & = E_{6 5} - E_{1 2} , \nnr
E^{Q_1} & = E_{4 5} - E_{1 8} , & E^{Q_2} & = E_{3 2} - E_{6 7} , & E^{Q_3} & = E_{3 6} - E_{2 7} , & E^{Q_4} & = E_{4 1} - E_{5 8} , \nnr
E^{P^1} & = E_{5 7} - E_{3 1} , & E^{P^2} & = E_{4 6} - E_{2 8} , & E^{P^3} & = E_{4 2} - E_{6 8} , & E^{P^4} & = E_{1 7} - E_{3 5} ,
\label{positiverootgenerators}
\end{align}
and 12 negative-root generators
\begin{align}
F_0 & = E_{7 4} - E_{8 3} ,& F_1 & = E_{7 8} - E_{4 3} ,& F_2 & = E_{5 2} - E_{6 1} ,& F_3 & = E_{5 6} - E_{2 1} , \nnr
F^{Q_1} & = E_{5 4} - E_{8 1} , & F^{Q_2} & = E_{2 3} - E_{7 6} , & F^{Q_3} & = E_{6 3} - E_{7 2} , & F^{Q_4} & = E_{1 4} - E_{8 5} , \nnr
F^{P^1} & = E_{7 5} - E_{1 3} , & F^{P^2} & = E_{6 4} - E_{8 2} ,& F^{P^3} & = E_{2 4} - E_{8 6} , & F^{P^4} & = E_{7 1} - E_{5 3} ,
\label{negativerootgenerators}
\end{align}
where $E_{i j}$ is the $8 \times 8$ matrix with 1 in the $(i, j)$ component, and zeros elsewhere.  Our generators $(E^{Q_I}, E^{P^I}, F^{Q_I}, F^{P^I})$ are related to the generators $(E_{q_\Lambda}, E_{p^\Lambda}, F_{q_\Lambda}, F_{p^\Lambda})$ of \cite{Bossard:2009we} by
\begin{align}
(E_{q_i}, E_{p^i}) & = (E^{P^i}, - E^{Q_i}) , & (E_{q_0}, E_{p^0}) & = (E^{Q_4}, E^{P^4}) , \nnr
(F_{q_i}, F_{p^i}) & = (F^{P^i}, - F^{Q_i}) , & (F_{q_0}, F_{p^0}) & = (F^{Q_4}, F^{P^4}) ,
\end{align}
whilst we use the same notation for the generators $H_\Lambda$, $E_\Lambda$ and $F_\Lambda$.

The generalized transpose $^\sharp$ is defined to act on the generators as
\begin{align}
H_\gL^\sharp & = H_\gL , & E_\gL^\sharp & = F_\gL , & F_\gL^\sharp & = E_\gL ,
\end{align}
and
\begin{align}
(E^{Q_I})^\sharp & = - F^{Q_I} , & (E^{P^I})^\sharp & = - F^{P^I} , & (F^{Q_I})^\sharp & = - E^{Q_I} , & (F^{P^I})^\sharp & = - E^{P^I} .
\end{align}
The following are elements of the eigenspace of  the involution $\tau(x) = - x^\sharp$ with eigenvalue $+1$:
\begin{align}
\label{kgenerators}
k_\gL & = E_\gL - F_\gL , & k^{Q_I} & = E^{Q_I} + F^{Q_I} , & k^{P^I} & = E^{P^I} + F^{P^I} ;
\end{align}
and the following have eigenvalue $-1$:
\begin{align}
p_\gL & = E_\gL + F_\gL , & p^{Q_I} & = E^{Q_I} - F^{Q_I} , & p^{P^I} & = E^{P^I} - F^{P^I} .
\end{align}
$k_\gL$, $p^{Q_I}$ and $p^{P^I}$ are compact, and $p_\gL$, $k^{Q_I}$ and $k^{P^I}$ are non-compact.  Equivalently, the generalized transpose $^\sharp$ adapted to the coset is
\ben
\label{generalized transpose}
A^\sharp = \eta A^T \eta^{-1} ,
\een
where the $8 \times 8$ matrix
\be
\eta = \textrm{diag} (-1, -1, 1, 1, -1, -1, 1, 1)
\ee
is the quadratic form preserved by $\mathfrak{sl}(2,\mathbb R)^4 = \mathfrak{so}(2,2)^2$. The explicit generators of the four commuting $\mathfrak{sl}(2,\mathbb R)$ subalgebras were detailed in \cite{Katsimpouri:2013wka,Banerjee:2014hza}.

The symmetric space $G/K$ can then be parametrized by the group element
\begin{align}
\cV & = \exp (- U H_0) \exp (\tf{1}{2} \textstyle \sum_i \gvf_i H_i) \exp ( - \sum_i \chi_i E_i) \exp [ - \sum_I (\gz^I E^{Q_I} + \wtd{\gz}_I E^{P^I})] \exp ( - \tf{1}{2} \gs E_0 ) \nnr
& = \exp (- U H_0) \exp [ - \tf{1}{2} \textstyle \sum_i (\log y_i) H_i ] \exp ( - \sum_i x_i E_i) \exp [  \sum_\gL (-\wtd \gz_\gL E_{q_\gL} + \gz^\gL E_{p^\gL})] \exp ( - \tf{1}{2} \gs E_0 ) .
\label{cosetelement}
\end{align}
The metric on $G/K$ is then the right-invariant metric obtained from the Maurer--Cartan 1-form $\theta = \df \cV \, \cV^{-1}$,
\begin{align}
\df s^2_{G/K} & = \Tr( P_*\; P_*) , & P_* & = \tf{1}{2} ( \theta + \theta^\sharp) .
 \label{dsp}
\end{align}
Equivalently, one can define the matrix
\be
\cM = \cV^\sharp \cV
\label{Mmatrix}
\ee
and the coset Lagrangian is then given by
\be
\label{cosetLagrangian}
- \tf{1}{2} G_{ab} \, \partial_\mu \varphi^a \, \partial^\mu \varphi^b \star_3 1 = -\tf{1}{8} \Tr [\star_3 (\mathcal M^{-1} \, \df \mathcal M ) \wedge  ( \mathcal M^{-1}\, \df \mathcal M )] .
\ee
Either way, we recover the 3-dimensional moduli space of \eq{scalarmetric}.

A group element $g$ acts as
\be
\cV \rightarrow k \cV g
\ee
where $k \in \SL(2,\mathbb R)^4$ is a local compensator, depending on the fields, defined to ensure that the coset element remains in Borel gauge, i.e.\ of the form \eq{cosetelement}.  Since $k^\sharp k = \bbI$, $\cM$ transforms as
\be
\cM \rightarrow g^\sharp \cM g ,
\ee
which is simpler than working with $\cV$, because the compensator is not required.


\subsection{Extracting 3-dimensional fields}



\subsubsection{Scalars}
\label{extractscalars}


The 3-dimensional scalars are determined from the matrix $\cM$ \eq{cosetelement}.  For our choice of $\so(4, 4)$ parameterization, they can be extracted from $\cM$ by inspection, using the following formulae.  The scalar $U$, which corresponds to the $g_{t t}$ component of the metric, is given by
\be
\expe{- 4U} = \cM_{3 3} \cM_{4 4} - \cM_{3 4}^2 .
\label{Ueqn}
\ee
The $i = 1$ dilaton and axion can be extracted from
\begin{align}
x_1 & = \frac{\cM_{3 4}}{\cM_{3 3}} , & y_1^{-1} & = \expe{2U} \cM_{3 3} .
\label{x1y1}
\end{align}
The remaining dilatons and axions are obtained from
\begin{align}
\fr{1}{y_2 y_3} & = \cM_{1 1} + \expe{4 U} (\cM_{3 3} \cM_{4 1}^2 + \cM_{4 4} \cM_{3 1}^2 - 2 \cM_{3 1} \cM_{3 4} \cM_{4 1}) , \nnr
\fr{x_2}{y_2 y_3} & =\cM_{1 6} + \expe{4 U} (\cM_{3 4} \cM_{4 1} \cM_{6 3} + \cM_{3 1} \cM_{3 4} \cM_{6 4} - \cM_{3 1} \cM_{4 4} \cM_{6 3} - \cM_{3 3} \cM_{4 1} \cM_{6 4}) , \nnr
\fr{x_3}{y_2 y_3} & = \cM_{1 2} + \expe{4 U} (\cM_{3 1} \cM_{3 2} \cM_{4 4} + \cM_{3 3} \cM_{4 1} \cM_{4 2} - \cM_{3 1} \cM_{3 4} \cM_{4 2} - \cM_{3 2} \cM_{3 4} \cM_{4 1}) , \nnr
\fr{x_3^2 + y_3^2}{y_2 y_3} & = \cM_{2 2} + \fr{\cM_{3 2}^2}{\cM_{3 3}} + \expe{4 U} \fr{(\cM_{3 2} \cM_{3 4} - \cM_{3 3} \cM_{4 2})^2}{\cM_{3 3}} .
\label{scalarfields}
\end{align}
The electromagnetic scalars $\gz^I$ and $\wtd{\gz}_I$ are obtained from
\begin{align}
\expe{- 4U} {\gz}^1 & = \cM_{3 5} \cM_{3 4}-\cM_{4 5} \cM_{3 3} , & \expe{- 4U} \wtd{\gz}_1 & = \cM_{3 1} \cM_{4 4} - \cM_{4 1} \cM_{3 4} , \nnr
\expe{- 4U} {\gz}^2 & =  \cM_{4 2} \cM_{3 4}-\cM_{3 2} \cM_{4 4} , & \expe{- 4U} \wtd{\gz}_2 & = \cM_{6 4} \cM_{3 3} - \cM_{6 3} \cM_{3 4} , \nnr
\expe{- 4U} {\gz}^3 & = \cM_{6 3} \cM_{4 4} - \cM_{6 4} \cM_{3 4} , & \expe{- 4U}\wtd{\gz}_3 & = \cM_{3 2} \cM_{3 4} - \cM_{4 2} \cM_{3 3} , \nnr
\expe{- 4U} \gz^4 & = \cM_{3 1} \cM_{3 4} - \cM_{4 1} \cM_{3 3} ,& \expe{- 4U} \wtd{\gz}_4 & = \cM_{3 5} \cM_{4 4} - \cM_{4 5} \cM_{3 4} .
\label{electroscalars}
\end{align}
The scalar $\gs$, dual to the Kaluza--Klein vector, is
\begin{align}
\gs & = \fr{2 \cM_{3 8}}{\cM_{3 3}} + \frac{\expe{4 U}}{\cM_{3 3}}  (\cM_{3 3} \cM_{3 5} \cM_{4 1} + \cM_{3 1} \cM_{3 3} \cM_{4 5} + 2 \cM_{3 2} \cM_{3 4} \cM_{6 3} - \cM_{3 3} \cM_{4 2} \cM_{6 3} \nnr
& \quad - \cM_{3 2} \cM_{3 3} \cM_{6 4} - 2 \cM_{3 1} \cM_{3 4} \cM_{3 5}) \nnr
& = \fr{2 \cM_{3 8}}{\cM_{3 3}} - \gz^4 \wtd{\gz}_4 - \gz^1 \wtd{\gz}_1+\gz^2 \wtd{\gz}_2+\gz^3 \wtd{\gz}_3  +2 x_1 \wtd\gz_2 \wtd\gz_3 - 2 x_1 \gz^4 {\gz}^1.
\end{align}
With the exception of $U$, the scalars do not depend on the overall factor in $\cM$ but only on ratios of entries of $\cM$, and in calculations it can be more practical to rescale $\cM$ by a convenient factor.


\subsubsection{Gauge fields}
\label{extractgauge}


Three-dimensional gauge fields can be reconstructed from the 3-dimensional scalars using the dualizations \eq{d1} and \eq{d2}.  It is easier, however, to perform these dualizations initially in terms of the seed solution, and act with the solution generating technique on the gauge fields directly.  This prevents the dualization of complicated expressions.  For $STU$ supergravity, this approach was noted in \cite{Gal'tsov:2008sh}.

From \eq{cosetLagrangian}, the coset matrix $\cM$ obeys the equation of motion $\df (\cM^{-1} \star_3 \df \cM) = 0$. Therefore, we can define the matrix of one-forms $\cN$ as
\be
\df \cN = \cM^{-1} \star_3 \df\cM .
\ee
The coset transformations act on $\cN$ as
\be
\cN \rightarrow g^{-1} \cN g .
\ee

The matrix $\cM^{-1} \, \df \cM$ is a combination of all 28 $\mathfrak{so} (4, 4)$ generators with coefficients that depend on the 3-dimensional scalars.  Some of these coefficients are directly related to 1-form potentials.  In particular, we have
\begin{align}
\df \cN = \cM^{-1} \, \star_3 \df \cM & = \df \gw_3 \, F_0 + \sum_I (\df A_\three^I \, F^{P^I} - \df \wtd{A}_{I \three} \, F^{Q_I}) + \ldots \nnr
& = \df \gw_3 \, F_0 + \sum_\gL (\df \widetilde{A}_{\Lambda \three} \, F_{p^\gL} + \df A^\gL_{\three} \, F_{q_\gL}) + \ldots ,
\end{align}
where the Kaluza--Klein 1-form, gauge fields and dual gauge fields are related to 3-dimensional scalars through \eq{d1} and \eq{d2}.  The dots stand for the terms involving the remaining generators, whose coefficients involve more complicated dependence on the 3-dimensional scalars.  From \eq{negativerootgenerators}, one can extract the Kaluza--Klein 1-form
\be
\label{exomega}
\gw_3 = \cN_{7 4} ,
\ee
and the 3-dimensional electromagnetic 1-forms
\begin{align}
{A}^1_{\three} & = \cN_{7 5} , &{A}^2_{\three} & = \cN_{6 4} , &A^3_{\three} & = \cN_{2 4} , & A^4_{\three} & = \cN_{7 1} , \nnr
\widetilde{A}_{1 \three} & = \cN_{8 1} , & \widetilde{A}_{2 \three} & = \cN_{7 6} , & \widetilde{A}_{3 \three} & = \cN_{7 2} , & \wtd{A}_{4 \three} & = \cN_{8 5} .
\label{gaur}
\end{align}


\subsection{Conserved charges}


Consider solutions that are asymptotically flat, or more generally asymptotically Taub--NUT, with vanishing scalars at infinity.  Taub--NUT spacetime is asymptotically flat at spatial infinity, in the sense that its metric has the appropriate fall-off, so charges may be defined at spatial infinity.  For the metric ansatz, we assume that $\df s^2_3$ is asymptotically euclidean, and take $r$ to be the usual radial coordinate. More precisely, we assume that
\be
\df s_{3}^2 = \df r^2 + (r^2 - 2 m r) (\df \theta^2 + \sin^2 \theta \, \df \phi^2) + O(r^{-2})\df r^2 + O(r^0) \, \df \theta^2 + O(r^0) \, \df \phi^2 , 
\label{ds3}
\ee
where $m$ is a constant. The asymptotic behavior of a solution gives 10 independent conserved charges at first order in the asymptotic radial expansion around Minkowski: mass $M$, NUT charge $N$, 4 electric charges $Q_I$, and 4 magnetic charges $P^I$. There is also the angular momentum $J$ defined at second order in the radial expansion. We define $Q_I$ and $P^I$ to be associated with $A^I$.  There are also 6 scalar charges, dilaton charges $\Sigma_i$ and axion charges $\Xi_i$, but they are not independent for the solutions that we consider.  These 16 charges are encoded in the first-order asymptotic behavior of the 16 3-dimensional scalars $\{ U, \sigma, \gz^I, \wtd{\gz}_I, x_i, y_i \}$, using the reduction ansatzes and dualizations of Section \ref{reduction3d}. The angular momentum $J$ appears in the second-order asymptotic behavior of $\sigma$. 

More precisely, we assume that we have the expansions at infinity
\begin{align}
\expe{2 U} & = 1 - \fr{2 M}{r} + O (r^{-2}) , & \zeta^I & = \frac{Q_I}{r} + O(r^{-2}) , & \gvf_i & = \fr{\Sigma_i}{r} + O(r^{-2}) , \nnr
\omega_3 & = \left(2 N \cos \theta +2J \frac{\sin^2\theta}{r} + O(r^{-2})\right)\df\phi , & \wtd \zeta_I & = \frac{P^I}{r} + O(r^{-2}) , & \chi_i & = \fr{\Xi_i}{r} + O(r^{-2}) .
\label{expsc}
\end{align}
Then $M$ is the canonical Arnowitt--Deser--Misner mass and $J$ is the canonical angular momentum obtained by the standard Komar integral.  We have fixed the gauge so that $\gz^I$ and $\wtd{\gz}_I$ vanish at infinity.

Our convention for the 3-dimensional and 4-dimensional volume forms are $\epsilon_{r\theta \phi} > 0$ and $\epsilon_{t r \theta \phi} > 0$, so that as $r \rightarrow \infty$,\footnote{The 4-dimensional orientation is the same as in \cite{Chong:2004na}, but the opposite of \cite{Virmani:2012kw}.}
\begin{align}
\star_3 1 & \sim r^2 \sin \theta \, \df r \wedge \df \theta \wedge \df \phi , & \star 1 & \sim r^2 \sin \theta \, \df t \wedge \df r \wedge \df \theta \wedge \df \phi .
\label{orientation}
\end{align}

Dualizing $\omega_3$, we have
\be
\star_3 \df \omega_3 = - \frac{2 N}{r^2} \, \df r -\frac{1}{2} \df \bigg( \frac{4J \cos\theta + c}{r^2}\bigg) + O(r^{-3}),
\ee
where $c$ is a constant. The duality relation \eqref{sigmadual} then implies that
\begin{align}
\sigma & = -\frac{4N}{r} + \frac{4J \cos\theta + c}{r^2}+O(r^{-3}).\label{exps}
\end{align}
Therefore, the charges are
\begin{align}
M & = - \lim_{r \rightarrow \infty} (r U) , & Q_I & = \lim_{r \rightarrow \infty} (r {\gz}^I) , & \Sigma_i & = \lim_{r \rightarrow \infty} (r \gvf_i) , & J & = \lim_{r \rightarrow \infty} \bigg( \frac{r(\omega_{3 \phi} - 2N \cos\theta)}{2\sin^2\theta} \bigg) , \nnr
N & = - \frac{1}{4} \lim_{r\rightarrow \infty}(r\sigma) , & P^I & = \lim_{r \rightarrow \infty} (r \wtd\gz_I) , & \Xi_i & = \lim_{r \rightarrow \infty} (r \chi_i) .
\end{align}
For comparison with other duality frames, it is useful to define electromagnetic charges $\widetilde{Q}^I$ and $\widetilde{P}_I$ corresponding to $\widetilde{F}_I$, analogous to the electromagnetic charges $Q_I$ and $P^I$ corresponding to $F^I$.  These electromagnetic charges are related by
\be
(Q_I, P^I) = (-\widetilde P_I, \widetilde Q^I) .
\ee
Charges for $A^0$ are related to charges for $A^4$ by
\begin{align}
(Q_0, P^0) & = (- \wtd{Q}^4, - \wtd{P}_4) , &  (\widetilde Q^0 , \widetilde P_0)  & = (Q_4, P^4) .
\label{QPsymplectic}
\end{align}


\subsection{Charge matrices}


The charge matrix $\cQ$ is defined by a $1/r$ expansion of the matrix $\cM$ as
\be
\cM = \mathbb{I} + \fr{\cQ}{r} +  \fr{\cQ^{(2)}}{r^2} + O (r^{-3}) .
\label{Mexpand}
\ee
Using the definition of $\cM$ in terms of the 3-dimensional scalars and the expansions \eqref{expsc} and \eqref{exps}, the charge matrix is expressed in terms of physical charges as
\begin{align}
\cQ & = 2 M H_0 + 2 N p_0 - \sum_{I = 1}^4 (Q_I p^{Q_I} + P^I p^{P_I}) + \sum_{i = 1}^3 (\Sigma_i H_i - \Xi_i p_i) \nnr
& = 2 M H_0 + 2 N p_0 + \sum_{\Lambda = 0}^3 (- Q_\Lambda p_{p^\Lambda} + P^\Lambda p_{q_\Lambda}) + \sum_{i = 1}^3 (\Sigma_i H_i - \Xi_i p_i) .
\end{align}
From $\cQ$ alone, one may therefore read off the charges without knowing full details of the solution.  Since a group element $g$ acts as $\cM \rightarrow g^\sharp \cM g$, to preserve asymptotic flatness at spatial infinity we should have $g^\sharp g = \bbI$.  For $S^3$ supergravity, the charge matrix has been studied before in \cite{Kim:2010bf}.

Using the generators of \eq{Cartangenerators}, \eq{positiverootgenerators} and \eq{negativerootgenerators}, we have
\be
\frac{1}{4} \Tr (\cQ^2) = 4 (M^2 + N^2) - \sum_{I = 1}^4 [ (Q_I)^2 + (P^I)^2 ] +\sum_{i = 1}^3 ( \Sigma_i^2 + \Xi_i^2 ).
\label{eq:Q2}
\ee
This quantity is invariant under transformations that preserve asymptotic flatness at spatial infinity.

The angular momentum does not appear in the charge matrix $\cQ$, since it enters the $\cM$ expansion \eq{Mexpand} in $\cQ^{(2)}$, at subleading order $1/r^2$.  Using the expansions \eqref{expsc}-\eqref{exps}, one can show that
\be
\cQ^{(2)} = (-2 J \cos\theta + a_0) p_0 + \dots ,
\label{Q2d}
\ee
where $a_0$ is a constant and the dots are the other terms proportional to the Cartan generators $H_\Lambda$ and the Lie algebra generators $p_i,p_{q_\Lambda},p_{p^\Lambda}$ which all have eigenvalue $-1$ under the $\tau$ involution. 

In \cite{Andrianopoli:2012ee} (see also \cite{Andrianopoli:2013kya, Andrianopoli:2013jra}), it was proposed to define the charge matrix integral $\cQ_{\p_\phi}$ as
\be
\cQ_{\p_\phi} \equiv - \fr{3}{8 \pi} \int_{S^2_\infty} \! (\p_\phi)_{\mu} \cM^{-1}\p_\nu \cM \, \df x^\mu \wedge \df x^\nu .
\ee
This may be written as
\be
\cQ_{\p_\phi} = -\fr{3}{4} \int_0^\pi \! \df \theta \, \sin^2\theta \, \p_\theta \cQ^{(2)} = -2J p_0 + \dots ,
\ee
where we used the 3-dimensional line element \eqref{ds3} at the first step and \eqref{Q2d} at the second step. The angular momentum can therefore be extracted from $\cQ_{\p_\phi}$.  The quantity 
\be
\tfrac{1}{16} \Tr (\cQ_{\p_\phi}^2) = J^2 + \dots
\label{Qpsi}
\ee
contains the angular momentum square and is invariant under the action of transformations that preserve asymptotic flatness at spatial infinity.


\section{Charging up the black holes}
\label{charBH}


We apply the solution generating technique to the specific example of the Ricci-flat Kerr--Taub--NUT spacetime \cite{demianskinewman} to obtain dyonic rotating black holes. The resulting solutions of supergravity will in general carry 11 independent parameters, consisting of mass, NUT charge, angular momentum, 4 electric charges and 4 magnetic charges.  It is convenient to keep the NUT charge on the same footing as the mass, which allows for an $\SO(2)$ symmetry that simplifies the solution. When discussing asymptotically flat black holes, we are free to restrict the solution to a 10-parameter family by solving the final zero NUT charge constraint.  This constraint is a linear equation in terms of the NUT charge of the initial seed Kerr--Taub--NUT black hole and is therefore straightforwardly solved.


\subsection{Seed solutions}


We present here the initial seed solutions used in the solution generating technique.


\subsubsection{Taub--NUT seed solution}


Static solutions are obtained by starting with the Taub--NUT spacetime, whose metric is
\ben
\label{TaubNUT}
\df s^2 = - \fr{r^2 - 2 m r - n^2}{r^2 + n^2} (\df t + 2 n \cos \gq \, \df \phi)^2 + \fr{r^2 + n^2}{r^2 - 2 m r - n^2} \, \df r^2 + (r^2 + n^2) (\df \gq^2 + \sin^2 \gq \, \df \phi^2) ,
\een
where $m$ is the mass and $n$ is the NUT charge.  By Kaluza--Klein reduction on the $t$ coordinate, it may be expressed in terms of 3-dimensional fields as
\begin{align}
\expe{- 2 U}& = \fr{r^2 + n^2}{r^2 - 2 m r - n^2} , & \gw_3 & =  2 n \cos \gq \, \df \phi , \nnr
\df s^2_3 & = \df r^2 + (r^2 - 2 m r - n^2) (\df \gq^2 + \sin^2 \gq \, \df \phi^2) .
\end{align}
By Hodge dualizing $\gw_3$, using the orientation \eq{orientation}, we obtain the 3-dimensional scalar
\ben
\gs = - \fr{4 n (r - m)}{r^2 + n^2} .
\een
Since this is a Ricci-flat metric, all other 3-dimensional scalars are trivial.  It is convenient to define the rescaled matrix $\overline \cM = (r^2 - 2 m r - n^2) \cM$, which has polynomial entries.


\subsubsection{Kerr--Taub--NUT seed solution}


Our seed for rotating black holes is the Kerr--Taub--NUT solution \cite{demianskinewman}, which can be written as
\be
\df s^2 = - \fr{R}{r^2 + u^2} \bigg( \df \overline t - \fr{\overline{a}^2 - u^2}{\overline{a}} \, \df \overline{\phi} \bigg) ^2 + \fr{U}{r^2  + u^2} \bigg( \df \overline t - \fr{r^2 + \overline{a}^2}{\overline{a}} \, \df \overline{\phi} \bigg) ^2 + (r^2 + u^2) \bigg( \fr{\df r^2}{R} + \fr{\df u^2}{U} \bigg) ,
\ee
where\footnote{The function $U$ defined here should not be confused with $U$ defined in \eqref{metricansatz}. It should be clear to the reader which definition is valid depending on the context.}
\begin{align}
\label{RU}
R & = r^2 + \overline{a}^2 - 2 m r , & U & = \overline{a}^2 - u^2 + 2 n u .
\end{align}
Standard Boyer--Lindquist-like coordinates and parameters come from defining the coordinates $(t, \theta, \phi)$ by
\begin{align}
\frac{\phi}{a} & = \frac{\overline \phi }{\overline a} , & t & = \overline t + \frac{2n^2}{\overline a}\overline \phi , & u & = n+a \cos \gq ,\label{rep0}
\end{align}
where the new angular parameter $a$ and Kaluza--Klein 1-form $\omega_3$ are defined by
\begin{align}
\overline{a}^2 & = a^2 - n^2, & \df t + \omega_3 & = \df \overline t + \overline \omega_3 .\label{rep}
\end{align}
To recover the Taub--NUT solution \eq{TaubNUT}, then take $a \rightarrow 0$.  Note that if $a = 0$, then $\overline a^2 = -n^2$ leads to an imaginary rotation parameter $\overline a$, but this is not a physical feature since it can be removed by the reparametrization \eqref{rep}.  In Kaluza--Klein form \eq{metricansatz}, the Kerr--Taub--NUT solution can be written as
\begin{align}
\df s^2_3 & = \fr{R U}{\overline a^2} \, \df \overline \gf^2 + (R - U) \bigg( \fr{\df r^2}{R} + \fr{\df u^2}{U} \bigg) , & \expe{- 2 U} & = \fr{r^2 + u^2}{R - U} , \nnr
\overline \gw_3 & = \fr{(r^2 + \overline{a}^2) U - (\overline{a}^2 - u^2) R}{\overline{a} (R - U)} \, \df \overline{\phi} =  \fr{2 (m r U + n u R)}{\overline{a} (R - U)} \, \df \overline{\phi} .
\end{align}
By Hodge dualizing $\gw_3$, using the orientation \eq{orientation}, we obtain the 3-dimensional scalar
\be
\sigma = \fr{4 (m u  -  n r)}{r^2 + u^2} .
\ee
The non-trivial 3-dimensional scalars are $\expe{- 2 U}$ and $\sigma$.  We also have $\gz^I = \wtd{\gz}_I = 0$, for $I = 1, 2, 3, 4$, and $x_i = 0$, $y_i = 1$, for $i = 1, 2, 3$.  It is convenient to define, for the Kerr--Taub--NUT solution, the rescaled matrix
\be
\overline{\cM} \equiv (R - U) \cM ,
\label{defMProt}
\ee
since its entries are polynomials rather than rational functions.  Specifically, its entries are quadratic in $r$ and $u$,
\be
\overline{\cM} =
\left( \begin{smallmatrix}
R-U & 0 & 0 & 0 & 0 & 0 & 0 & 0 \\
0 & R-U & 0 & 0 & 0 & 0 & 0 & 0 \\
0 & 0 & r^2+u^2 & 0 & 0 & 0 & 0 &2(mu-nr)\\
0 & 0 & 0 & r^2+u^2 & 0 & 0 &-2(mu-nr) & 0 \\
0 & 0 & 0 & 0 & R-U & 0 & 0 & 0 \\
0 & 0 & 0 & 0 & 0 & R-U & 0 & 0 \\
0 & 0 & 0 &-2(mu-nr) & 0 & 0 &(r-2m)^2+(u-2n)^2 & 0 \\
0 & 0 &2(mu-nr) & 0 & 0 & 0 & 0 & (r-2m)^2+(u-2n)^2
\end{smallmatrix}
\right)
.
\label{MKerrNUT}
\ee
The static limit is obtained in the same way as discussed earlier.

The matrix of one-forms $\cN$ takes the form $\cN = \cN_\phi \, \df \phi$.  By definition, the components $\cN_\phi$ obey
\begin{align}
\p_u \cN_\phi & = -\frac{R}{a} \cM^{-1}\p_r \cM, & \p_r \cN_\phi & = \frac{U}{a} \cM^{-1}\p_u \cM.
\end{align}
These are solved by (up to a gauge choice)
\be
\cN_\phi = \gw_{3 \phi} (F_0 + E_0) - \fr{4 (m^2 U + n^2 R)}{a (R - U)} E_0+ \fr{2 (m u R - n r U)}{a (R - U)} H_0 .
\ee


\subsection{Addition of charges}
\label{addcharges}


We act on the Kerr--Taub--NUT matrix $\cM_{\textrm{KTN}}$ with the group element
\ben
g = \exp \bigg( - \sum_I \gc_I k^{P^I} \bigg) \exp \bigg( - \sum_I \gd_I k^{Q_I}\bigg) .
\label{generator}
\een
The generators $k^{Q_I}$ and $k^{P^I}$ are given in \eq{kgenerators}.  $\gd_I$ are electric charge parameters, and $\gc_I$ are magnetic charge parameters.  The generator $k$ is explicitly
\begin{align}
\label{kexplicit}
k & =
\begin{pmatrix}
c_{\gc 1} c_{\gc 4} & 0 & s_{\gc 1} c_{\gc 4} & 0 & s_{\gc 1} s_{\gc 4} & 0 & - c_{\gc 1} s_{\gc 4} & 0 \\
0 & c_{\gc 2} c_{\gc 3} & 0 & - c_{\gc 2} s_{\gc 3} & 0 & s_{\gc 2} s_{\gc 3} & 0 & s_{\gc 2} c_{\gc 3} \\
s_{\gc 1} c_{\gc 4} & 0 & c_{\gc 1} c_{\gc 4} & 0 & c_{\gc 1} s_{\gc 4} & 0 & - s_{\gc 1} s_{\gc 4} & 0 \\
0 & - c_{\gc 2} s_{\gc 3} & 0 & c_{\gc 2} c_{\gc 3} & 0 & - s_{\gc 2} c_{\gc 3} & 0 & - s_{\gc 2} s_{\gc 3} \\
s_{\gc 1} s_{\gc 4} & 0 & c_{\gc 1} s_{\gc 4} & 0 & c_{\gc 1} c_{\gc 4} & 0 & - s_{\gc 1} c_{\gc 4} & 0 \\
0 & s_{\gc 2} s_{\gc 3} & 0 & - s_{\gc 2} c_{\gc 3} & 0 & c_{\gc 2} c_{\gc 3} & 0 & c_{\gc 2} s_{\gc 3} \\
- c_{\gc 1} s_{\gc 4} & 0 & - s_{\gc 1} s_{\gc 4} & 0 & - s_{\gc 1} c_{\gc 4} & 0 & c_{\gc 1} c_{\gc 4} & 0 \\
0 & s_{\gc 2} c_{\gc 3} & 0 & - s_{\gc 2} s_{\gc 3} & 0 & c_{\gc 2} s_{\gc 3} & 0 & c_{\gc 2} c_{\gc 3}
\end{pmatrix}
\nnr
& \quad
\times
\begin{pmatrix}
c_{\gd 1} c_{\gd 4} & 0 & 0 &- c_{\gd 1} s_{\gd 4} & s_{\gd 1} s_{\gd 4} & 0 & 0 &  s_{\gd 1} c_{\gd 4} \\
0 & c_{\gd 2} c_{\gd 3} &- s_{\gd 2} c_{\gd 3} & 0 & 0 & s_{\gd 2} s_{\gd 3} &  c_{\gd 2} s_{\gd 3} & 0 \\
0 & -s_{\gd 2} c_{\gd 3} & c_{\gd 2} c_{\gd 3} & 0 & 0 &- c_{\gd 2} s_{\gd 3} & - s_{\gd 2} s_{\gd 3} & 0 \\
-c_{\gd 1} s_{\gd 4} & 0 & 0 & c_{\gd 1} c_{\gd 4} & -s_{\gd 1} c_{\gd 4} & 0 & 0 & - s_{\gd 1} s_{\gd 4} \\
s_{\gd 1} s_{\gd 4} & 0 & 0 & -s_{\gd 1} c_{\gd 4} & c_{\gd 1} c_{\gd 4} & 0 & 0 &  c_{\gd 1} s_{\gd 4} \\
0 & s_{\gd 2} s_{\gd 3} &- c_{\gd 2} s_{\gd 3} & 0 & 0 & c_{\gd 2} c_{\gd 3} &  s_{\gd 2} c_{\gd 3} & 0 \\
0 & c_{\gd 2} s_{\gd 3} & - s_{\gd 2} s_{\gd 3} & 0 & 0 & s_{\gd 2} c_{\gd 3} & c_{\gd 2} c_{\gd 3} & 0 \\
 s_{\gd 1} c_{\gd 4} & 0 & 0 & - s_{\gd 1} s_{\gd 4} &  c_{\gd 1} s_{\gd 4} & 0 & 0 & c_{\gd 1} c_{\gd 4}
\end{pmatrix}
.
\end{align}
We use the notation $s_{\gd I} = \sinh \gd_I$, $c_{\gd I} = \cosh \gd_I$, $s_{\gd I \ldots J} = s_{\gd I} \ldots s_{\gd J}$, $c_{\gd I \ldots J} = c_{\gd I} \ldots c_{\gd J}$, and similarly for $\gc$ instead of $\gd$.

This choice of group element is motivated by the 4-fold symmetry of the gauge fields $F^I$, and by examining the resulting charge matrix when acting on a simple uncharged solution such as the Schwarzschild solution.  Asymptotic flatness at spatial infinity, which means that the scalars become trivial at infinity, implies that $k^\sharp k = \bbI$.  The generators $k_i$ do not alter the charge matrix of Schwarzschild, and furthermore leave the Schwarzschild solution invariant, up to a gauge transformation.  The generator $k_0$ rotates the mass into a NUT charge; the group element $k = \expe{\beta k_0}$ gives the Taub--NUT solution with mass $M = m \cos (2 \beta)$ and NUT charge $N = m \sin (2 \beta)$.  This leaves the generators $k^{Q_I}$ and $k^{P^I}$ that we use.  The new matrix
\be
\cM = k^\sharp \cM_{\textrm{KTN}} k ,
\ee
with the generalized transpose $^\sharp$ defined in \eq{generalized transpose}, encodes the 16 3-dimensional scalars, which can be extracted using the formulae of Section \ref{extractscalars}.

In particular, the $O(r^{-1})$ part of $\cM$ determines a new charge matrix $\cQ$, from which we can read off the asymptotic charges.  Since Taub--NUT and Kerr--Taub--NUT differ in $\cM$ at order $O(r^{-2})$, the rotating and non-rotating cases share the same charge matrix.

We obtain the mass and NUT charge,
\begin{align}
M & = m \mu_1 + n \mu_2 , & N & = m \nu_1 + n \nu_2 ,
\label{MassNUT}
\end{align}
where
\begin{align}
\label{mudef}
\mu_1 & = 1 + \sum_I \bigg( \fr{s_{\gd I}^2 + s_{\gc I}^2}{2} - s_{\gd I}^2 s_{\gc I}^2 \bigg) + \fr{1}{2} \sum_{I, J} s_{\gd I}^2 s_{\gc J}^2 , & \mu_2 & = \sum_I s_{\gd I} c_{\gd I} \bigg( \fr{s_{\gc I}}{c_{\gc I}} c_{\gc 1 2 3 4} - \fr{c_{\gc I}}{s_{\gc I}} s_{\gc 1 2 3 4}\bigg) ,
\end{align}
and
\begin{align}
\label{nudef}
\nu_1 & = \sum_I s_{\gc I} c_{\gc I} \bigg( \fr{c_{\gd I}}{s_{\gd I}} s_{\gd 1 2 3 4} - \fr{s_{\gd I}}{c_{\gd I}} c_{\gd 1 2 3 4} \bigg) , & \nu_2 = \iota - D
\end{align}
where
\begin{align}
\iota &= c_{\gd 1 2 3 4}c_{\gc 1 2 3 4}+s_{\gd 1 2 3 4} s_{\gc 1 2 3 4}+ \sum_{I < J} c_{\gd 1 2 3 4} \fr{s_{\gd I J}}{c_{\gd I J}} \fr{c_{\gc I J}}{s_{\gc I J}} s_{\gc 1 2 3 4} , \nnr
D &= c_{\gd 1 2 3 4}s_{\gc 1 2 3 4}+s_{\gd 1 2 3 4}c_{\gc 1 2 3 4} + \sum_{I < J} c_{\gd 1 2 3 4} \fr{s_{\gd I J}}{c_{\gd I J}} \fr{s_{\gc I J}}{c_{\gc I J}} c_{\gc 1 2 3 4}.
\end{align}
For asymptotically flat solutions, we cancel the NUT charge \eqref{MassNUT} by setting $n = n_0$ where
\be
\label{n0}
n_0 \equiv - m \frac{\nu_1}{\nu_2} .
\ee

The electric and magnetic charges admit elegant expressions in terms of derivatives of the mass and NUT charge with respect to $\delta_I$,
\begin{align}
Q_I & = 2\frac{\p M}{\p \delta_I} , &
P^I & = -2 \frac{\p N}{\p \delta_I} .
\label{charges2}
\end{align}
Equivalently,
\begin{align}
\label{charges}
Q_I & = m \rho_I^1 + n \rho_I^2 , & P^I & = m \pi^I_1 + n \pi^I_2 ,
\end{align}
where
\begin{align}
\rho_I^1 & = 2 \frac{\p \mu_1}{\p \delta_I} , &
\rho_I^2 & = 2 \frac{\p \mu_2}{\p \delta_I} , &
\pi^I_1 & = -2 \frac{\p \nu_1}{\p \delta_I} , &
\pi^I_2 & = -2 \frac{\p \nu_2}{\p \delta_I} .
\end{align}
These explicit coefficients are
\begin{align}
\rho_I^1 & = 2 s_{\gd I} c_{\gd I} \bigg( 1 - s_{\gc I}^2 + \sum_{J \neq I} s_{\gc J}^2 \bigg) , & \rho_I^2 & = 2 (1 + 2 s_{\gd I}^2) \bigg( \fr{s_{\gc I}}{c_{\gc I}} c_{\gc 1 2 3 4} - \fr{c_{\gc I}}{s_{\gc I}} s_{\gc 1 2 3 4} \bigg) ,
\label{rhodef}
\end{align}
and
\begin{align}
\pi^I_1 & = 2 \bigg[ s_{\gc I} c_{\gc I} (c_{\gd 1 2 3 4} - s_{\gd 1 2 3 4}) + \sum_{J \neq I} s_{\gc J} c_{\gc J} \bigg( c_{\gd 1 2 3 4} \fr{s_{\gd I J}}{c_{\gd I J}} - s_{\gd 1 2 3 4} \fr{c_{\gd I J}}{s_{\gd I J}} \bigg) \bigg] , \nnr
\pi^I_2 & = - 2 \bigg\{ ( c_{\gc 1 2 3 4} - s_{\gc 1 2 3 4} ) \bigg( c_{\gd 1 2 3 4} \fr{s_{\gd I}}{c_{\gd I}} - s_{\gd 1 2 3 4} \fr{c_{\gd I}}{s_{\gd I}} \bigg)  \nnr
& \qquad + \sum_{J \neq I} \bigg[ c_{\gc 1 2 3 4} \fr{s_{\gc I J}}{c_{\gc I J}} \bigg( \fr{c_{\gd J}}{s_{\gd J}} s_{\gd 1 2 3 4} - \fr{s_{\gd J}}{c_{\gd J}} c_{\gd 1 2 3 4} \bigg) + s_{\gc 1 2 3 4} \fr{c_{\gc I J}}{s_{\gc I J}} \bigg( \fr{s_{\gd J}}{c_{\gd J}} c_{\gd 1 2 3 4} - \fr{c_{\gd J}}{s_{\gd J}} s_{\gd 1 2 3 4} \bigg) \bigg] \bigg\} .
\label{pidef}
\end{align}
The angular momentum can be read from \eqref{Q2d} and is
\be
J = (\nu_2 m - \nu_1 n )a ,
\label{Ang6}
\ee
where $\nu_1,\nu_2$ are defined in \eqref{nudef}.


\subsection{Reconstruction of the 4d solution}


We can determine the full 4-dimensional solution by extracting the 3-dimensional scalars and gauge fields, using the formulae of Sections \ref{extractscalars} and \ref{extractgauge}.  The solution can then be simplified after lengthy algebraic manipulations and using the insights of previously known subcases.  The procedure of identifying patterns and relationships among the various functions appearing in the solution is the most non-trivial part of the solution generating process.  Here, we describe how to obtain the 4-dimensional fields, and then in Section \ref{gensol} we summarize the solutions in the simplest presentation that we found.


\subsubsection{Non-rotating, no NUT}


For the static case with no NUT charge, the solution generating technique gives a 4-dimensional spherically symmetric metric of the form
\be
\df s^2 = - \fr{r^2 - 2 m r - n_0^2}{W_0} \df t^2 + W_0 \bigg( \fr{\df r^2}{r^2 - 2 m r -n_0^2} + \df \theta^2 + \sin^2 \theta \, \df \phi^2 \bigg) .
\ee
$W_0^2(r)$ is a quartic polynomial in $r$ that can be written down concisely from the components of $\cM$ using \eq{Ueqn}, namely
\be
W_0^2(r) = \overline{\cM}_{33} \overline{\cM}_{44} - \overline{\cM}_{34}^2 .
\ee

The electromagnetic scalars $\gz^I$ and $\wtd{\gz}_I$ of \eq{electroscalars} encode the gauge fields $A^I$. The scalars $\gz^I$ are related by appropriate permutation of the indices $I = 1, 2, 3, 4$, and similarly for $\wtd{\gz}_I$.  We dualize the 3-dimensional scalars $\sigma$, $\gz^I$ and $\wtd{\gz}_I$ to 3-dimensional vectors, using \eq{d1} and \eq{d2}, to obtain the $\df \phi$ coefficients of $A^I$, $\wtd{A}_{I}$ and $\gw_3$. This is straightforward for spherically symmetric solutions with no NUT charge. In this case, $\gw_3 = 0$ and $\zeta^I$, $\wtd{\gz}_I$ only depend on $r$. Therefore, equation \eqref{zetadual} implies that $\df \wtd A_{I \three} = \wtd{P}_I(r) \sin \theta \, \df \theta \wedge \df \phi$ for some functions $\wtd{P}_I(r)$.  Integrability implies that $F_I$ are constants, which implies that $\wtd A_{I \three}$ are given in terms of the magnetic charges as $\wtd A_{I\three} = \wtd{P}_I \cos\theta \, \df \phi $. The gauge fields $A^I_{\three}$ are then most easily obtained by electromagnetic duality.

The 4-dimensional dilatons and axions are simply the 3-dimensional scalars derived from \eq{scalarfields}.  The scalar fields $x_i, y_i$, are obtained from \eq{x1y1} and \eq{scalarfields}.  The easiest way to obtain them is to read off $x_1$ and $y_1$ from \eq{x1y1}, and then, from symmetry arguments, obtain $x_2$, $x_3$, $y_2$ and $y_3$ by permutation of indices.


\subsubsection{General rotating}


In the general rotating case, the solution generating technique will give a 4-dimensional metric of the form
\ben
\df s^2 = - \fr{R - U}{W} (\df t + \omega_{3\phi} \, \df \phi)^2 + W \bigg( \fr{\df r^2}{R} + \frac{\df u^2}{U} + \frac{R U}{a^2 (R - U)} \df \phi^2 \bigg) ,
\label{genmetric}
\een
where $R(r)$ and $U(u)$ are defined in \eqref{RU} and $W^2 (r,u)$ is a quartic polynomial in $r$ and $u$ that can be obtained from
\be
W^2(r,u) = \overline{\cM}_{33} \overline{\cM}_{44} - \overline{\cM}_{34}^2 .
\ee
Here we define $a^2 = \overline a^2 + n^2$ as in \eqref{rep}. The Kaluza--Klein 1-form $\omega_3$ can be obtained from \eqref{exomega}. The scalars can be obtained from the same procedure as in the static case.  Rather than dualizing electromagnetic scalars, the 4-dimensional gauge fields $\wtd{A}_1$, and $A^4$ are more conveniently obtained from the matrix $\cN$ as \eqref{gaur} and \eqref{vectoransatz}. The other gauge fields $\wtd{A}_2$, $\wtd{A}_3$, $\wtd{A}_4$ and $A^1$, $A^2$, $A^3$ can then be obtained by appropriate permutation of indices.


\section{Summary of general charged black holes}
\label{gensol}


In this section, we summarize the explicit expressions for the general black hole solutions that we have constructed.


\subsection{Static black hole}
\label{stsec}


A general asymptotically flat, static generating solution for $\cN = 8$ supergravity was obtained in \cite{Cvetic:1995kv}.  It is parameterized by a mass and 5 independent electromagnetic parameters, which are 6 electromagnetic charges with one constraint in order to cancel the NUT charge.  Here, we present an 9-parameter asymptotically flat, static solution with 4 independent electric and 4 independent magnetic charges, including the explicit matter fields, which generalizes the seed solution of \cite{Cvetic:1995kv}.  A NUT charge can also be included.  Starting from this seed solution, one may then follow the procedure of \cite{Cvetic:1995kv} and generate, using U-dualities, the static asymptotically flat solution of $\cN = 8$ supergravity with 56 electromagnetic parameters.  Extreme, asymptotically flat, static black holes were studied in \cite{Cvetic:1995bj, Behrndt:1996hu, Gimon:2007mh, Bellucci:2008sv, Dall'Agata:2010dy, Bossard:2012ge, Ortin:2012gg}.

Including NUT charge, the solution is parameterized by 10 constants: mass parameter $m$, NUT parameter $n$, electric charge parameters $\delta_I$ and magnetic charge parameters $\gamma_I$, for $I=1,2,3,4$. The mass and NUT charges are defined in \eqref{MassNUT} and the NUT charge can be cancelled by fixing $n = n_0$ defined in \eqref{n0}. The electric charges $Q_I$ and magnetic charges $P^I$ are given by \eq{charges}.  The orientation is given by \eq{orientation}.


\subsubsection{Metric}


The metric can be written as
\be
\df s^2 = -\frac{R_0(r)}{W_0(r)}(\df t+2N \cos\theta \, \df\phi)^2 + W_0(r) \left( \frac{dr^2}{R_0(r)} + \df \theta^2 + \sin^2\theta \, \df \phi^2  \right),\label{generalmetric}
\ee
where
\begin{align}
R_0(r) & = r^2 -2 m r -n^2 , & W_0^2(r) & = R_0^2(r)+2 R_0(r) (2M r+V )+(L(r)+2 N n)^2 .
\end{align}
Here $M, N$ are the mass and NUT charge defined earlier in \eqref{MassNUT},
\be
L(r) = \lambda_1 r+\lambda_0
\label{Lfunction}
\ee
is a linear function in $r$, and the three remaining constants $\lambda_0,\;\lambda_1,\, V$ are
\begin{align}
\lambda_1 & = 2(m \nu_2 -n \nu_1 ) , & \lambda_0 & = 4(m^2+n^2)D , & V & = 2(-\mu_2 m+\mu_1 n)n + 2(m^2+n^2)C,
\end{align}
where all quantities have been defined earlier in \eqref{mudef} and \eqref{nudef}, except $C$, given by
\begin{align}
C & = 1 +   \sum_I (s_{\gd I}^2 c_{\gc I}^2 + s_{\gc I}^2 c_{\gd I}^2) +   \sum_{I < J} (s_{\gd I J}^2 + s_{\gc I J}^2)  +   \sum_{I \neq J} s_{\gd I}^2 s_{\gc J}^2 +  \sum_I \sum_{J < K} (s_{\gd I}^2 s_{\gc J K}^2 + s_{\gc I}^2 s_{\gd J K}^2) \nnr
& \quad + 2  \sum_{I < J} \bigg( s_{\gd 1 2 3 4} c_{\gd 1 2 3 4} \frac{s_{\gc I J}}{c_{\gd I J}} \frac{c_{\gc I J}}{s_{\gd I J}}  + s_{\gd 1 2 3 4}^2 \frac{s_{\gc I J}^2}{s_{\gd I J}^2} + s_{\gd I J} s_{\gc I J} c_{\gd I J} c_{\gc I J} + s_{\gd I J}^2 s_{\gc I J}^2 \bigg) - \nu_1^2 - \nu_2^2 .\label{coefC}
\end{align}
The metric is asymptotically flat when $n = n_0$ given in \eqref{n0}, which cancels the NUT charge $N = 0$. A global coordinate system is then achieved when the angular coordinates have the standard ranges $\theta \in [0,\pi]$, $\phi \sim \phi + 2\pi$. 


\subsubsection{Gauge fields}


The gauge fields and dual gauge fields are
\begin{align}
A^I & = \gz^I(r)\, (\df t +2 N \cos\theta \, \df \phi) + P^I \cos \gq \, \df \phi , &\wtd{A}_I & = \tzeta_I(r) \, (\df t +2 N \cos\theta \, \df \phi) - Q_I \cos \gq \, \df \phi ,
\end{align}
where it turns out that one can write the scalars $\gz^I(r)$ in terms of the master function $W_0(r)$ as
\be
\zeta^I = \frac{1}{2 W_0^2}\frac{\p W_0^2}{\p \delta_I} = \frac{1}{W_0^2(r)} \bigg[ R(r) \bigg( Q_I r + \frac{\p V}{\p \delta_I} \bigg) + (L(r) +2 N n) \bigg( \frac{\p L(r)}{\p \delta_I} - P^I n \bigg) \bigg] .
\label{zetat1}
\ee
In the case without NUT charge, one needs to take the derivative with generic $n$ first, then set $n=n_0$ in the result. The dual scalars $\tzeta_I(r)$ are
\be
\wtd{\zeta}_I = \frac{R(r) (P^I r + \widetilde V_I) + (L(r) + 2 N n)(\widetilde L_I(r) + Q_I n )}{W_0^2 (r)} ,
\ee
where $\widetilde L_I(r)$ is a linear function and $\widetilde V_I$ a constant, given by
\begin{align}
\label{linearfunctions}
\widetilde{L}_I(r) & = (m \rho_I^2 - n \rho_I^1)r - 4 (m^2+n^2) \widetilde D_I, & \widetilde V_I & = (n \pi^I_1 - m \pi^I_2) n + 2(m^2+n^2) \widetilde C_I ,
\end{align}
with
\begin{align}
\wtd{D}_I & = \frac{s_{\gc I}}{c_{\gc I}} c_{\gc 1 2 3 4} s_{\gd I}^2 - \frac{c_{\gc I}}{s_{\gc I}} s_{\gc 1 2 3 4} c_{\gd I}^2 , \nnr
\wtd{C}_I & = (s_{\gd 1 2 3 4} - c_{\gd 1 2 3 4}) \wtd{C}_{I I} + 2 s_{\gc I} c_{\gc I} s_{\gd 1 2 3 4} \bigg( 2 +  \sum_K s_{\gc K}^2 \bigg) + \sum_{J \neq I} \bigg( c_{\gd 1 2 3 4} \frac{s_{\gd I J}}{c_{\gd I J}} - s_{\gd 1 2 3 4} \frac{c_{\gd I J}}{s_{\gd I J}} \bigg) \wtd{C}_{I J} \nnr
& \quad + 2  \sum_{J \neq I} s_{\gc J} c_{\gc J} \bigg(  \frac{ s_{\gd I J}}{ c_{\gd I J}} c_{\gd 1 2 3 4} (s_{\gc I}^2 + s_{\gc J}^2)  - \frac{c_{\gd I J}}{s_{\gd I J}} s_{\gd 1 2 3 4}\sum_{K \neq I, J} s_{\gc K}^2 \bigg) , \nnr
\wtd{C}_{I J} & = 2 (1 + 2 s_{\gd I}^2) s_{\gc 1 2 3 4} \bigg[ \bigg( 2 + \sum_{K \neq J} \fr{1}{s_{\gc K}^2} \bigg) s_{\gc 1 2 3 4} \frac{c_{\gc J}}{s_{\gc J}}  - (1 + 2 s_{\gc J}^2) \frac{c_{\gc 1 2 3 4}}{s_{\gc J} c_{\gc J}} \bigg] \nnr
& \quad + 2 s_{\gd I}^2 s_{\gc J} c_{\gc J} \bigg( 1 +  \sum_K s_{\gc K}^2 \bigg) .
\label{defDCI}
\end{align}


\subsubsection{Scalar fields}


The scalar fields are\footnote{Note added in v4: Typo corrected. We thank Tucker Manton and Cindy Keeler.}
\begin{align}
\expe{\varphi_i} & = \frac{r^2 + u^2 + g_i}{W} , & \chi_i & = \frac{f_{i}}{r^2 + u^2 + g_i} ,
\end{align}
where\footnote{Note added in v4: Typo corrected. We thank Tucker Manton and Cindy Keeler.}
\begin{align}
f_{i} & = 2 (m r + n u) \xi_{i 1} + 2 (m u - n r) \xi_{i 2} + 4 (m^2 + n^2) \xi_{i 3} , \nnr
g_{i} & = 2 (m r + n u) \eta_{i 1} +  2 (m u - n r) \eta_{i 2} + 4 (m^2 + n^2) \eta_{i 3} .
\end{align}
The coefficients $\xi_{i1}$, $\xi_{i2}$ and $\xi_{i3}$ for $i = 1$ are
\begin{align}
\xi_{11} & = [ (s_{\delta 123}c_{\delta 4} - c_{\delta 123}s_{\delta 4} )s_{\gamma_1}c_{\gamma_1} + (1 \leftrightarrow 4) ] - ( (1,4) \leftrightarrow (2,3)) , \nnr
\xi_{12} & = [\tfrac{1}{2}(c_{\delta 23}s_{\gamma 14}+c_{\gamma 14} s_{\delta 23 })(c_{\delta 14}c_{\gamma 23} + s_{\gamma 23}s_{\delta 14})  + s_{\gd 1} s_{\gc 4} c_{\gd 4} c_{\gc 1} (s_{\gd 2} s_{\gc 2} c_{\gd 3} c_{\gc 3} + s_{\gd 3} s_{\gc 3} c_{\gd 2} c_{\gc 2})\nnr
& \quad + (1 \leftrightarrow 4)] - ( (1,4) \leftrightarrow (2,3)) , \nnr
\xi_{13} & = [( s_{\delta 1 3 4} c_{\delta 2} c_{\gamma 2}^2 + c_{\delta 1 3 4} s_{\delta 2} s_{\gamma 2}^2 ) s_{\gamma_3} c_{\gamma_3} + (2 \leftrightarrow 3) ] - ( (1,4) \leftrightarrow (2,3)) ,
\label{xicoeffs}
\end{align}
and the coefficients $\eta_{i1}$, $\eta_{i2}$ and $\eta_{i3}$ for $i = 1$ are
\begin{align}
\eta_{1 1} & = s_{\gd 2}^2 + s_{\gd 3}^2 + s_{\gc 1}^2 + s_{\gc 4}^2 + (s_{\gd 2}^2 + s_{\gd 3}^2) (s_{\gc 1}^2 + s_{\gc 4}^2) + (s_{\gd 2}^2 - s_{\gd 3}^2) (s_{\gc 3}^2 - s_{\gc 2}^2) , \nnr
\eta_{1 2} & = 2 s_{\gd 2} c_{\gd 2} (c_{\gc 2} s_{\gc 1 3 4} - s_{\gc 2} c_{\gc 1 3 4}) + (2 \leftrightarrow 3) , \nnr
\eta_{1 3} & = 2 s_{\gd 2 3} c_{\gd 2 3} (s_{\gc 2 3} c_{\gc 2 3} + s_{\gc 1 4} c_{\gc 1 4})+ s_{\gd 2 3}^2 (1 + \textstyle \sum_I s_{\gc I}^2 ) + (s_{\gd 2}^2 + s_{\gd 3}^2+2 s_{\delta 23}^2) (s_{\gc 1 4}^2 + s_{\gc 2 3}^2)\nnr
& \quad + s_{\gd 2}^2 s_{\gc 2}^2 + s_{\gd 3}^2 s_{\gc 3}^2 + s_{\gc 1 4}^2 .
\label{etacoeffs}
\end{align}
The results for $i = 2$ and $i = 3$ are obtained by respectively interchanging indices $1 \leftrightarrow 2$ and $1 \leftrightarrow 3$.


\subsection{Rotating black hole}


The general rotating solution depends on 11 independent parameters: the mass, NUT and rotation parameters ($m$, $n$, $a$); and electric ($\delta_I$) and magnetic ($\gamma_I$) charge parameters. The mass and NUT charges are defined in \eqref{MassNUT} and the NUT charge can be cancelled by fixing $n = n_0$ defined in \eqref{n0}. The electric charges $Q_I$ and magnetic charges $P^I$ are given by \eq{charges} and the angular momentum is given in \eq{Ang6}.  The orientation is given by \eq{orientation}.


\subsubsection{Metric}


The metric of the general solution is
\begin{align}
\df s^2 & = - \fr{R - U}{W} (\df t + \omega_3 )^2  + W \bigg( \fr{\df r^2}{R} + \frac{\df u^2}{U} + \frac{R U}{a^2 (R - U)} \, \df \phi^2 \bigg) ,
\label{genmetric2}
\end{align}
where $R$ and $U$ are the quadratic functions
\begin{align}
R(r) & = r^2 -2 m r + a^2 - n^2 , & U(u) & = a^2 - (u - n)^2 .
\label{RUgeneral}
\end{align}
The master function $W$ and the Kaluza--Klein 1-form $\omega_3$ can be expressed as
\begin{align}
W^2 & = (R -  U)^2  + (2 N u + L)^2 + 2 (R - U) ( 2 M r + V ) ,  \nnr
\omega_3 & = \frac{2 N (u-n) R + U (L +2 N n)}{a (R - U)} \, \df \phi\label{W2o}
\end{align}
in terms of $R(r)$, $U(u)$ and two linear functions $L(r)$ and $V(u)$ given by
\begin{align}
L(r) & = 2(-n \nu_1 + m \nu_2 ) r + 4(m^2 +n^2)D, & V(u) & = 2(n \mu_1 - m \mu_2)u +2 (m^2+n^2) C,
\label{LV}
\end{align}
where $\nu_1$, $\nu_2$, $\mu_1$, $\mu_2$ and $D$ have been defined in \eqref{mudef} and \eqref{nudef} and $C$ has been defined in \eqref{coefC}.  The static limit is obtained by setting $u = n+a\cos\theta$ and taking $a \rightarrow 0$. Then $\omega_3 = 2N\cos\theta \, \df \phi$, and the solution reduces to the static solution presented previously.  The expression of $W$ and $\omega_3$ solely in terms of $R$, $U$ and linear functions gives an elegant form of the metric.


\subsubsection{Gauge fields}


Astonishingly, the gauge fields can be expressed in the elegant form
\be
A^I = - W \frac{\p }{\p \delta_I} \bigg( \frac{\df t + \omega_3}{W}  \bigg) ,
\label{AIW}
\ee
which makes manifest that the gauge fields $A^I$ can be built solely from functions already appearing in the metric. In terms of 3-dimensional fields, we have the equivalent relations
\be
 A^I = \zeta^I (\df  t + \gw_3) + A_{\three}^I,
\ee
where
\begin{align}
\zeta^I & = \frac{1}{2 W^2} \fr{\pd}{\pd \gd_I} ( W^2 ) =\frac{1}{W^2} \bigg[ (R - U) \bigg( Q_I r + \frac{\p V}{\p \delta_I} \bigg) + (L +2 N u) \bigg( \frac{\p L}{\p \delta_I} - P^I u \bigg) \bigg] , \nnr
A_{\three}^I & = -\frac{\p}{\p \delta_I} \omega_3 =  \bigg[ P^I (u - n) + \fr{U}{R - U} \bigg( P^I u - \frac{\p L}{\p \delta_I} \bigg) \bigg] \, \fr{\df \phi}{a} .
\end{align}

The dual gauge fields are
\be
\wtd{A}_I = \widetilde{\zeta}_I (\df  t + \gw_3) + \wtd{A}_{I \three} ,
\ee
where
\begin{align}
\wtd{A}_{I \three} & = - \bigg( Q_I (u - n) + \fr{U( Q_I u + \widetilde L_I )}{R - U}  \bigg) \, \fr{\df \phi}{a} , \nnr
\wtd{\zeta}_I & = \frac{1}{W^2} \left( (R - U) (P^I r + \widetilde V_I) + (L + 2 N u)(\widetilde L_I + Q_I u ) \right) ,
\end{align}
where $\widetilde L_I(r)$, $\widetilde V_I(u)$ are the linear functions
\begin{align}
\label{linearfunctions2}
\widetilde{L}_I(r) & = (m \rho_I^2 - n \rho_I^1)r - 4 (m^2+n^2) \widetilde D_I, &\widetilde V_I(u) & = (n \pi^I_1 - m \pi^I_2) u + 2(m^2+n^2) \widetilde C_I .
\end{align}
The coefficients $\rho_I^1$, $\rho_I^2$, $\pi^I_1$ and $\pi^I_2$ are defined in \eq{rhodef} and \eq{pidef}.  The coefficients $\widetilde D_I$, $\widetilde C_I$ are defined in \eqref{defDCI}.  We have not found an elegant expression for $\wtd{A}_I$ analogous to \eqref{AIW}. The asymmetry between $A^I$ and $\wtd{A}_I$ originates from the choice of $\SO(4,4)$ group element \eqref{generator}, which does not have symmetry under interchange of $\gd_I$ and $\gc_I$.


\subsubsection{Scalar fields}


The scalar fields are
\begin{align}
\expe{\varphi_i} & = \frac{r^2 + u^2 + g_i}{W} , & \chi_i & = \frac{f_{i}}{r^2 + u^2 + g_i} ,
\end{align}
where
\begin{align}
f_{i} & = 2 (m r + n u) \xi_{i 1} + 2 (m u - n r) \xi_{i 2} + 4 (m^2 + n^2) \xi_{i 3} , \nnr
g_{i} & = 2 (m r + n u) \eta_{i 1} +  2 (m u - n r) \eta_{i 2} + 4 (m^2 + n^2) \eta_{i 3} ,
\end{align}
and the coefficients $\xi_{i1},\,\xi_{i2},\,\xi_{i3},\,\eta_{i1},\,\eta_{i2},\,\eta_{i3}$ are the same as the static coefficients \eqref{xicoeffs} and \eq{etacoeffs}.


\section{Physical quantities}
\label{phys}


In this section, we restrict to asymptotically flat solutions, which have vanishing NUT charge, $N = 0$, by setting $n = n_0$ given by \eq{n0}, unless otherwise stated.  Note that derivatives with respect to $\delta_I$ must be done before setting $n = n_0$.


\subsection{Thermodynamics}
\label{thermo}


In this subsection, we explicitly reinstate the 4-dimensional Newton constant $G$.  Recall from Section \ref{addcharges} that the charge matrix provides the mass $M$ in \eqref{MassNUT}, and electric charges $Q_I$ and magnetic charges $P^I$ in \eq{charges}.  We normalize the electromagnetic charges as
\begin{align}
\overline{Q}_I & = \frac{1}{4 G} Q_I , & \overline{P}^I & = \frac{1}{4 G} P^I .
\end{align}
The angular momentum $J$ is obtained from another charge matrix in \eqref{Ang6}.  Canonical methods then associate the mass to $\p_t$, the angular momentum to $-\p_\phi$, the electric charges $\overline{Q}_I$ associated with $A_I$ to the gauge parameter $\Lambda_I = -1$, and similarly magnetic charges $\overline{P}^I$ associated with $\wtd{A}_I$.  To recapitulate, we have
\begin{align}
M & = \fr{m}{G} \bigg( \mu_1 - \fr{\nu_1 \mu_2}{\nu_2} \bigg) , &J& = \frac{m a}{G} \fr{(\nu_1^2 + \nu_2^2)}{\nu_2} , \label{J} \nnr
\overline{Q}_I & = \fr{m}{2 G} \bigg( \fr{\pd \mu_1}{\pd \gd_I} - \fr{\nu_1}{\nu_2} \fr{\pd \mu_2}{\pd \gd_I} \bigg) , & \overline{P}^I & = \fr{m}{2 G} \bigg( \fr{\nu_1}{\nu_2} \fr{\pd \nu_2}{\pd \gd_I} - \fr{\pd \nu_1}{\pd \gd_I} \bigg) ,
\end{align}
where $\mu_1$, $\mu_2$, $\nu_1$ and $\nu_2$ are given in \eq{mudef} and \eq{nudef}. 

The black hole has outer and inner horizons at $r = r_\pm$, the roots of the radial function $R(r)$.  The angular velocity  $\Omega_+$ at the outer horizon is determined by the Killing vector
\be
\xi^\mu \, \p_\mu =\p_t + \Omega_+ \, \p_\phi
\ee
that becomes null at the horizon, and is
\be
\Omega_+ = \frac{a}{L(r_+)} ,
\ee
where $L$ is given in \eq{Lfunction}.  The entropy and temperature are
\begin{align}
S_+ & = \frac{\pi}{G}L(r_+), & T_+ & = \frac{R'(r_+)}{4\pi L(r_+)} = \frac{r_+ - m}{2\pi L(r_+)}.
\end{align}
In the static case, the function $W(r_+,u)$ defined in \eqref{W2o} reduces to $L(r_+)$, and so these quantities can be expressed in terms of $W$.  The electric potential $\Phi_+^I = \xi_+ ^\mu A^I_{\mu}$ and magnetic potential $\Psi^+_I = \xi_+^\mu \widetilde{A}_{I \mu}$ at the horizon are
\begin{align}
\Phi_+^I & =\Omega_+ A_{\three \phi}^{I}(r_+) = \frac{1}{L} \bigg( \frac{\p L}{\p \delta_I} - n_0 P^I \bigg) \bigg| _{r = r_+} , & \Psi^+_I & =\Omega_+ \wtd A_{\three \phi}^{I}(r_+) = \frac{\widetilde L_I + n_0 Q_I}{L} \bigg|_{r=r_+} ,
\end{align}
where $\wtd{L}_I$ is given in \eq{linearfunctions}.  These quantities obey the first law of thermodynamics
\be
\label{firstlaw}
\delta M = T_+ \, \delta S_+ + \Omega_+ \, \delta J + \Phi^I_+ \, \delta \overline{Q}_I + \Psi_I^ + \, \delta \overline{P}^I,
\ee
and the Smarr relation
\be
\label{Smarr}
M = 2 T_+ S_+ + 2 \Omega_+ J + \Phi^I_+ \overline{Q}_I + \Psi_I^+ \overline{P}^I .
\ee
Technically, the Smarr relation follows from non-trivial identities obeyed by the parameters,
\begin{align}
\sum_I \left( \rho_{I}^2 \pi^I_1 - \rho_I^1 \pi_2^I \right) & = 8(\mu_1 \nu_2 -\mu_2 \nu_1 -\iota - D), \nnr
\sum_I \left( Q_I \frac{\p D}{\p \delta_I} - P^I \widetilde D_I \right) & = 4D(\mu_1 +1)m+(4D\mu_2 +2\nu_1)n_0 .
\end{align}


\subsection{Cayley hyperdeterminant}
\label{add}


For regular, static, extremal black holes of $\cN = 8$ supergravity, the entropy is expressed in terms of the electromagnetic charges as \cite{Kallosh:1996uy}\footnote{Our convention for the normalization of $\lozenge$ differs by a factor of 4 from \cite{Kallosh:1996uy}.}
\be
S_+ =2 \pi \sr{ |\lozenge |} ,
\ee
where $\lozenge$ is the Cartan--Cremmer--Julia $\textrm{E}_{7 (7)}$ quartic invariant.  See \cite{Gunaydin:2000xr} for further details of the definitions of $\lozenge$.  Specializing to $STU$ supergravity, the $\textrm{E}_{7 (7)}$ quartic invariant $\lozenge$ reduces to an $\SL(2, \bbR)^3$-invariant, the Cayley hyperdeterminant $\Delta$.  Consequently, the entropy reduces to (see e.g.\ \cite{Duff:2006uz})
\be
S_+ = 2 \pi \sr{| \Delta |} ,
\ee
where the hyperdeterminant is
\be
\Delta (Q_I, P^I)  = \frac{1}{16} \bigg( 4 (Q_1 Q_2 Q_3 Q_4 + P^1 P^2 P^3 P^4) + 2 \sum_{J < K} Q_J Q_K P^J P^K - \sum_J (Q_J)^2 (P^J)^2 \bigg) .
\label{quarticinvariant}
\ee
Some special cases of $\Delta$ are: all gauge fields equal ($Q_I = Q$, $P^I = P$), with $\Delta = \tf{1}{4} (Q^2 + P^2)^2$; only electric charges ($P^I = 0$), with $\Delta = \tf{1}{4} Q_1 Q_2 Q_3 Q_4$; only one non-vanishing gauge field ($Q_I = P^I = 0$ for $I = 2, 3, 4$), with $\Delta = - \tf{1}{16} (Q_1)^2 (P^1)^2$; and pairwise equal gauge fields ($(Q_1, P^1) = (Q_4, P^4)$ and $(Q_2, P^2) = (Q_3, P^3)$), with $\Delta = \tf{1}{4} (Q_1 Q_2 + P^1 P^2)^2$.
	
The hyperdeterminant is invariant under permutations of the four gauge fields.  It is also manifestly invariant under $\SL (2, \bbR)^3$ upon rewriting as
\be
\Delta = \tf{1}{32} \gep^{a b} \gep^{c d} \gep^{a' b'} \gep^{c' d'} \gep^{a'' c''} \gep^{b'' d''} \gc_{a a' a''} \gc_{b b' b''} \gc_{c c' c''} \gc_{d d' d''} ,
\ee
where $\gep^{ab}=\gep^{[ab]}$, $\gep^{0 1} = 1$, and the components $\gamma_{a a' a''}$ are\footnote{Note added in v4: This equation was corrected in accordance with \cite{Compere:2015roa}.}
\begin{align}
(\gc_{000}, \gc_{111}) & = (P^4,-Q_4) , & (\gc_{100}, \gc_{011}) & = (Q_1,-P^1) , \nnr
(\gc_{010}, \gc_{101}) & = (Q_2,-P^2) , & (\gc_{001}, \gc_{110}) & = (Q_3,-P^3) .
\end{align}
The sets of indices $(a, b, c, d)$, $(a', b', c', d')$ and $(a'', b'', c'', d'')$ each correspond to different copies of $\SL(2, \bbR)$.  Using Schouten identities such as $\gep^{a [b} \gep^{c d]} = 0$, the hyperdeterminant may be rewritten as
\begin{align}
\Delta & = \tf{1}{32} \gep^{a' b'} \gep^{c' d'} \gep^{a'' b''} \gep^{c'' d''} \gep^{a c} \gep^{b d} \gc_{a a' a''} \gc_{b b' b''} \gc_{c c' c''} \gc_{d d' d''} \nnr
& = \tf{1}{32} \gep^{a'' b''} \gep^{c'' d''} \gep^{a b} \gep^{c d} \gep^{a' c'} \gep^{b' d'} \gc_{a a' a''} \gc_{b b' b''} \gc_{c c' c''} \gc_{d d' d''} ,
\end{align}
so the hyperdeterminant is invariant when the three copies of $\SL(2, \bbR)$ are cycled.  Since each expression is also manifestly invariant under interchange of two copies of $\SL(2, \bbR)$, the hyperdetermiant is invariant under the triality symmetry of permuting the three copies of $\SL(2, \bbR)$.

For the general NUT-free, non-extremal black hole solution that we derived, the hyperdeterminant can be expressed in terms of the parameters $m$, $\gd_I$ and $\gc_I$ as
\be
\Delta = \fr{m^4 (\nu_1^2 + \nu_2^2)^2 (4 \iota\, D - \nu_1^2)}{\nu_2^4} ,
\ee
where $\nu_1$, $\nu_2$, $\iota$ and $D$ are given in \eq{nudef}.


\subsection{Inner horizon thermodynamics}


Associating thermodynamic quantities to the inner horizon of a black hole is an old idea \cite{Curir:1979aa, Curir:1979bb, Okamoto:1992aa}, but the physical interpretation of these quantities remains unclear.  Two  particularly interesting inner horizon quantities are the ``temperature''  $T_-$ and ``entropy'' $S_-$ which are defined from geometrical quantities at the horizon, through $S_- = A_-/4$ and $T = \kappa_-/2 \pi$, where $A_-$ is the area of the inner horizon (defined with a particular orientation) and $\kappa_-$ is the surface gravity corresponding to the null generator  $\xi_-^\mu \, \pd_\mu = \pd_t + \Omega_- \, \pd_\phi$ of the inner horizon.  All inner horizon thermodynamic quantities $T_-$, $S_-$, $\Omega_-$, $\Phi^I_-$ and $\Psi^-_I$ are those defined at the outer horizon, but with $r_+$ replaced by $r_-$. It is then easy to see that
\be
S_- T_- \leq 0,
\ee
which makes the physical interpretation of $T_-$ and $S_-$ unclear.  We emphasize that\footnote{Noted added in v4: We defined $T_-=(r_- -m)/(2\pi L(r_-))$ and $S_-=\pi L(r_-)/G$. In whole generality, the surface gravity of the future (resp. past) inner horizon is negative (resp. positive). Our convention for defining $T_-$ matches the future inner surface gravity when $L(r_-) > 0$ and the past inner surface gravity when $L(r_-)<0$. In the subsequent work \cite{Cvetic:2018dqf}, one defines instead $T_-=(r_- -m)/(2\pi |L(r_-)|)$ and $S_-=\pi |L(r_-)|/G$, i.e. the temperature is matched to the future inner surface gravity and the entropy to the geometrical inner horizon area.} 
\be
S_- = \frac{\pi}{G}L(r_-)
\ee
is not necessarily non-negative, and therefore whether the negative sign in $S_- T_-$ comes from $T_-$ or $S_-$ depends on the particular solution.

The inner horizon thermodynamic quantities also obey the first law of thermodynamics and the Smarr relation,
\begin{align}
\delta M & = T_- \, \delta S_- + \Omega_- \, \delta J + \Phi^I_- \, \delta \overline{Q}_I + \Psi_I^- \, \delta \overline{P}^I, \nnr
M & = 2 T_- S_- + 2 \Omega_- J + \Phi^I_- \overline{Q}_I + \Psi_I^- \overline{P}^I.
\end{align}
There are relations between the outer and inner entropies, temperatures and angular velocities,
\be
\fr{S_-}{S_+} = \fr{T_+}{- T_-} = \fr{\Omega_+}{\Omega_-} = \fr{L(r_-)}{L(r_+)} ,
\label{innerouterrelations}
\ee
generalizing formulae known for the Kerr solution \cite{Okamoto:1992aa}. We also notice the relations 
\be
S_+ = \frac{\pi^2}{3}c_J \frac{-2 T_-}{\Omega_- - \Omega_+} = \frac{\pi^2}{3} c_{Q_I} \frac{-2 T_-}{\Phi_-^I - \Phi_+^I} = \frac{\pi^2}{3} c_{P^I} \frac{-2 T_-}{\Psi^-_I - \Psi^+_I} ,
\label{Spf}
\ee
for each $I = 1, 2, 3, 4$, where we define the ``central charges''
\begin{align}
c_J & = 6 \fr{\pd \Delta_J}{\pd J} , & c_{Q_I} & = 6 \fr{\pd \Delta_J}{\pd \overline{Q}_I} , & c_{P^I} & = 6 \fr{\pd \Delta_J}{\pd \overline{P}^I} , \label{cc}
\end{align}
and 
\be
\Delta_J = \Delta + J^2 . 
\ee
The quantities \eqref{cc} can be obtained as central charges of a Virasoro algebra for the class of extremal fast and slow rotating black holes, as discussed in Sections \ref{maxrot} and \ref{secslow}.  In the case of general non-extremal black holes, there is no known derivation of these central charges from a Virasoro algebra.  The first relation in \eqref{Spf} can also be written using \eq{innerouterrelations} as
\be
8 \pi^2 J = \Omega_+ S_+ \bigg( \fr{1}{T_+} + \fr{1}{T_-} \bigg) .
\ee
The thermodynamics of the inner horizon has been considered in higher-derivative theories in \cite{Castro:2013pqa}.


\subsection{Product of horizon areas}


The product of the two horizon areas is independent of the mass and quantized in terms of the angular momentum and electromagnetic charges as 
\be
\frac{A_+ A_-}{64\pi^2 G^2} =  J^2 + \Delta (Q_I, P^I) = \Delta_J.
\label{prodarea}
\ee
Some special cases have been considered previously: the 4-charge Cveti\v{c}--Youm black hole \cite{Cvetic:2010mn}; the dyonic Kerr--Newman black hole \cite{Visser:2012zi} and the dyonic black hole of Kaluza--Klein theory \cite{Cvetic:2013eda}.  Since the metric is unaltered by U-dualities, this result generalizes to black holes of $\cN = 8$ supergravity with 28 electric and 28 magnetic charges by replacing the hyperdeterminant $\Delta$ with the quartic $\textrm{E}_{7 (7)}$-invariant $\lozenge$.

A natural interpretation of the product of areas formula is given in terms of auxiliary left and right ``entropies'' 
\begin{align}
S_L & \equiv \tfrac{1}{2}(S_+ + S_-) , & S_R & \equiv \tfrac{1}{2}(S_+ - S_-) ,\label{SL}
\end{align}
which are clearly non-negative. The cases where $S_- <0$ are then rephrased as cases where $S_R > S_L$. The product formula becomes a level-matching condition,
\be
S_L^2 - S_R^2 = 4\pi^2 (J^2+\Delta ) .
\label{prod1}
\ee

Generalizing a result of Einstein gravity \cite{Ansorg:2008bv}, in Einstein--Maxwell theory, it has been shown \cite{Ansorg:2009yi, Hennig:2009aa} (see \cite{Ansorg:2010ru} for a review) that universally
\ben
A_+ A_- = (8 \pi J)^2 + (4 \pi Q^2)^2 ,
\een
for any electrically charged stationary axisymmetric black hole with surrounding matter.  Furthermore, there are inequalities involving the area $\mathcal A$ of a smooth stable axisymmetric marginally outer trapped surface \cite{Hennig:2008zy, Clement:2011np, Clement:2012vb}, for example
\be
\mathcal A^2 \geq (8 \pi J)^2 + (4 \pi Q^2)^2 .
\ee
These types of inequalities are reviewed in \cite{Dain:2011mv}.  The inequalities can generalize to Einstein--Maxwell--dilaton theory, in particular to Kaluza--Klein theory \cite{Yazadjiev:2012bx}.  We expect that these results further generalize using the appropriate quartic invariant in the charges, to the $STU$ model as
\begin{align}
A_+ A_- & = (8 \pi J)^2 + (8 \pi)^2 \Delta , & \mathcal A^2 & \geq (8 \pi J)^2 + (8 \pi)^2 \Delta ,
\end{align}
and to $\cN = 8$ supergravity as
\begin{align}
A_+ A_- & = (8 \pi J)^2 + (8 \pi)^2 \lozenge , & \mathcal A^2 & \geq (8 \pi J)^2 + (8 \pi)^2 \lozenge .
\end{align}


\subsection{Non-extremal entropy and \texorpdfstring{$F$}{F}-invariant}


The non-extremal black hole entropy can be rewritten in the Cardy form \cite{Cvetic:1996kv,Cvetic:1997uw,Cvetic:2011dn}
\be
S_+ = 2 \pi \big( \sqrt{\Delta + F} + \sqrt{-J^2 + F} \big) ,
\label{Cardyentropy}
\ee
where
\be
F(M, Q_I, P^I) = \fr{m^4 (\nu_1^2+\nu_2^2)^3}{\nu_2^4} .
\label{FF}
\ee
Indeed, the equality $\Omega_+/T_+= -\Omega_-/T_-$ \eq{innerouterrelations} implies that $\p S_L / \p J =0$ after using the first law at the outer and inner horizons and the definition of $S_L$ in \eqref{SL}. Differentiating \eqref{prod1} with respect to $J$, one has $\p (S_R^2) / \p J = - 8 \pi^2 J$.  Then integrating gives $S_R = 2\pi \sqrt{-J^2 + F}$. The constant of integration $F$ is fixed by the actual value of $S_R$ to be \eq{FF}. Using \eqref{prod1}, we deduce that $S_L = 2\pi \sqrt{\Delta + F}$. The result for $S_+ = S_L+S_R$ follows.

Since the entropy, the quartic invariant and $J$ are all $\textrm{E}_{7(7)}$-invariant, $F$ admits an $\textrm{E}_{7(7)}$-invariant generalization, depending also on the moduli.  We will therefore refer to $F$ defined in \eqref{FF} as the $F$-invariant.


\subsection{BPS bound}


For the general rotating black hole, from \eq{Mimatrix} we have
\be
\mathcal M_i (r,u) = \fr{1}{W}
\begin{pmatrix}
r^2 + u^2 + g_i & f_i \\
f_i & (W^2 + f_i^2)/(r^2 + u^2 + g_i)
\end{pmatrix}
.
\ee
At infinity, we find the identity since all scalar moduli are trivial,
\be
\mathcal M_i = \bbI + O(r^{-1}).
\ee
In generality, we define the moduli-dependent $\SL(2,\mathbb R)^3$ invariant
\begin{align}
M_2 & = \tfrac{1}{16} \gamma_{a a' a''} [ (\cM_1^{-1})^{a b} (\cM_2^{-1})^{a' b'} (\cM_3^{-1})^{a'' b''} - (\cM^{-1}_1)^{a b} \gep^{a' b'} \gep^{a'' b''} - \gep^{a b} (\cM^{-1}_2)^{a' b'} \gep^{a'' b''} \nnr
& \qquad - \gep^{a b} \gep^{a' b'} (\cM^{-1}_3)^{a'' b''} ] \gamma_{b b' b''} ,
\end{align}
which, for trivial moduli evaluated at infinity, is
\begin{align}
M_2^\infty & = \tfrac{1}{16} \gamma_{a a' a''} (\delta^{a b} \delta^{a' b'} \delta^{a'' b''} - \delta^{a b} \gep^{a' b'} \gep^{a'' b''} - \gep^{a b} \delta^{a' b'} \gep^{a'' b''} - \gep^{a b} \gep^{a' b'} \delta^{a'' b''}) \gamma_{b b' b''} \nnr
& = \frac{1}{16} \sum_{I, J} (Q_I Q_J + P^I P^J) . 
\label{M2inf}
\end{align}
The quantity $M^\infty_2 = |Z(P,Q,z_\infty)|^2$ is also the modulus of the central charge of the $\cN = 2$ algebra \cite{Ceresole:1995ca}
\be
Z(P,Q,z,\overline z) = \frac{1}{\sqrt{2}} \expe{K(z,\overline z)/2}(X^\Lambda(z) Q_\Lambda - F_\Lambda(z) P^\Lambda)
\ee
where $K=-\log{(-8 y_1 y_2 y_3)}$ is the K\"ahler potential of the $STU$ model and $F_\Lambda =\p_\Lambda F$. We have the Bogomolny bound on the square mass,
\be
M^2 \geq M^\infty_2.
\ee


\subsection{Quadratic mass formula}


We define the moduli-dependent symplectic invariants \cite{Ceresole:1995ca}
\begin{align}
I_2(r,u) & = -\frac{1}{4} (\widetilde{P}^\Lambda, \widetilde{Q}_\Lambda) 
\begin{pmatrix}
\Imag \mathcal N + \Real \mathcal N (\Imag \mathcal N)^{-1} \Real \mathcal N  & - \Real \mathcal N (\Imag \mathcal N)^{-1} \\
-(\Imag \mathcal N)^{-1} \Real\mathcal N & (\Imag \mathcal N)^{-1}
\end{pmatrix}
\begin{pmatrix}
\widetilde{P}^\Lambda \\
\widetilde{Q}_\Lambda
\end{pmatrix}
, \nnr
J_2(r,u) & = \frac{1}{4} (\widetilde{P}^\Lambda, \widetilde{Q}_\Lambda)
\begin{pmatrix}
\Imag F + \Real F (\Imag F)^{-1}\Real F & - \Real F (\Imag F)^{-1} \\
-(\Imag F)^{-1} \Real F & (\Imag F)^{-1}
\end{pmatrix}
\begin{pmatrix}
\widetilde{P}^\Lambda \\
\widetilde{Q}	_\Lambda
\end{pmatrix}
,
\end{align}
where $F_{\Lambda \Sigma} = \p_\Lambda \p_\Sigma F$ and $F=-X^1 X^2 X^3/X^0$ is the prepotential of the $STU$ model. Here, asymptotic flatness at spatial infinity fixes the scalar moduli at infinity as $x_i=0$, $y_i =1$, at $r = \infty$. The invariants read at infinity
\begin{align}
I_2^\infty \equiv I_2(\infty,u)  & = \frac{1}{4}\sum_I [ \mathcal (Q_I)^2 + (P^I)^2 ] , \label{I2inf} \nnr
J_2^\infty \equiv J_2(\infty,u) & = \frac{1}{4}\sum_I [ (Q_I)^2 + (P^I)^2 ] -\frac{1}{8}\sum_{I,J} \left( P_I P_J + Q_I Q_J \right),
\end{align}
where we used \eq{QPsymplectic}.

For any $\cN =2$ model,
\begin{align}
|Z|^2 + |Z_i|^2 & = I_2^\infty, & -|Z|^2 + |Z_i|^2 & = J_2^\infty,
\end{align}
where $Z$ is the central charge and $Z_i = D_i Z$ is the K\"ahler derivative of the central charge.  Therefore, $J_2^\infty$ can simply be expressed as $J_2^\infty = I_2^\infty - 2 M_2^\infty$.

It is useful to define the invariant
\be
S^\infty_2= \tfrac{1}{4} G_{AB}\p_r \Phi^A \p_r \Phi^B |_{r=\infty} .
\label{S2}
\ee
For the $STU$ model, we have
\be 
S^\infty_2 = \frac{1}{4}\sum_i ( \Sigma_i^2 + \Xi_i^2 ) .
\ee

It was observed by Gibbons \cite{Gibbons:1982ih} that for static configurations, the black hole mass obeys the condition
\be
M^2 + N^2 + S^\infty_2 = I_2^\infty + 4 S_+^2 T_+^2.
\ee
This relation was interpreted in \cite{Gibbons:1996af} as the statement that the total self-force on the black hole due to the attractive self-force of gravity and the scalar fields is not exceeded by the repulsive self-force due to the gauge fields, and vanishes only at extremality.  For static black holes of Einstein--Maxwell theory and the Einstein--Maxwell--dilaton--axion theory \eq{EMDALagrangian}, similar relations were derived using the 3-dimensional coset model in \cite{Heusler:1997sd}, and further generalized in \cite{Breitenlohner:1998cv}.  

We find that when rotation is present, the relation generalizes to
\be
M^2 + N^2 + S^\infty_2 = I_2^\infty+4 S_+^2 \bigg( T_+^2+\frac{\Omega_+^2}{4\pi^2} \bigg) .
\label{bal}
\ee
The angular velocity leads to an additional repulsive centrifugal force.

In fact, the last term on the right-hand side can also be written in terms of seed parameters $m,n$ or quantities defined at the inner horizon as
\be
4 G^2 S_+^2 \left( T_+^2+\frac{\Omega_+^2}{4\pi^2} \right) = m^2+n^2 = 4 G^2 S_-^2 \left( T_-^2+\frac{\Omega_-^2}{4\pi^2} \right).\label{bal2}
\ee
Using the latter relation, the identity \eqref{bal} amounts to the statement that the quantity $\Tr (\cQ^2)$ defined in \eqref{eq:Q2} is invariant under coset model transformations and therefore has the same value on the seed and final solutions. The identity therefore follows from a conservation law associated with the 3d coset model.



\section{Non-extremal special cases}
\label{nonextremal}


The general solution that we have constructed unifies many solutions in the literature.  We now show how these are special cases of our general solution.  We first describe non-extremal special solutions while some extremal limits will be discussed in Section \ref{part}. In all cases, the black hole entropy is given by \eqref{Cardyentropy} in terms of the angular momentum, the quartic invariant $\Delta$ \eqref{quarticinvariant} and the $F$-invariant \eqref{FF}. 


\subsection{Dyonic Kerr--Newman--Taub--NUT}


If $\gd_I = \gc_I = 0$, then all electromagnetic charges vanish.  This gives the Ricci-flat Kerr--Taub--NUT solution, which we used as the starting point of the solution generating technique.

More generally, if $\gd_I = \gd$ and $\gc_I = \gc$ for all gauge fields, then all electric charges are equal and all magnetic charges are equal.  This gives the dyonic Kerr--Newman--Taub--NUT solution \cite{demianskinewman} of Einstein--Maxwell theory \eq{EMLagrangian}.  The conserved charges are
\begin{align}
M & = m \cosh (2 \gd) \cosh (2 \gc) + n \sinh (2 \gd) \sinh (2 \gc) , \nnr
N & = n \cosh (2 \gd) \cosh (2 \gc) - m \sinh (2 \gd) \sinh (2 \gc) , \nnr
Q \equiv Q_I & = m \sinh (2 \gd) \cosh (2 \gc) + n \sinh (2 \gc) \cosh (2 \gd) , \nnr
P \equiv P^I & = m \sinh (2 \gc) \cosh (2 \gd) - n \sinh (2 \gd) \cosh (2 \gc) ,\nnr
J&= a \, M,
\end{align}
and the quartic invariant is
\be
\Delta = \tfrac{1}{4}(Q^2 + P^2)^2 .
\ee

Specializing to the dyonic Kerr--Newman solution, we set $n = n_0$, so that $N = 0$.  Then the $F$-invariant is 
\be
F = M^2 (M^2 - Q^2 - P^2).
\ee


\subsection{Kaluza--Klein black hole}


If $\gd_I = \gc_I = 0$ for $I = 2, 3, 4$ and $N = 0$, then we have the asymptotically flat, dyonic, rotating black hole \cite{Rasheed:1995zv, Matos:1996km, Larsen:1999pp} (see also \cite{Matos:2000ai}) of Kaluza--Klein theory \eq{KKLagrangian}.  The conserved charges are
\begin{align}
M & = \tf{1}{2} m (c_{\gd 1}^2 c_{\gc 1}^2 - 1) , & Q_1 & = \fr{2 m s_{\gd 1} (c_{\gd 1}^2 + s_{\gd 1}^2 s_{\gc 1}^2)}{c_{\gd 1}} , & J & = \fr{m a c_{\gc 1} (c_{\gd 1}^2 + s_{\gd 1}^2 s_{\gc 1}^2)}{c_{\gd 1}} , \nnr
N & = 0 , & P^1 & = \fr{2 m s_{\gc 1} c_{\gc 1}}{c_{\gd 1}} .
\label{DKK}
\end{align}
The quartic invariant and $F$-invariant are
\begin{align}
\Delta & = -\frac{1}{16} (Q_1)^2 (P^1)^2 , & F & = m^4 \frac{c_{\gamma_1}^2}{c^4_{\delta_1}}(c^2_{\delta_1} + s^2_{\delta_1}s^2_{\gamma_1})^3 ,
\end{align}
but the $F$-invariant is not easily expressed in terms of the conserved charges.  For this purpose, we define the monotonic function
\be
H(\psi) = 2 \cos \psi \, \cos(\psi/3) + 6 \sin \psi \, \sin(\psi/3) - 2 ,
\ee
where $0 \leq \psi \leq \pi/2$.  We take
\be
\sin^2 \psi (M, Q_1, P^1) = \frac{54 M^2 [(Q_1)^2 - (P^1)^2]^2}{[8 M^2 + (Q_1)^2 + (P^1)^2 ]^3} ,
\ee
which satisfies $0 \leq \psi \leq \pi/2$ for regular black hole configurations obeying $4 M \geq [(Q_1)^{2/3}+(P^1)^{2/3}]^{3/2}$.  Then, after some lengthy algebra, we obtain 
\be
F = [M^2 - \tfrac{1}{4}(Q_1)^2] [M^2-\tfrac{1}{4}(P^1)^2] + \tfrac{1}{3} \{ M^2 + \tfrac{1}{8} [(Q_1)^2 + (P^1)^2] \}^2 H (\psi (M, Q_1, P^1)) .
\label{FKK}
\ee
For this class of solutions, the triality invariance reduces to the $\mathbb Z_2$ invariance $Q_1 \rightarrow P^1$, $P^1 \rightarrow -Q_1$, under which $F$ is manifestly invariant. The function $H(\psi)$ was found by first expanding $F$ in terms of the sum and difference of squares of electric and magnetic charges in a perturbation series. The Taylor coefficients of the function $H(\psi)$ were then recognized as belonging to a hypergeometric series using an algorithm for integer sequence recognition\footnote{The algorithm can be found at \url{http://www.oeis.org}.}, then simplified in terms of trigonometric functions. The final result was then tested numerically. Finally, note that when $P^1=0$ we have
\be
F = \tfrac{1}{64} [ 32 M^4 - 40 M^2 (Q_1)^2 - (Q_1)^4 +4M (4M^2 + 2 (Q_1)^2)^{3/2} ] .
\ee
We therefore obtained a novel expression for the entropy of the Kaluza--Klein black hole 
\be
S_+ = 2 \pi \bigg( \sqrt{ F - \tfrac{1}{16} (Q_1)^2 (P^1)^2} + \sqrt{F-J^2 } \bigg) 
\ee
where $F$ is given in \eqref{FKK}, which could be used to study its thermodynamics further.


\subsection{Four electric charges (\texorpdfstring{Cveti\v{c}}{Cvetic}--Youm)}


If $\gc_I = 0$ and $n=0$, then the NUT charge vanishes, $N = 0$, and we have the asymptotically flat, 4-charge Cveti\v{c}--Youm solution \cite{Cvetic:1996kv}.  The full explicit solution, including expressions for the gauge fields, was given in \cite{Chong:2004na}.  If we include non-vanishing $n$, then we recover the Kerr--Taub--NUT solution with 4 electric charges given in \cite{Chong:2004na}\footnote{We swap parameters $\gd_1$ and $\gd_2$, and correct a typographical error in the sign of $\chi_2$ for the solution with NUT charge presented there.}.

In our parametrization, $\mu_2 = \nu_1 = 0$.  The conserved charges are
\begin{align}
M & = \frac{m}{4}\sum_I \cosh (2 \delta_I) , & N & = n (c_{\gd 1 2 3 4} - s_{\gd 1 2  3 4} ), \nnr
Q_I & = m \sinh (2 \delta_I) , & P^I & = 2 n (c_{\gd 1} s_{\gd 2 3 4} - s_{\gd 1} c_{\gd 2 3 4}) .
\end{align}
The NUT charge can be set to zero by setting $n = 0$, which we assume from now on. The angular momentum is then
\be
J = m a (c_{\gd 1 2 3 4} - s_{\gd 1 2 3 4} ) .
\ee
The quartic invariant \eq{quarticinvariant} and $F$-invariant \eq{FF} are
\begin{align}
\Delta & = \frac{1}{4} Q_1 Q_2 Q_3 Q_4 , \nnr
F & = \frac{1}{8} \bigg( m^4 - 4\Delta + \prod_I \sqrt{m^2+Q_I^2} + m^2 \sum_{I < J} \sqrt{m^2+Q_I^2}  \sqrt{m^2+Q_J^2} \bigg) .
\end{align}
We have not found a closed form expression for the $F$-invariant in terms of physical charges only. 

Let us also present the metric in our notation. The master function \eqref{W2o} takes the almost factorized form
\be
W^2(r,u) = (r^2 - 2 m r+u^2)(r^2 + 2(2M-m)r+u^2) + 4 \nu_2^2 m^2 r^2.
\ee
The metric is then given by
\ben
\df s^2 = - \fr{r^2-2mr +u^2}{W(r,u)} (\df t + \omega_{3} )^2 + W(r,u) \bigg( \fr{\df r^2}{R(r)} + \frac{du^2}{a^2 -u^2} + \frac{R(r) (a^2-u^2) }{a^2 (r^2-2m r +u^2)} \df \phi^2 \bigg) ,\label{genmetric3}
\een
where $R(r)=r^2 - 2 mr+a^2$ and the Kaluza--Klein 1-form is
\be
\omega_3 = \frac{2\nu_2 m(a^2-u^2)r}{a(r^2-2 m r+u^2)} \, \df \phi .
\ee


\subsection{\texorpdfstring{$-\im X^0 X^1$}{-i X0 X1} supergravity black hole}


If we set the electric and magnetic charges pairwise equal, which is equivalent to $(\gd_1, \gc_1) = (\gd_4, \gc_4)$ and $(\gd_2, \gc_2) = (\gd_3, \gc_3)$, then we have the dyonic rotating black hole  \cite{LozanoTellechea:1999my} of $-\im X^0 X^1$ supergravity \eq{iX0X1Lagrangian}. The dyonic Kerr--Newman--Taub--NUT is recovered upon setting $Q_2 = Q_1$, $P^2 = P^1$. 

The solution is substantially simpler in this truncation.  To simplify the solution and physical quantities, it is convenient to define
\begin{align}
\gD r_1 & = m [\cosh (2 \delta_1) \cosh (2 \gamma_2) - 1] + n \sinh (2 \delta_1) \sinh (2 \gamma_1) , \nnr
\gD r_2 & = m [\cosh (2 \delta_2) \cosh (2 \gamma_1) - 1] + n \sinh (2 \delta_2) \sinh (2 \gamma_2) , \nnr
\gD u_1 & = n [\cosh (2 \delta_1) \cosh (2 \gamma_2) - 1] - m \sinh (2 \delta_1) \sinh (2 \gamma_1) ,\nnr
\gD u_2 & = n [\cosh (2 \delta_2) \cosh (2 \gamma_1) - 1] - m \sinh (2 \delta_2) \sinh (2 \gamma_2) ,
\end{align}
and
\begin{align}
r_1 & = r + \Delta r_1 , & r_2 & = r + \Delta r_2 , & u_1 & = u + \Delta u_1 , & u_2 & = u + \Delta u_2 .
\end{align}
Then $W = r_1 r_2 + u_1 u_2$ and the metric takes the simplified form
\begin{align}
\df s^2 & = - \fr{R}{W} \bigg( \df t - \fr{a^2 - u_1 u_2 + (\Delta u_1 + n) (\Delta u_2 + n)}{a} \, \df \phi \bigg) ^2 + \fr{W}{R} \, \df r^2 \nnr
& \quad + \fr{U}{W} \bigg( \df t - \fr{r_1 r_2 + a^2 + (\Delta u_1 + n) (\Delta u_2 + n)}{a} \, \df \phi \bigg) ^2 + \fr{W}{U} \, \df u^2 .
\end{align}
The gauge fields and duals are
\begin{align}
A^1 & = \fr{Q_1 r_2}{W} \bigg( \df t - \fr{a^2 - u_1 u_2 + (\Delta u_1 + n) (\Delta u_2 + n)}{a} \, \df \phi \bigg) \nnr
& \quad - \fr{P^1 u_2}{W} \bigg( \df t - \fr{r_1 r_2 + a^2 + (\Delta u_1 + n) (\Delta u_2 + n)}{a} \, \df \phi \bigg) + \fr{(\Delta u_2 + n)}{2 a} \fr{\pd (\Delta u_1)}{\pd \gd_1} \, \df \phi ,
\end{align}
and
\begin{align}
\wtd{A}_1 & = \fr{P^1 r_1}{W} \bigg( \df t - \fr{a^2 - u_1 u_2 + (\Delta u_1 + n) (\Delta u_2 + n)}{a} \, \df \phi \bigg) \nnr
& \quad + \fr{Q_1 u_1}{W} \bigg( \df t - \fr{r_1 r_2 + a^2 + (\Delta u_1 + n) (\Delta u_2 + n)}{a} \, \df \phi \bigg) + \fr{(\Delta u_1 + n)}{2 a} \fr{\pd (\Delta r_1)}{\pd \gd_1} \, \df \phi ,
\end{align}
with $A^2$ and $\wtd{A}_2$ obtained by interchanging $1 \leftrightarrow 2$.  Here, the partial derivatives with respect to $\delta_I$ are performed after setting $(\gd_1, \gc_1) = (\gd_4, \gc_4)$ and $(\gd_2, \gc_2) = (\gd_3, \gc_3)$.  The non-trivial scalar fields are
\begin{align}
\expe{\gvf_1} & = \fr{r_2^2 + u_2^2}{W} , & \chi_1 & = \fr{r_2 u_1 - r_1 u_2}{r_2^2 + u_2^2} .
\end{align}
Using a linear coordinate transformation of the coordinates $t$ and $\phi$, and a gauge transformation, the metric and gauge fields may be written in the simpler form
\be
\df s^2 = - \fr{R}{W} (\df \tau + u_1 u_2 \, \df \psi)^2 + \fr{U}{W} (\df \tau - r_1 r_2 \, \df \psi)^2 + W \bigg( \fr{\df r^2}{R} + \fr{\df u^2}{U} \bigg) ,
\ee
and
\begin{align}
A^1 & = \fr{Q_1 r_2}{W} (\df \tau + u_1 u_2 \, \df \psi) - \fr{P^1 u_2}{W} (\df \tau - r_1 r_2 \, \df \psi) , \nnr
\wtd{A}_1 & = \fr{P^1 r_1}{W} (\df \tau + u_1 u_2 \, \df \psi) + \fr{Q_1 u_1}{W} (\df \tau - r_1 r_2 \, \df \psi) .
\end{align}
Guided by this simplified form of the solution, asymptotically AdS generalizations in gauged supergravity were obtained in \cite{Chow:2013gba}.

The parameters for the mass and NUT charge are
\begin{align}
\nu_1 & = - \mu_2 = - \tf{1}{2} [\sinh (2 \gd_1) \sinh (2 \gc_1) + \sinh (2 \gd_2) \sinh (2 \gc_2)] , \nnr
\nu_2 & = \mu_1 = \tf{1}{2} [\cosh (2 \gd_1) \cosh (2 \gc_2) + \cosh (2 \gd_2) \cosh (2 \gc_1)] .
\end{align}
The conserved charges are therefore
\begin{align}
M & = m + \tf{1}{2} (\Delta r_1 + \Delta r_2) , & N & = n + \tf{1}{2} (\Delta u_1 + \Delta u_2) , \nnr
Q_1 & = \frac{\p M}{\p \delta_1} = \frac{1}{2}\frac{\p (\Delta r_1)}{\p \delta_1}, & P^1 & = - \frac{\p N}{\p \delta_1} = - \frac{1}{2}\frac{\p (\Delta u_1)}{\p \delta_1}, \nnr
Q_2 & = \frac{\p M}{\p \delta_2} = \frac{1}{2}\frac{\p (\Delta r_2)}{\p \delta_2}, & P^2 & = - \frac{\p N}{\p \delta_2} = - \frac{1}{2}\frac{\p (\Delta u_2)}{\p \delta_2} , \nnr
J & = M a .
\end{align}
Since the gauge fields are set pairwise equal before taking $\gd_I$ derivatives, there is a factor of 2 difference for the electromagnetic charges compared with the general formulae \eq{charges2}.

Setting the NUT charge to zero, we obtain the quartic and $F$-invariants 
\begin{align}
\Delta & = (\tfrac{1}{2} I^\infty_2 - M^\infty_2)^2=  \tfrac{1}{4}(Q_1 Q_2 + P^1 P^2)^2, \nnr
F & = (M^2 -\tfrac{1}{2} I^\infty_2 )^2-\Delta = (M^2 - M^\infty_2) (M^2 +M^\infty_2 -I^\infty_2) 
\end{align}
in terms of other invariants defined in \eqref{M2inf} and \eqref{I2inf}, and which read here
\begin{align}
I^\infty_2 & = \tfrac{1}{2} [(Q_1)^2 +(P^1)^2+(Q_2)^2 +(P^2)^2] , & M^\infty_2 & =  \tfrac{1}{4} [(Q_1+P^1)^2+(Q_2+P^2)^2] .
\end{align}

If $(\gd_1, \gc_1) = (\gd_4, \gc_4)$ and $\gd_2 = \gd_3 = \gc_2 = \gc_3 = 0$, then we have the Einstein--Maxwell--dilaton--axion solution of \cite{Galtsov:1994pd}, which is labelled by its conserved charges $Q_1,P^1,M,J$.


\subsection{Reduction of the black string of minimal 5d supergravity}


If $\gd_2 = \gd_3 = \gd_4$, $\gc_2 = \gc_3 = \gc_4$ and $P^1 = N = 0$, then we have the Kaluza--Klein reduction of the most general asymptotically Kaluza--Klein homogeneous 5-dimensional black string of minimal $\cN=1$ 5d supergravity \cite{Compere:2010fm}. The solution is labelled by its conserved charges $M,J,Q_1,Q_2,P^2$. The charge $Q_1$ is the momentum along the string in the Kaluza--Klein direction while $Q_2$ and $P^2$ are the 4-dimensional electromagnetic charges. 


\subsection{One dyonic gauge field and two magnetic gauge fields}


If $P^4 = Q_2 = Q_3 = Q_4 =0$ and $N=0$, then we have an analytic continuation of the Kaluza--Klein black hole solution with two additional magnetic charges of \cite{Giusto:2007tt}.


\section{Extremal black holes}
\label{part}


An extremal black hole is characterized by the property that its Hawking temperature vanishes. All extremal black holes enjoy the attractor mechanism, which states that at the horizon all scalar moduli reach an extremum value, which is solely a function of the electromagnetic charges and angular momentum carried by the black hole. In terms of 3-dimensional coset models, extremal black holes lie on nilpotent orbits of the symmetry algebra of the coset model. 

There are a number of different extremal solutions that may be obtained as limits of our general non-extremal solution\footnote{See \cite{Andrianopoli:2013kya,Andrianopoli:2013jra} for developments on a limiting procedure for relating non-extremal to extremal coset orbits.}. We will not attempt a classification but simply present 3 extremal limits of general interest that lead to black holes with finite area: the $1/8-$Bogomolny--Prasad--Sommerfield (BPS) static black hole, the extremal fast rotating black hole and the extremal slow rotating black hole which includes as a subcase the regular static extremal non-BPS black hole.


\subsection{Static 1/8--BPS limit}
\label{18bps}

Supersymmetric black holes of $\cN = 8$ supergravity which are $1/2$--BPS or $1/4$--BPS have zero area in the supergravity regime, see e.g. \cite{Arcioni:1998mn}. Instead, the $1/8$--BPS black holes have finite area. Such black holes can be generated through U-dualities from a $1/8$--BPS black hole of the $STU$ model, as constructed in \cite{Cvetic:1995bj,Behrndt:1996hu,Bertolini:1998mt,Bertolini:1999je}. In this section, we will show how the $1/8$--BPS black hole can be obtained as a specific extremal limit of the non-extremal solution. 


In the static case $a=0$, we take the limit $\eps \rightarrow 0$ while scaling
\begin{align}
m & \sim \eps^2, & \delta_I & \sim \eps^0, & \expe{\gamma_I} & \sim \eps^{-1} .
\label{BPSl}
\end{align}
The solution admits 4 independent electric and 4 independent magnetic charges. The mass saturates the BPS bound
\be
M^2 = M_2^\infty
\ee
where $M_2^\infty$ is defined in \eqref{M2inf}, which indicates that the solution is supersymmetric. The $F$-invariant is zero in the limit. The quartic invariant is non-negative, $\Delta \geq 0$, and the entropy \eq{Cardyentropy} is
\be
S_+ = 2\pi \sqrt{\Delta},
\ee
which reproduces the known entropy formula \cite{Duff:2006uz}. Since the area is generically non-vanishing, the black hole is $1/8$--BPS. The scalar fields also obey the particular property
\be
S_2^\infty = I_2^\infty - M_2^\infty\, 
\ee
where these quantities have been defined in Section \ref{phys}.

The metric \eqref{generalmetric} takes the isotropic form
\be
\df s^2 = -r^2 W_0^{-1}(r) \, \df t^2 + W_0(r) r^{-2}(\df r^2 + r^2 \, \df \Omega^2),
\ee
and the scalar fields admit a non-trivial radial profile interpolating between the attractor values at the horizon and trivial values at infinity, as imposed by asymptotic flatness. Up to U-dualities, the black hole is expected to reduce to the one discussed in \cite{Cvetic:1995bj,Behrndt:1996hu,Bertolini:1998mt,Bertolini:1999je}.


\subsection{Extremal fast rotating solution}
\label{maxrot}


The extremal, fast rotating solution is achieved for $a = \sqrt{m^2+n_0^2}$.  The solution admits 4 independent electric and 4 independent magnetic charges as well as angular momentum. There is a degenerate horizon at $r = r_+ = r_- = m$. Using \eqref{bal} and \eqref{bal2}, the mass obeys the remarkable formula
\be
M^2 = I_2^\infty - S^\infty_2 + a^2.
\ee
In our parametrization, we have $J/a = m (\nu_1^2+\nu_2^2)/\nu_2$. Therefore, the $F$-invariant can be evaluated as
\be
F = J^2.
\ee
The entropy \eq{Cardyentropy} then becomes
\be
S_+ = 2 \pi \sqrt{\Delta + J^2}.
\ee
The entropy of the generic extremal rotating black hole is therefore independent of scalar moduli in general, since it is only a functional of the quartic invariant and the angular momentum. This is a feature of the attractor mechanism.

Angular momentum breaks supersymmetry. In the BPS limit \eqref{BPSl}, $a \rightarrow 0$ and $J/a \sim \eps^0$, then $J \rightarrow 0$ and all conserved quantities coincide with those of Section \ref{18bps}. Therefore, one can also consider the BPS limit as a special limit of the extremal fast rotating solution.

The near-horizon limit is defined as
\begin{align}
t & \rightarrow r_0 \lambda^{-1} t , & r & \rightarrow r_+ + \lambda r_0 r , & \phi & \rightarrow \phi +\Omega_+^{\textrm{ext}} \lambda^{-1}r_0 t ,
\end{align}
and
\begin{align}
A^I & \rightarrow A^I - \Phi^I_{+,\textrm{ext}} \lambda^{-1} r_0 \, \df t , & \widetilde{A}_I & \rightarrow \widetilde{A}_I - \Psi^+_{I, \textrm{ext}} \lambda^{-1} r_0 \, \df t ,
\end{align}
where $\lambda \rightarrow 0$, $\Omega_+^{\textrm{ext}}$, $\Phi^I_{+,\textrm{ext}}$,$\Psi^+_{I,\textrm{ext}}$ are the chemical potentials at extremality and $r_0$ is an overall constant that we choose to be $r_0^2 = L(r_+)$.  The near-horizon metric is
\be
\df s^2 = W_+ \bigg( -r^2 \, \df t^2 + \frac{\df r^2}{r^2} + \frac{\df u^2}{U} + \Gamma^2 (\df \phi + k r \, \df t)^2 \bigg) ,
\ee
where $W_+ (u) = W(r_+, u)$, and
\begin{align}
\Gamma^2 (u) & = \fr{L(r_+)^2 U(u)}{a^2 W_+^2(u)} , & k & = 2( m \nu_2 -n_0 \nu_1 )\Omega_+ = \frac{2 \pi J}{S_+} .
\end{align}
The near-horizon gauge fields are
\begin{align}
A^I & = f^I (\df \phi + k r \, \df t) + \frac{e^I}{k} \, \df \phi , & \widetilde A_I & = \widetilde f_I (\df \phi + k r \, \df t) + \frac{\widetilde e_I}{k} \, \df \phi  ,
\end{align}
where
\begin{align}
f^I(u) & = - \frac{L(r_+)}{a} \bigg( \zeta^I(r_+, u) + \frac{ \nu_1 \pi^I_1 + \nu_2 \pi^I_2}{ 2 (\nu_1^2 + \nu_2^2)} \bigg) , & e^I & = 2 (m \nu_2 - n_0 \nu_1)\Phi_+^I-n_0 \pi^I_1+m \pi^I_2, \nnr
\widetilde f_I(u) & = - \frac{L(r_+)}{a} \bigg( \widetilde \zeta_I(r_+, u) - \frac{\nu_1 \rho_I^1 + \nu_2 \rho_I^2}{2 (\nu_1^2 + \nu_2^2)} \bigg) , & \widetilde e_I & = 2 (m \nu_2 - n_0 \nu_1) \Psi^+_I+n_0 \rho_I^1-m \rho_I^2 .
\end{align}
The geometry has the expected enhanced $\SL(2,\mathbb R) \times \textrm{U} (1)$ symmetry \cite{Kunduri:2007vf} and the expected functional form \cite{Kunduri:2011zr}. In the BPS limit, $k=0$ and the geometry reduces to AdS$_2 \times S^2$.

Following  the Kerr/CFT conjecture \cite{Guica:2008mu}, the entropy is reproduced by Cardy's formula
\be
S_+ = \tfrac{1}{3} \pi^2 c_J T_J
\ee
for a chiral sector of a CFT with central charge and temperature
\begin{align}
c_J & = 12 J , & T_J & = \frac{1}{2\pi k }.
\end{align}
We expect that boundary conditions exist when a Virasoro algebra acts as asymptotic symmetry algebra, as in all known subcases (see \cite{Compere:2012jk} for references).  A distinct description of the entropy is in terms of Cardy's formula
\be
S_+ = \tf{1}{3} \pi^2 c_{Q_1} T_{Q_1}
\ee
for a chiral sector of a CFT with central charge and temperature
\begin{align}
c_{Q_1} & = 24 \frac{\p \Delta}{\p Q_1} , & T_{Q_1} & = \frac{1}{2\pi e^1} ,
\end{align}
which generalizes \cite{Hartman:2008pb, Chen:2013rb}. More explicitly,
\be
c_{Q_1} = 6 Q_2 Q_3 Q_4 +  3 P^1 (P^2 Q_2 + P^3 Q_3 + P^4 Q_4 - P^1 Q_1) .
\ee
There are similar expressions corresponding to the other electromagnetic charges.


\subsection{Extremal slow rotating and non-BPS static limit}
\label{secslow}


An extremal limit with slow rotation is defined as
\begin{align}
m & \sim \eps^2 m, & n & \sim \eps n, & a & \sim \eps a, & \expe{\gamma_1} & \sim \eps^{-1} \expe{\gamma_1}
\label{nonBPSl}
\end{align}
with $\eps \rightarrow 0$ and the remaining parameters ($\gamma_2,\gamma_3,\gamma_4,\delta_I$, $I=1,2,3,4$) unscaled. The non-BPS static limit is defined analogously but with $a=0$. There are four distinct limits depending on the choice of $\gamma_I$, $I=1,2,3,4$ that is blown up. By permutation symmetry, all limits lead to the same metric. We expect that the four different limits are related by field redefinitions of the charging parameters without changing the physics. Since $n_0 =O(\eps)$, one can set the NUT charge to zero by setting the final $n = n_0$. Besides angular momentum, the solution admits 4 independent electric and 4 independent magnetic charges. 

In the limit \eqref{nonBPSl} the temperature $T_+$ and angular velocity $\Omega_+$ vanish. The horizon is located at $r =0$. In the limiting procedure $r_+ = - r_- +O(\eps^2)$, which implies that $S_- = -S_+$. From \eqref{prodarea}, we deduce that $J^2 +\Delta \leq 0$, the $F$-invariant is $F = - \Delta$ and the entropy \eq{Cardyentropy} \textbf{}becomes
\be
S_+ = 2\pi \sqrt{-\Delta -J^2}.
\ee
Since $\Delta \leq 0$, there are no BPS solutions with finite area in this class. One can explicitly check that the mass obeys
\be
M^2 =  I_2^\infty - S^2_\infty
\ee
and, in particular, it does not depend upon the angular momentum $J$. Upon setting to zero all magnetic charges, the solution reduces to the extremal slow rotating four-charge extremal solution studied in Section 5 of  \cite{Bena:2009ev}\footnote{Note that contrary to the claim of \cite{Bena:2009ev}, at least five independent electromagnetic charges are necessary to obtain a generating solution.}.

Regular extremal static non-BPS black holes with 8 independent electromagnetic charges are obtained by setting $J=0$. One such class of black holes labeled by 2 independent parameters was obtained in \cite{Song:2011ii}. In that case, we have the charge assignments $\widetilde P_1 = 1$, $\widetilde P_2 = \widetilde P_3$, $\widetilde Q_2 = \widetilde Q_3$ and $P^4=0$. 

The general metric can be obtained by taking the limit \eqref{nonBPSl}. It turns out that the functions $L(r)$ and $V(u)$ defined in \eqref{LV} blow up as $L=O(\eps^{-1})$, $V = O(\eps^{-2})$. Therefore, the form of the $W^2$ and $\omega_3$ functions is not adapted to the description of the extremal slow rotating limit. However, these functions are finite in the limit, and we find
\begin{align}
W^2 & = r^4 + 4 Mr^3 + (M^2 b_1 + b_2 J \cos\theta ) r^2 + b_3 M^3 r -4J^2 \cos^2\theta -4 \Delta, & \omega_3 & = \frac{2J}{r} \sin^2 \theta \, \df \phi ,
\label{WnonBPS}
\end{align}
where $b_1,b_2,b_3$ only depend on the charging parameters $(\delta_I,\gamma_I)$, $I=1,2,3,4$. We have $b_1 \geq 0,b_3 \geq 0$. The form of $\omega_3$ is exceptionally simple and only depends on the physical angular momentum.  Since $J^2 \leq -\Delta$, $W^2$ is indeed positive near $r=0$, which is the location of the extremal horizon.  Finally, the metric is
\be
\df s^2 = - \fr{r^2}{W(r,\theta)} \bigg( \df t + \frac{2J}{r} \sin^2 \theta \, \df \phi \bigg) ^2 + \fr{W(r, \theta)}{r^2} [\df r^2 + r^2 (\df \theta^2 + \sin^2 \theta \, \df \phi^2)] .
\ee
The matter fields can be obtained from the limit and we will not display them here.

In the near-horizon limit, we replace
\begin{align}
t & \rightarrow \lambda^{-1} r_0 \sqrt{-\Delta-J^2} t , & r & \rightarrow \lambda r_0 r,
\label{nhl}
\end{align}
and
\begin{align}
A^I & \rightarrow A^I - \df ( \Phi^I_+  \lambda^{-1} r_0 \sqrt{-\Delta-J^2} t), & \widetilde A_I & \rightarrow \widetilde A_I - \df ( \Psi_I^+ \lambda^{-1} r_0 \sqrt{-\Delta-J^2} t) ,
\label{gaugetransform}
\end{align}
and then take $\lambda \rightarrow 0$.  For convenience, we fix $r_0 = \sqrt{2}$ for convenience, we obtain
\begin{align}
\df s^2 & = W_+ \bigg( -r^2 \, \df t^2 + \fr{\df r^2}{r^2} + \df \theta^2 + \Gamma^2 (\df \phi - k r \, \df t)^2 \bigg)
\end{align}
with
\begin{align}
W_+ & = 2\sqrt{-\Delta -J^2 \cos^2\theta}, & k & = \frac{J}{\sqrt{-\Delta+J^2}} , &
\Gamma^2 & = \sin^2 \theta \frac{- \Delta - J^2}{- \Delta - J^2 \cos^2\theta}.
\end{align}
The geometry only depends upon the quartic invariant and the angular momentum and it admits the expected enhanced $\SL(2,\mathbb R) \times \textrm{U}(1)$ symmetry \cite{Kunduri:2007vf}.
  The gauge fields in the near-horizon limit can be most easily obtained by taking the extremal limit with slow rotation \eqref{nonBPSl}  followed by the near-horizon limit \eqref{nhl} of the expression \eqref{AIW}. In order to evaluate the latter expression, we need to keep $n$ general, and take $n=n_0$ only after taking the derivative with respect to $\delta_I$. We first note that
$C  = -\nu_1^2 +O(\eps^{-2})$ and $\Phi_+^I = \p_{\delta_I}\log \nu_1 +O(\eps)$. Then, we obtain after the limit $\eps \rightarrow 0$,
\begin{align}
W^2 & = 4 \nu_1^2 n^2 (n^2 - a^2 \cos^2\theta)+O(\lambda) , & \omega_3 & = - \frac{2 a n \nu_1}{\lambda r_0 r} \sin^2\theta \, \df \phi+O(\lambda^0).
\end{align}
After using \eqref{charges2}, we get $\xi^I = \p_{\delta_I}\log \nu_1$ and $A^I = P^I \cos\theta \, \df \phi - \p_{\delta_I}\log \nu_1 \, \omega_3$, which results finally in
\begin{align}
A^I & = f^I (\df \phi - k r \, \df t)- \frac{e^I}{k} \, \df \phi , & f^I & = \frac{P^I (-\Delta-J^2) (\widehat \pi^I + J \cos \theta)}{J(-\Delta -J^2 \cos^2\theta)} , & e^I & = \frac{P^I \widehat \pi^I}{\sqrt{-\Delta-J^2}} , \nnr
\widetilde A^I & = \widetilde f_I (\df \phi - k r \, \df t) - \frac{\widetilde e_I}{k} \, \df \phi , & \widetilde f_I & = \frac{Q_I (-\Delta-J^2) (\widehat \rho_I -J\cos\theta)}{J(-\Delta -J^2 \cos^2\theta)}, & \widetilde e_I & = \frac{Q_I  \widehat \rho_I }{\sqrt{-\Delta-J^2}} .
\end{align}
after performing the gauge transformation as indicated in \eqref{gaugetransform}. After analysis, we find
\begin{align}
\widehat \pi^I & = -2 \frac{1}{P^I}\frac{\p \Delta}{\p Q_I}, & \widehat \rho_I & = -2 \frac{1}{Q_I}\frac{\p \Delta}{\p P^I}.
\end{align}

Following the Kerr/CFT conjecture \cite{Guica:2008mu}, the entropy is reproduced by Cardy's formula
\be
S_+ = \tfrac{1}{3} \pi^2 c_J T_J
\ee
for a chiral sector of a CFT with central charge and temperature
\begin{align}
c_J & = 12 J , & T_J & = \frac{1}{2\pi k }.
\end{align}
Eight other Cardy formulae hold,
\be
S_+ = \tf{1}{3} \pi^2 c_{Q_I} T_{Q_I} = \tf{1}{3} \pi^2 c_{P^I} T_{P^I} ,
\ee
one for each electric or magnetic charge, with central charges and temperatures
\begin{align}
c_{Q_I} & = - 6 \frac{\p \Delta}{\p \overline Q_I}, & T_{Q_I} & = \frac{1}{2\pi e^I}, \nnr
c_{P^I} & = - 6 \frac{\p \Delta}{\p \overline P^I}, & T_{P^I} & = \frac{1}{2\pi \widetilde e_I}.
\end{align}


\subsubsection{Kaluza--Klein black hole}


Let us present the details of the extremal slow rotating solution in the case where the only non-zero electromagnetic charges are $Q \equiv Q_1$, $P \equiv P^1$, corresponding to the charging parameters $\delta \equiv \delta_1$ and $\gamma \equiv \gamma_1$. This is a 4-dimensional solution of Kaluza--Klein theory, about by reduction of the 5-dimensional Einstein gravity \cite{Rasheed:1995zv,Matos:1996km,Larsen:1999pp} (see also \cite{Jana:2013gda}).

Extremality fixes the mass in terms of the electromagnetic charges. In our parametrization, we find
\begin{align}
M & = \frac{m \expe{2\gamma} \cosh^2 \delta}{8} , & \overline Q & = \frac{m \expe{2\gamma} \sinh^3\delta}{8 \cosh \delta} , & \overline P & = \frac{m \expe{2\gamma}}{8\cosh \delta} ,
\end{align}
which satisfy
\be
M^{2/3} = \overline Q^{2/3} + \overline P^{2/3} .
\ee
Let us assume for simplicity and without loss of generality that $ \overline Q \geq 0$, $ \overline P \geq 0$. Then we have the factorization $W^2 = W_Q W_P$, where
\begin{align}
W_Q & = r^2 + 4 \overline Q^{2/3} \sqrt{\overline Q^{2/3} + \overline P^{2/3}} r + 8 \overline Q^{1/3} \overline P^{-1/3} ( \overline Q \, \overline P -J \cos\theta) , \nnr
W_P & = r^2 + 4 \overline P^{2/3} \sqrt{\overline Q^{2/3} + \overline P^{2/3}}r +8 \overline P^{1/3} \overline Q^{-1/3} ( \overline Q \, \overline P + J \cos\theta) .
\end{align}
Note that reversing the spacetime orientation would lead to a change $J \rightarrow -J$ in $W_Q$ and $W_P$ as a consequence of the equations of motion.  

The coefficients in the gauge fields in the near-horizon limit are given by
\be
\widehat \pi^1 = \widehat \rho_1 = \sqrt{-\Delta} = \overline Q \, \overline P .
\ee
At the horizon $r=0$, the scalar moduli reduce to
\begin{align}
x_i & = 0, & y_2=y_3 = \fr{1}{y_1} & = \frac{\overline P^{2/3} (\overline Q \, \overline P + J \cos\theta)}{\overline Q^{2/3} (\overline Q \, \overline P - J \cos\theta)} .
\end{align}

Introducing $\psi \sim \psi+2\pi$, one can reconstruct a 5-dimensional Ricci-flat metric as
\begin{align}
\df s^2_5 & = f^2(\theta)  \bigg[ R_\psi \df \psi - \frac{\overline P r}{\overline Q\, \overline P -J\cos\theta } \bigg( \df t + \frac{J}{r} \sin^2\theta \, \df \phi  \bigg) + \overline P \cos\theta \, \df \phi \bigg] ^2 \nnr
& \qquad + \frac{ G(\theta)}{2 f(\theta)} \bigg[ - \fr{r^2}{G^2(\theta)} \bigg( \df t + \frac{J}{r}\sin^2\theta \, \df \phi \bigg) ^2  + \frac{\df r^2}{r^2} + \df \theta^2 + \sin^2\theta \, \df \phi^2 \bigg] ,
\end{align}
where $R_\psi$ is arbitrary and 
\begin{align}
f(\theta) & =  \left( \frac{\overline Q}{\overline P}\right)^{1/3} \left( \frac{\overline Q\, \overline P - J \cos\theta }{\overline Q \, \overline P + J\cos\theta }\right)^{1/2} , & G(\theta) & = \sqrt{\overline Q^2 \overline P^2 - J^2 \cos^2\theta} .
\end{align}
The near-horizon metric is obtained by replacing $r \rightarrow \lambda r$, $t \rightarrow t/\lambda$, and then taking the limit $\lambda \rightarrow 0$; it falls into the classification of \cite{Kunduri:2008rs}. The geometry of the horizon is globally $S^3$. The metric can be put in the form
\be
\df s^2 = \Gamma(\theta) \bigg( -r^2 \, \df \overline t^2 + \frac{\df r^2}{r^2} + \df \theta^2 + \sum_{A,B=1}^2 \gamma_{AB}(\theta) (\df \phi^A - k^A r \, \df \overline t) (\df \phi^B - k^B r \, \df \overline t) \bigg) ,
\ee
where $\overline t = t/(\overline Q^2 \, \overline P^2 -J^2)^{1/2}$, $\phi^1 = \phi$, $\phi^2 = R_\psi \psi$,
\begin{align}
\Gamma(\theta) & = \frac{ \overline P^{1/3} (\overline Q \, \overline P + J \cos\theta)}{2 \overline Q^{1/3}} , & k^1 & = \frac{J}{(\overline Q^2 \, \overline P^2 - J^2)^{1/2}}, & k^2 = \frac{\overline Q \, \overline P^2}{(\overline Q^2 \, \overline P^2 - J^2)^{1/2}} ,
\end{align}
and
\be
\gamma_{A B} =
\fr{1}{(\overline Q \, \overline P + J \cos \theta)^2}
\begin{pmatrix}
\overline Q^2 \overline P^2 - J^2 \cos^2 \theta + (\overline Q \, \overline P \cos \theta - J)^2 & 2 \overline Q (\overline Q \, \overline P \cos\theta - J) \\
2 \overline Q (\overline Q \, \overline P \cos\theta - J) & 2 \overline Q (\overline Q \, \overline P - J \cos\theta)/\overline P
\end{pmatrix}
.
\ee
It admits an $\SL(2,\mathbb R) \times \textrm{U}(1)^2$ symmetry.  The Killing vectors $\xi_{-1} = \p_t$, $\xi_0 = r \, \pd_r - t \, \pd_t$ and
\be
\xi_1 = \bigg( \frac{1}{2r^2} + \frac{\overline t^2}{2} \bigg) \, \p_{\overline t} -\overline t r \, \pd_r + \frac{k^1}{r} \, \pd_\phi + \frac{k^2}{R_\psi r} \, \pd_\psi
\ee
satisfy the $\SL(2, \bbR)$ commutators $[\xi_0, \xi_1] = - \xi_1$, $[\xi_0, \xi_{-1}] = \xi_{-1}$ and $[\xi_{-1}, \xi_1] = - \xi_0$.

Following the Kerr/CFT conjecture \cite{Guica:2008mu}, the entropy is reproduced by either of Cardy's formulae
\be
S_+ = \tfrac{1}{3} \pi^2 c_J T_J = \tfrac{1}{3} \pi^2 c_Q T_Q
\ee
for a chiral sector of CFTs with central charges and temperatures
\begin{align}
c_J & = 12 J , & T_J & = \frac{1}{2\pi k^1 },\\
c_Q &= - 6 \frac{\p \Delta}{\p \overline Q}, & T_Q & = \frac{1}{2\pi k^2 },
\end{align}
as obtained in \cite{Azeyanagi:2008kb} (see also \cite{Li:2010ch}).


\section{Killing tensors and separability}
\label{Killingtensors}


It is well-known that the Kerr solution possesses various types of Killing tensors.  These tensors are related to the integrability of geodesic motion, and the separability of the Klein--Gordon equation and the Dirac equation.  Black hole solutions of $\cN = 8$ supergravity also involve metrics that possess various types of Killing tensors as we will now demonstrate.  Using \eqref{W2o}, the metric \eqref{genmetric2} can be written in the form
\begin{align}
\label{separablemetric}
\df s^2 & = - \fr{R - U}{W} \, \df t^2 - \fr{(L_u R + L_r U)}{a W} \, 2 \, \df t \, \df \phi  + \fr{(W_r^2 U - W_u^2 R)}{a^2 W} \, \df \phi^2 + W \bigg( \fr{\df r^2}{R} + \fr{\df u^2}{U} \bigg) ,
\end{align}
where
\be
\label{separableW}
W^2 = (R - U) \bigg( \frac{W_r^2}{R} - \frac{W_u^2}{U} \bigg) + \frac{(L_u R + L_r U)^2}{R U} .
\ee
Its determinant is $\sr{-g} = W $. For the black hole solution, the functions $R(r)$ and $U(u)$ are given in \eq{RUgeneral}, and 
\begin{align}
L_r (r)& = L + 2 N n , & W_r^2 (r)& = R^2+4 M r R +(L+2 N n)^2 , \nnr
L_u (u)& = 2 N(u-n), & W_u^2 (u)& = U^2 - 2 U V +4 N^2(u-n)^2,
\end{align}
where $L(r)$ and $V(u)$ are given in \eqref{LV}.

Henceforth, in this section we consider a more general class of metrics of the form \eq{separablemetric}.  We generalize so that: $R$, $W_r$ and $L_r$ are arbitrary functions of $r$; $U$, $W_u$ and $L_u$ are arbitrary functions of $u$; and $W$ satisfies \eq{separableW}.  There are two conformally related metrics of interest: the usual Einstein frame metric $\df s^2$, and the string frame metric\footnote{Note added in v4: The convenient choice of frame used here should strictly be called ``string frame'' in only certain subcases, such as with a single non-zero electric charge.  In general, there are several possible string frames, depending on which dilaton combination is identified with a string coupling constant.}
\be
\df \wtd{s}^2 = \fr{r^2 + u^2}{W} \, \df s^2 ,
\label{stringframemetric}
\ee
whose inverse $(\pd / \pd \wtd{s})^2$ is given by
\be
(r^2+u^2) \bigg( \fr{\pd}{\pd \wtd{s}} \bigg) ^2 = R \, \pd_r^2 + U \, \pd_u^2 + \bigg(\fr{W_u^2}{U} - \fr{W_r^2}{R} \bigg) \, \pd_t^2  - a \bigg( \fr{L_r}{R} + \fr{L_u}{U} \bigg) \, 2 \, \pd_t \, \pd_\phi +a^2 \bigg( \fr{1}{U} - \fr{1}{R} \bigg) \, \pd_\phi^2 .
\ee

Let us recall some definitions of Killing tensors.  A (rank-2) Killing--St\"{a}ckel tensor is a symmetric tensor $K_{\mu \nu} = K_{(\mu \nu)}$ that satisfies $\nabla_{(\mu} K_{\nu \rho)} = 0$.  A (rank-2) conformal Killing--St\"{a}ckel tensor is a symmetric tensor $Q_{\mu \nu} = Q_{(\mu \nu)}$ that satisfies $\nabla_{(\mu} Q_{\nu \rho)} = q_{(\mu} g_{\mu \nu)}$ for some $q_\mu$, given in 4 dimensions by $q_\mu = \tf{1}{6} (\pd_\mu Q{^\nu}{_\nu} + 2 \nabla_\nu Q{^\nu}{_\mu})$.

For black hole solutions of supergravity, usually only the string frame metric admits Killing tensors, whereas the Einstein frame metric usually only admits conformal Killing tensors \cite{{Chow:2008fe}}.  Here we note that in general, the string frame metric has a Killing--St\"{a}ckel tensor given by
\begin{align}
\wtd{K}^{\mu \nu} \, \pd_\mu \, \pd_\nu & = \fr{1}{r^2 + u^2} \bigg[ \bigg( \fr{u^2 W_r^2}{R} + \fr{r^2 W_u^2}{U} \bigg) \, \pd_t^2 - a \bigg( \fr{u^2 L_r}{R} + \fr{r^2 L_u}{U} \bigg) \, 2 \, \pd_t \, \pd_\phi + a^2 \bigg( \fr{r^2}{U} - \fr{u^2}{R}\bigg) \, \pd_\phi^2 \nnr
& \qquad - u^2 R \, \pd_r^2 + r^2 U \, \pd_u^2 \bigg] .
\end{align}
It is generically irreducible, i.e.\ not a linear combination of the metric and products of Killing vectors.  In general, if a metric $\df \wtd{s}^2$ possesses a Killing--St\"{a}ckel tensor $\wtd{K}_{\mu \nu}$, then for any conformally related metric $\df s^2$ there is an induced conformal Killing--St\"{a}ckel tensor with components given by $Q^{\mu \nu} = \wtd{K}^{\mu \nu}$; see e.g.\ \cite{Rani:2003br}.  In particular, the string frame Killing--St\"{a}ckel tensor induces a conformal Killing--St\"{a}ckel tensor for the Einstein frame metric.  Note that the existence of a conformal frame admitting a Killing--St\"{a}ckel tensor is a more restrictive condition than the existence of a conformal Killing--St\"{a}ckel tensor in Einstein frame.  This conformal Killing--St\"{a}ckel tensor was identified for the subcases with 4 electric charges in \cite{Chow:2008fe}, and for the non-extremal rotating Kaluza--Klein black hole in \cite{Keeler:2012mq}.

If we specialize to $L_r = W_r$ and $L_u = W_u$, then we can write without` loss of generality $W = W_r + W_u$.  Then the Einstein frame metric is of the form
\be
\df s^2 = - \fr{R}{W_r + W_u} \bigg( \df t + \fr{W_u}{a} \, \df \phi \bigg)^2  + \fr{U}{W_r + W_u} \bigg( \df t - \fr{W_r}{a} \, \df \phi \bigg) ^2 + (W_r + W_u) \bigg( \fr{\df r^2}{R} + \fr{\df u^2}{U} \bigg) .\label{sp}
\ee
This class of metrics has been studied in detail \cite{Chow:2013gba}, and has the property that both the string frame and Einstein frame metrics possess Killing--Yano tensors with torsion.  It implies that both the Einstein and string frame metrics admit Killing--St\"ackel tensors. The class of metrics includes the general black hole metric truncated to $-\im X^0 X^1$ supergravity, by setting the gauge fields pairwise equal, say $(\gd_1, \gc_1) = (\gd_4, \gc_4)$ and $(\gd_2, \gc_2) = (\gd_3, \gc_3)$.


\subsection{Geodesics}


The Killing tensor in string frame guarantees the complete integrability of geodesic motion in this frame, which we now demonstrate explicitly.  In string frame, the Hamilton--Jacobi equation for geodesic motion is
\be
\fr{\pd S}{\pd \lambda} + \fr{1}{2} \wtd{g}^{\mu \nu} \, \pd_\mu S \, \pd_\nu S = 0 ,
\ee
where $S$ is Hamilton's principal function, $\pd_\mu S = p_\mu = \df x_\mu / \df \gl$, $p_\gl$ are momenta conjugate to $x^\mu$, and $\gl$ is an affine parameter.  Consider the ansatz
\be
S = \tf{1}{2} \mu^2 \gl - E t + L \phi + S_r (r) + S_u (u) .
\ee
The constants $p_t = - E$ and $p_\phi = L$ are momenta conjugate to the ignorable coordinates $t$ and $\phi$, related to energy and angular momentum.  The particle mass is $\mu$, so that $p^\mu p_\mu = - \mu^2$.  The components $(r^2 + u^2) \wtd{g}^{\mu \nu}$ are additively separable into functions of $r$ and of $u$, and so the Hamilton--Jacobi equation is additively separable.  Explicitly, we have
\be
\bigg( \fr{W_u^2}{U} - \fr{W_r^2}{R} \bigg) E^2 + 2 a \bigg( \fr{L_r}{R} + \fr{L_u}{U} \bigg) E L + a^2 \bigg( \fr{1}{U} - \fr{1}{R} \bigg) L^2 + R \bigg( \fr{\df S_r}{\df r} \bigg) ^2 + U \bigg( \fr{\df S_u}{\df u} \bigg) ^2 + \mu^2 (r^2 + u^2) = 0 ,
\label{HamiltonJacobi}
\ee
and so
\begin{align}
\fr{\df S_r}{\df r} & = \fr{1}{R} \sr{W_r^2 E^2 - 2 a L_r E L + a^2 L^2 - (C + \mu^2 r^2) R} , \nnr
\fr{\df S_u}{\df u} & = \fr{1}{U} \sr{- W_u^2 E^2 - 2 a L_u E L - a^2 L^2 + (C - \mu^2 u^2) U} ,
\end{align}
where $C$ is a separation constant.  We then determine $r(\gl)$ and $u (\gl)$ by integrating
\begin{align}
\fr{\df r}{\df \gl} & = \wtd{g}^{r r} p_r = \fr{R}{r^2 + u^2} \fr{\df S_r}{\df r} , & \fr{\df u}{\df \gl} & = \wtd{g}^{u u} p_u = \fr{U}{r^2 + u^2} \fr{\df S_u}{\df u} .
\end{align}
Finally, we determine $t (\gl)$ and $\phi (\gl)$ by integrating
\begin{align}
\fr{\df t}{\df \gl} & = \wtd{g}^{t t} p_t + \wtd{g}^{t \phi} p_\phi = \fr{E}{r^2 + u^2} \bigg( \fr{W_r^2}{R} - \fr{W_u^2}{U} \bigg) - \fr{a L}{r^2 + u^2} \bigg( \fr{L_r}{R} + \fr{L_u}{U} \bigg) , \nnr
\fr{\df \phi}{\df \gl} & = \wtd{g}^{t \phi} p_t + \wtd{g}^{\phi \phi} p_\phi = \fr{a E}{r^2 + u^2} \bigg( \fr{L_r}{R} + \fr{L_u}{U} \bigg) + \fr{a^2 L}{r^2 + u^2} \bigg( \fr{1}{U} - \fr{1}{R} \bigg) .
\end{align}
In Einstein frame, generically only the $\mu = 0$ massless Hamilton--Jacobi equation separates.


\subsection{Klein--Gordon equation}


Separability of the massless Klein--Gordon equation makes the analysis of \cite{Castro:2010fd} applicable to the general black hole of $\cN = 8$ supergravity, which will therefore admit hidden conformal symmetries in the near-horizon region. 

The massive Klein--Gordon equation for the Einstein frame metric is
\be
\square \Phi = \fr{1}{\sr{-g}} \pd_\mu (\sr{-g} g^{\mu \nu} \pd_\nu \Phi) = \mu^2 \Phi .
\ee
Consider the ansatz
\be
\Phi = \Phi_r (r) \Phi_u (u) \expe{\im (k \phi - \omega t)} .
\ee
Then the Klein--Gordon equation gives
\be
\mu^2 W = \fr{\omega^2 W_r^2 - 2 a \omega k L_r + a^2 k^2}{R} - \fr{\omega^2 W_u^2 + 2 a \omega k L_u + a^2 k^2}{U} + \fr{1}{\Phi_r} \fr{\df}{\df r} \bigg( R \fr{\df \Phi_r}{\df r} \bigg) + \fr{1}{\Phi_u} \fr{\df}{\df u} \bigg( U \fr{\df \Phi_u}{\df u} \bigg) .
\ee
In the particular case where the Einstein metric takes the form \eqref{sp}, such as for generic black holes of $-\im X_0 X_1$ supergravity, the  $\mu \neq 0$ massive Klein--Gordon equation in Einstein frame separates. Generically, there is separation only in the massless case $\mu = 0$, leading to
\begin{align}
& \fr{\df}{\df r} \bigg( R \fr{\df \Phi_r}{\df r} \bigg) + \bigg( \fr{\omega^2 W_r^2 - 2 a \omega k L_r + a^2 k^2}{R} + C \bigg) \Phi_r = 0 , \nnr
& \fr{\df}{\df u} \bigg( U \fr{\df \Phi_u}{\df u} \bigg) - \bigg( \fr{\omega^2 W_u^2 + 2 a \omega k L_u + a^2 k^2}{U} + C \bigg) \Phi_u = 0 ,
\end{align}
where $C$ is an integration constant.  Specializing to the black hole solutions we constructed, the radial equation has regular singular points at the locations of the horizons, $r = r_\pm$, and an irregular singular point at infinity, similar to what happens for the Kerr solution.  The solutions are Heun functions.  The angular equation involving $u$ can be analyzed similarly.


\section{Conclusion and further directions}
\label{conclusion}


We have constructed a generating solution for the most general stationary, asymptotically flat black hole of $\cN = 8$ supergravity. We checked that this black hole reduces in specific subcases to all previously known solutions of the $STU$ model with 4 independent (combinations of) electromagnetic charges \cite{Cvetic:1996kv, Chong:2004na, LozanoTellechea:1999my, Compere:2010fm, Giusto:2007tt}. Unlike many other treatments of $STU$ supergravity, we have emphasized the 4-fold permutation symmetry of the gauge fields in the 3-dimensional coset model, not just the triality symmetry.  We discussed several extremal limits of interest, but a comprehensive examination of all extremal limits of our solution remains to be done. The generic black hole that we constructed admits a conformal Killing--St\"{a}ckel tensor, and the massless Klein--Gordon equation separates, so we can deduce the presence of hidden conformal symmetries in the near-horizon region. 

The solution generating technique that we detailed is general and could be used for a wider class of stationary seed solutions, beyond the Kerr--Taub--NUT solution that we used.  Different choices of group element can be used, allowing for more general asymptotic behavior.  One example is the application to ``subtracted geometries'' \cite{Cvetic:2011hp, Cvetic:2011dn, Baggio:2012db}, obtained by solution generating techniques in \cite{Cvetic:2012tr, Virmani:2012kw}.  Another example is obtaining charged black holes in a magnetic Melvin universe \cite{Cvetic:2013roa}.  The techniques can also be applied to the various theories in 5 and 6 dimensions that we discussed.

The issue of black hole uniqueness has not been fully addressed (see \cite{Chrusciel:2012jk} for a recent review).  It was shown in \cite{Breitenlohner:1987dg} that, with certain assumptions, all charged black holes in coset models lie in the orbit of the Kerr black hole. These assumptions were clarified in \cite{Hornlund:2011ga} in the static case, where it was shown that all scalar fields should be regular on the horizon in order to apply the theorem of \cite{Breitenlohner:1987dg}. Clarifying the theorem of \cite{Breitenlohner:1987dg} in the stationary case seems a natural step to prove uniqueness. 

Under general assumptions, stationary 4-dimensional black holes are axisymmetric \cite{Hawking:1971vc}, so can be Kaluza--Klein reduced to 2 spatial dimensions.  For Einstein gravity and Einstein--Maxwell theory, inverse scattering techniques can then be used to generate solutions, as reviewed in \cite{BelinskiVerdagauer}.  These techniques can be generalized to certain theories of gravity coupled to matter, in particular supergravities.  They have been developed for the $S^3$ supergravity in \cite{Figueras:2009mc}, and more generally for the $STU$ supergravity in \cite{Katsimpouri:2013wka}.  One may gain more insights into the algebraic structure of the general black hole solution by deriving it using these techniques.

Inverting the relation between conserved charges and auxiliary parameters that parameterize the 4-dimensional fields would allow for expressing the entropy, or equivalently the $F$-invariant that we defined, in terms of physical charges.  Our formula for the entropy of a general non-extremal black hole is not manifestly invariant under $\SL(2, \bbR)^3$ or triality.  We therefore are unable to provide here a manifestly $\textrm{E}_{7 (7)}$-invariant entropy formula for non-extremal black holes in $\mathcal{N} = 8$ supergravity.  Even in the simpler case of Kaluza--Klein theory, i.e.\ reduction of 5-dimensional Einstein gravity, the $F$-invariant for the dyonic black hole takes an intricate form that we were able to present.  We leave this difficult algebraic problem for future investigations.

We checked that the first law of thermodynamics closes both at the outer and inner horizon and that the Smarr formula holds at the outer and inner horizons. We derived a generalization of the quadratic mass formula in the presence of rotation and NUT charge. We also presented some relationships between physical charges defined at the outer and inner horizon that generalize previously known subcases. A microscopic understanding of these relationships remains to be uncovered.

Extremal black holes have been of interest recently with regards to the Kerr/CFT conjecture and its generalizations. The general black hole admits two distinct extremal rotating limits, the fast and slow rotating cases. In each case, similar to each subset of solutions that has been previously studied under that viewpoint, we reproduced all expected properties of extremal black holes, such as the existence of an $\SL(2,\mathbb R) \times \textrm{U} (1)$ symmetric near-horizon region and the Cardy form of the entropy. We noted the property that the generic near-horizon metric of extremal slow rotating black holes only depends upon the angular momentum and the quartic invariant. These results, if combined with a general asymptotic symmetry group analysis, would allow a microscopic counting of these extremal black holes. 

Several avenues for microscopically accounting for the entropy of specific non-extremal black holes in $STU$ supergravity have been proposed \cite{Horowitz:1996ac, Emparan:2006it, Horowitz:2007xq, Castro:2010fd}. It would be very interesting to try to unify these approaches and propose a microscopic model for the general black hole.

We have given a generating solution for the most general black hole of maximal supergravity in four dimensions.  Without the complication of magnetic charges and with fewer gauge fields, the same had been done a long time ago for black holes in maximal supergravity in five dimensions \cite{Cvetic:1996xz} and higher dimensions \cite{Cvetic:1996dt}.  Black rings are a further class of exact solutions in five dimensions, and are known in Einstein gravity with two independent rotations \cite{Pomeransky:2006bd}.  Several charged generalizations are known; see \cite{Rocha:2013qya} for references. Using U-dualities, a generating black ring solution for maximal supergravity is expected to involve 21 parameters, including mass, 2 angular momenta, 3 electric monopole charges, and 15 dipole charges \cite{Larsen:2005qr}.  Its construction would be a formidable task, and even its truncation to the 5-dimensional $STU$ supergravity is not known.

Partial generalizations to asymptotically AdS black holes in the $\textrm{U}(1)^4$ truncation of maximal $\cN = 8$, $\SO(8)$ gauged supergravities (including the recently discovered one-parameter family of $\omega$-deformed theories \cite{Dall'Agata:2012bb, Lu:2014fpa}) have been found; see \cite{Chow:2013gba, Gnecchi:2013mja, Halmagyi:2013uza}.  The asymptotically flat solution presented here has been generalized in \cite{Chow:2013gba} to two classes of asymptotically AdS solutions: static solutions with 4 independent electric charges and 4 independent magnetic charges; and rotating solutions with pairwise equal gauge fields, generalizing the solution of $-\im X^0 X^1$ supergravity, which has 2 independent electric charges and 2 independent magnetic charges.  However, they are difficult to find, since the solution generating techniques of ungauged supergravity rely on hidden symmetries.  These symmetries of the bosonic theory are mostly broken in gauged supergravity by a scalar potential, in $STU$ supergravity from $\SL(2, \bbR)^3$ to $\SO(2)^3$ \cite{Cvetic:1999xp} (see also \cite{Halmagyi:2013uza}).  The most general AdS generalizations of our ungauged solutions remain to be found. One guide to finding these solutions is that they are expected to involve metrics that allow separability.  The class of metrics that we defined that admit a Killing--St\"ackel tensor in string frame might therefore be useful in this context.



\section*{Acknowledgments}


The work of D.C.\ was partially supported by the ERC Advanced Grant ``SyDuGraM'', by IISN-Belgium (convention 4.4514.08) and by the ``Communaut\'e Fran\c{c}aise de Belgique" through the ARC program. During a large part of the project, G.C.\ was supported by NSF grant 1205550 as a post-doctoral researcher at Harvard University. He is a Research Associate of the Fonds de la Recherche Scientifique F.R.S.-FNRS (Belgium) and is currently supported by the ERC Starting Grant 335146 ``HoloBHC''. We gratefully thank the Centro de Ciencias de Benasque Pedro Pascual for its warm hospitality and A.\ Lupsasca for his encouragements.

\end{document}